\documentclass[longauth]{aa} 

\usepackage{graphicx}
\usepackage{tablefootnote}

\usepackage{xcolor}
\usepackage[colorlinks=true, linkcolor=blue, citecolor=blue, urlcolor=blue]{hyperref}

\usepackage{caption}
\usepackage{txfonts}

\usepackage{soul}
\usepackage{xcolor}
\usepackage{amsmath}

\usepackage{supertabular,booktabs}

\usepackage{caption}
\usepackage{subcaption}
\usepackage{longtable}
\usepackage{rotating}
\usepackage{lscape}
\usepackage{amsmath}

\newcommand{\msun}{$M_{\odot}$}
\newcommand{\rsun}{$R_{\odot}$}
\newcommand{\lsun}{$L_{\odot}$}
\newcommand{\teff}{$T_\mathrm{eff}$}
\newcommand{\logg}{$\log g$}
\newcommand{\logy}{$\log n(\mathrm{He}) / n(\mathrm{H})$}

\newcommand{\kms}{km\,s$^{-1}$}

\newcommand{\atlas}{\textsc{Atlas}{\footnotesize12}}
\newcommand{\detail}{\textsc{Detail}}

\begin{document}

   \title{A 500 pc volume-limited sample of hot subluminous stars}

   \subtitle{III. The short-period binary population}

   \author{H. Dawson
          \inst{1}
          \and
          M. Dorsch\inst{1}
          \and
          J. Munday \inst{1,2}
          \and
          S. Geier\inst{1}
          \and
          F. Mattig \inst{1}
          \and
          M. Pritzkuleit\inst{1}
          \and          
          D. Benitez-Palacios\inst{4}
          \and
          M. {Vu{\v{c}}kovi{\'c}}\inst{4} 
          \and
          K. Deshmukh\inst{5}
          \and
          A. Bhat \inst{1}
          \and
          U. Heber\inst{3}
          \and
          I. Pelisoli\inst{2,1}
          \and
          R. Raddi \inst{6}
          \and
          P. Fernandez-Schlosser \inst{4}
          \and
          A. Durán-Reyes \inst{4}
          \and
          E. Arancibia-Rojas \inst{4}
          \and
          A. Bobrick \inst{7,8}
          \and
          V.~Schaffenroth \inst{9,1}
          \and
          G. T. Jones \inst{2,10}
}   
          
   \institute{Institute for Physics and Astronomy, University of Potsdam, Karl-Liebknecht-Str. 24/25, 14476 Potsdam, Germany\\
              \email{harry.b.a.dawson@gmail.com}
         \and
            Department of Physics, University of Warwick, Gibbet Hill Road, Coventry CV4 7AL, UK
         \and
            Dr. Remeis-Sternwarte and ECAP, Astronomical Institute, University of Erlangen-Nürnberg, Sternwartstr. 7, D-96049 Bamberg, Germany
         \and
             Instituto de Física y Astronomía, Universidad de Valparaíso, Gran Bretaña 1111, Playa Ancha, Valparaíso 2360102, Chile
         \and
            Institute of Astronomy, KU Leuven, Celestijnenlaan 200D, B-3001 Leuven, Belgium 
         \and
             Universitat Politècnica de Catalunya, Departament de Física, c/ Esteve Terrades 5, 08860 Castelldefels, Spain 
         \and 
             School of Physics and Astronomy, Monash University, Clayton, Victoria 3800, Australia
         \and
            ARC Centre of Excellence for Gravitational Wave Discovery -- OzGrav, Australia
         \and
            Thüringer Landessternwarte Tautenburg, Sternwarte 5, D-07778 Tautenburg, Germany
         \and
             Department of Physics \& Astronomy, Allen Building, 30A Sifton Rd, University of Manitoba, Winnipeg MB R3T 2N2, Canada
}

   \date{Received June 25, 2026 / Accepted August 13, 2026}
 
  \abstract
  {Hot subdwarf stars of spectral types O and B (sdO/B) in binaries form as the immediate products of substantial mass loss at or near the tip of the red giant branch and offer powerful constraints on binary-star evolutionary models. However, the details of some of their formation channels remain missing. Motivated by this, we present a comprehensive analysis of the short-period binary population in the 500~pc volume-limited sample of hot subdwarfs, enabled by precise parallax measurements from \emph{Gaia} data release 3 (DR3). In addition to the 45 previously known binaries within 500~pc, this work identifies 50 new single-lined radial-velocity variable systems, of which 34 have orbital solutions with periods between 0.07 and 22 days. We derive an overall short-period binary fraction of $34.7^{+2.8}_{-2.9}\%$ for the full sample of 301 hot subdwarfs within 500~pc. This value was corrected for orbital inclination and detection efficiency to account for observational incompleteness. The newly solved short-period binaries predominantly occupy the 1-20 day orbital period range, a parameter space previously under-represented in the literature. We identify five new reflection-effect systems, three ellipsoidal modulation systems, a newly solved HW~Vir system, and a new triple candidate. In line with previous work, the sdB and sdO binary fractions are very similar. The sdOB class, however, exhibits a binary fraction roughly half that of the sdB and sdO stars and have orbital periods of more than one day, suggesting that this population follows a different evolutionary pathway or evolutionary stage. Underluminous hot subdwarfs located below the canonical extreme horizontal branch in the Hertzsprung-Russell diagram show a binary fraction comparable to the remaining the sdB and sdO stars but exclusively host white dwarf (WD) companions. These WD companions also appear to be more massive than other WD companions in the sample and are mainly found on orbital periods of less than one day. Leveraging our complete binary population, we provide the first volume-complete estimates for the birthrates of sub-populations in the sample. We determine a Galactic merger rate of $2.5\pm1.5\times10^{-5}$yr$^{-1}$ for sdO/B binaries, sufficient to explain no more than $\sim12\%$ of the eHe-sdO population. The birthrates of two type~Ia supernova (SN~Ia) progenitor channels are also derived. From our sample, we infer that hot subdwarf binaries could account for up to $2.5^{+0.7}_{-0.5}$\% of the observed Galactic SN~Ia rate. 
  }


   \maketitle


\section{Introduction}

Hot subdwarf stars of spectral types O and B (sdO/B) occupy the extreme end of the horizontal branch \citep[EHB;][]{Greenstein_1974ApJS...28..157G, Newell_1973, Heber_1984} in the Hertzsprung-Russell diagram (HRD), a region between the main sequence (MS) and the white dwarf (WD) cooling track. They are understood to be core or shell helium-burning red-giant remnants that have lost nearly their entire hydrogen envelopes \citep{Heber_2009, Heber_2016_review, Heber_review_2024arXiv241011663H}.

Despite decades of study, the formation of hot subdwarfs is not yet fully understood. Radial-velocity (RV) surveys have established that at least $\sim$30\% reside in short-period binaries with orbital periods up to $\approx$30~days. These systems host companions that are either WDs or cool stellar or substellar objects such as M dwarfs (dMs) or brown dwarfs (BDs) \citep[e.g.][]{Schaffenroth_2022_1, Schaffenroth_2023_2}. This high binary fraction strongly implies common envelope (CE) evolution as a major formation channel. This can occur either through a single dynamically unstable CE phase, during which the companion becomes engulfed by the red-giant envelope, or through a two-stage process. The latter may involve an initial phase of stable Roche-lobe overflow followed by a subsequent CE phase, or two CE phases \citep{Han_2003, Podsiadlowski_2008ASPC..392...15P}. A handful of the resulting short-period systems qualify as candidate progenitors of type~Ia supernovae (SNe Ia) via double-detonation or double-degenerate merger channels \citep[see e.g. ][]{Geier_2013AA...551L...4G, Pelisoli_2021NatAs...5.1052P, Deshmukh_2024MNRAS.527.2072D}. However, they may contribute only a small amount towards the observed SN~Ia rate. 

Around 30\% of hot subdwarfs are found with long-period (500 to 1500~days) MS A-, F-, G-, or K-type companions, detected primarily through an infrared (IR) excess in their spectral energy distributions \citep[SEDs;][]{Stark_Wade_2003AJ....126.1455S, Vos_2018}. Such systems are well explained by a single episode of stable Roche-lobe overflow \citep{Han_2002, Chen_2013MNRAS.434..186C, Vos_2020AA...641A.163V}. 

Historically, $30-40\%$ of hot subdwarfs, however, have appeared to be single stars \citep{Napiwotzki_2001AN....322..411N, Geier_2022}. While single-star evolutionary pathways have been proposed \citep{Sweigart_1997, Battich_2018AA...614A.136B}, He-rich hot subdwarfs in particular are often explained by a merger scenario involving two He-core WDs \citep{Webbink_1984ApJ...277..355W, Yu_2021MNRAS.504.2670Y} or by hybrid mergers \citep[CO+He-WD;][]{Justham_2010}. The origin of apparently single, hydrogen-rich sdBs remains harder to account for theoretically, even though CE mergers involving red giant branch with dM companions \citep{Politano_2008}, or He-WD and low-mass MS mergers \citep{Clausen_2011} have been proposed. Although in most cases wider systems with small RV variability (RVV) cannot be ruled out, high-resolution spectroscopic follow-up has excluded companions in at least some systems \citep{Silvotti_2020arXiv200204545S}, leaving the formation of these objects an open question.

Progress in disentangling these formation channels has long been hampered by strong observational selection effects, especially driven by distance uncertainties. Dedicated RV surveys searching for short-period binaries began in earnest with \citet{Maxted_2001} and have since continued through the work of \citet{Copperwheat_2011MNRAS.415.1381C}, the MUCHFUSS project \citep{Geier_2011AA...526A..39G, Kupfer_2015}, and more recently using SDSS and LAMOST data \citep[e.g.][]{Geier_2022, He_2025AA...693A.121H}. Large-scale ground-based photometric surveys such as ATLAS and ZTF, together with space-based photometry from TESS, have greatly increased the number of short-period hot subdwarf binaries identified through binary signatures in their light curves (LCs), with sensitivities extending to periods of several days \citep[see e.g.][]{Schaffenroth_2021MNRAS.501.3847S,Kupfer_2021MNRAS.505.1254K, Schaffenroth_2022_1, Schaffenroth_2023_2}. 
The resulting compilations \citep{Geier_2020, Culpan_2022} are therefore subject to complex and often poorly quantified selection biases, with a strong preference for short-period systems.

The \emph{Gaia} mission has fundamentally changed this picture. Precise parallaxes from Gaia's early data release 3 \citep[EDR3; ][]{Gaia_mission_2016AA...595A...1G} enable the conversion of atmospheric parameters -- derived over decades of spectroscopic studies -- into fundamental stellar properties, thereby placing hot subdwarfs on the physical HRD. This enabled the determination of mass distributions for different hot subdwarf subtypes \citep{Luo_2021, Latour_2026AA...705A.248L, Heber_HQS_2026A&A...708A.115H}, from flux-limited samples. The excellent precision of Gaia parallaxes also made it possible to construct the first well-defined, volume-limited sample of hot subdwarfs within 500~pc \citep[Paper I; ][]{Dawson_2024AA...686A..25D}, substantially reducing the selection effects inherent to flux-limited surveys. From this sample, we derived a local space density significantly lower than that predicted by binary population synthesis models \citep{Han_2002, Han_2003, Clausen_2012ApJ...746..186C, Nicolas_2_2025PASA...42...12R}, although the most recent detailed models \citep{Vos_2020AA...641A.163V} derived rates that are close (to within a few 10\%) to the observed ones. A subsequent atmospheric and kinematic analysis of the same volume-limited sample \citep[Paper II; ][]{Dawson_2026AA...707A...6D} revealed a mass distribution somewhat inconsistent with theoretical predictions that most sdO/B stars descend from low-mass progenitors, motivating a volume-complete characterisation of the short-period binary fraction as a crucial next step.

Addressing this gap is the focus of the present work, which serves as the third instalment of the 500~pc survey.
Short-period hot subdwarf binaries can be identified through RV variations and characteristic photometric modulations in their LCs. These include ellipsoidal deformation and Doppler boosting (for white dwarfs or other compact companions), reflection effects (for cool MS or substellar companions), or eclipses. Constraining the short-period binary properties in our sample provides stringent tests of binary evolution models, in particular the poorly constrained CE evolution. The spectroscopic observations and data reduction procedures are described in Sect.~\ref{target_selection}. The analysis methods, including RV measurements, photometric analysis, and the determination of orbital solutions, are presented in Sect.~\ref{methods}. Our results are provided in Sect.~\ref{results}, followed by a discussion and comparison with other observational studies in Sect.~\ref{discussion}. Finally, in Sect.~\ref{birthrates}, we present birthrate estimates for several sub-populations within our sample, including their inferred contributions to the Galactic merger rate and the SN~Ia rate.

\section{Target selection, observations, and data reduction}
\subsection{Target selection}
\label{target_selection}
Our 500~pc sample presented in Paper~I was selected from a catalogue of candidates for hot subluminous stars \citep{Culpan_2022} based on precise \emph{Gaia} colours and absolute magnitudes, with detailed follow-up observations and atmospheric analysis presented in Paper~II. In this third study, we performed a binary analysis of the single-lined hot subdwarfs in the same sample, using all the individual medium- to high-resolution spectra acquired across Papers I and II for which reliable RVs can be measured.\footnote{One of the identified RV-variable systems, however, exhibits an IR excess, suggesting the presence of an early dM, or late K-type companion \citep[see e.g. GALEX J2205-3141 in][]{Schaffenroth_2023_2}. This system, is not included in the short-period binary fractions presented in this paper but is discussed in Sect.~\ref{highlighted_systems} as a possible triple candidate.}

We excluded five stars from our analysis as they turned out to be pre-extremely low-mass white dwarfs (pre-ELM WDs): EVR-CB-001 \citep{Ratzloff_2019ApJ...883...51R}, [PS72]97 \citep{Kosakowski_2023ApJ...950..141K}, 2MASS J02150626+0155041 \citep{Kosakowski_2023ApJ...950..141K}, CPD-20\,1123 \citep{Naslim_2012MNRAS.423.3031N, Kupfer_2015, Dawson_2026AA...707A...6D}, and Gaia EDR3 4318061098980872960. The last object is newly identified as a pre-ELM in this work. Although excluded from the subsequent analysis provided here, its orbital solution is presented in the appendix (see Appendix~\ref{new_sdA}). The final number of systems analysed in this paper is 253. 

\subsection{Observing strategy and acquired spectra}
\label{observing_strategy}
Our observing strategy was to first identify RVV for each target. Archival spectra were acquired from online databases, including LAMOST \citep{LAMOST_2022yCat.5156....0L}, the ESO archive\footnote{http://archive.eso.org/eso/eso\_archive\_main.html}, and MAST\footnote{https://archive.stsci.edu/} and then fitted to derive RV measurements. Identified binaries were then followed-up and solved, whereas for apparently non-RVV stars we acquired multi-epoch high-resolution spectra to ascertain their nature. 

At least three epochs were taken of each target: two separated by 2-5 hours and a third by at least one day. This cadence maximises sensitivity to short-period binaries, as the vast majority of close companions to sdO/B stars have orbital periods shorter than 10 days, with a typical period of order one day and RV amplitudes ranging from $\sim30\,$\kms\ to $\gtrsim 450\,$\kms \citep[e.g.][]{Schaffenroth_2022_1, Copperwheat_2011MNRAS.415.1381C, Kupfer_2015}. Extensive observational follow-up to solve for the identified close binaries has been conducted since August 2023 and is summarised in table~1 in Paper~II. More recent medium-resolution spectra were acquired using the Goodman spectrograph at the Southern Astrophysical Research Facility (SOAR/Goodman; programmes 2026A-730421 and 2025B-972608), and the Alhambra Faint Object Spectrograph and Camera at the Nordic Optical Telescope (NOT/ALFOSC; programme 71-201). Using a $1.0^{\prime \prime}$ slit with the 930 1/mm grating on SOAR, and a $0.5^{\prime \prime}$ slit with Grism 18 on the NOT, each setup achieves spectral resolutions of 3.2~\AA\ and 2.2~\AA, respectively, and covers the optical range from approximately 3600 to 5200~\AA. Spectra were also acquired using the ESO Faint Object Spectrograph and Camera 2 at the New Technology Telescope (EFOSC2/NTT). Paired with Grism~19 with a 1.0" slit the wavelength range 4441 to 5114~\AA\ is covered with a spectral resolution of 1.5~\AA. In addition to S/N and resolution, the fitted RV precisions ($\sigma_{v_\mathrm{rad}}$) are, in practice, limited by the broad and relatively sparse hydrogen Balmer and helium lines characteristic of hot subdwarfs, as well as the absence of strong metal lines in these hot stars. The fitted RV precisions per instrument used in this programme are summarised in Table~\ref{tab:rv_accuracy}. For more details on the fitted RV precisions in this work, see Sect.~\ref{radial_velocity_uncertainties}. 

The follow-up medium- and high-resolution spectroscopy of apparently non-RVV systems was carried out using the High-Efficiency and High-Resolution Mercator Echelle Spectrograph \citep[HERMES;][]{hermes_2011AA...526A..69R} on the 1.2 m Mercator telescope, the FIbre-fed Echelle Spectrograph (FIES) at the Nordic Optical Telescope (NOT), and X-shooter on the Very Large Telescope (VLT). HERMES provides a spectral resolving power of $R = 85~000$, while FIES offers $R = 25~000$ in its low-resolution mode. For our hot and compact stars with their few and broad spectral lines, this corresponds to fitted RV precisions of around $1$ \kms\ and $2$ \kms for HERMES and FIES, respectively. Using slit widths of $0.5^{\prime\prime}$ and $0.7^{\prime\prime}$ for the UVB and VIS arms of X-shooter, respectively, we achieve $R$=9861/18340 and precisions down to $\sim5~$\kms. 

In total, our spectroscopic dataset comprises 3850 spectra of 253 single-lined hot subdwarf stars. 160 ($\sim63\%$) stars have at least three medium- to high-resolution data (HERMES, FIES, X-shooter, UVES, or FEROS). The data were reduced using a combination of PyRAF procedures \citep{PYRAF_2012ascl.soft07011S}, a Python-based implementation of IRAF \citep[Image Reduction and Analysis Facility;][]{IRAF_1986SPIE..627..733T}, the \textsc{Molly} package \citep{Marsh1989optimalExtraction,Marsh2019Molly}, available GitHUB pipelines \citep{Mattig2025_MIDIR}, and instrument-specific pipelines. All reduction methods include bias and flat-field corrections and wavelength calibration. 

\subsection{Archival RV measurements}
\label{further_observational_material}
Where archival spectra were unavailable (Sect.~\ref{observing_strategy}), RV measurements of appropriate quality ($\lesssim \pm20$~\kms) based on spectra with sufficient resolution and precision were drawn directly from the literature \citep{Saffer_1994ApJ...432..351S,Moran_1999MNRAS.304..535M,Randall_2005ApJ...633..460R,Edelmann_2005AA...442.1023E,Ostensen_2010MNRAS.408L..51O,Naslim_CPD-20_2012MNRAS.423.3031N, Schaffenroth_2013AA...553A..18S,Schaffenroth_PHL457_2014AA...570A..70S,Sener_2014MNRAS.440.2676S,Kawka_2015,Schneider_2018, Latour_2018AA...609A..89L,Baran_2019MNRAS.489.1556B}.  
This effort resulted in 1193 additional RV measurements spanning nearly three decades, bringing the total number of epochs used in the subsequent analysis to 5106. The addition of published epochs significantly extends the time baseline of the observations, improving the orbital solutions of previously known binaries. When RV-shifts were identified when combining literature RV measurements with newly acquired spectra, follow-up spectra were secured to avoid false positives. New binary detections are therefore reported only on the basis of spectra acquired within our programme. Literature measurements were incorporated solely to improve the final orbital solution where necessary.

\section{Methods}
\label{methods}
\subsection{Synthetic spectra and radial-velocity measurements}
\label{methods_synthetic_spectra_and_radial_velocities}
We used the same model atmospheres and automated fitting procedure as described in Paper~II. The atmospheric parameters derived in Paper II (\teff, \logg, and \logy), which used the entire observed wavelength range, also included RV as a free parameter. These values were adopted for this study. Any further spectra obtained between the instalments were analysed in the same way. A $\chi^{2}$ minimisation technique was applied, as described by \citet{Irrgang_2014}. The model grid is based on \atlas\  \citep{Kurucz_1996} atmospheres and \detail\ \citep{Giddings1981} departure coefficients. The Surface code \citep{ButlerGiddings85} was used to synthesise the model spectra. The details of this hybrid local thermal equilibrium (LTE) / non-LTE approach can be found in \citet{Przybilla_2011JPhCS.328a2015P}. The specific model grids, so-called '$2^{nd}$ generation Bamberg models', used in this work are described in detail in \citet{Heber_HQS_2026A&A...708A.115H}.

\subsection{Radial-velocity uncertainties}
\label{radial_velocity_uncertainties}
For each instrumental setup, a systematic RV uncertainty was estimated. Where available, this was taken as the root mean square of the wavelength calibration residuals obtained during data reduction. For high-resolution spectrographs such as HERMES, UVES, and FIES, the intrinsic wavelength calibration accuracy is at the level of a few metres per second, well below the systematic floor introduced by our fitting methodology. For these instruments, the adopted systematic uncertainty therefore reflects limitations in the spectral fitting rather than the instrument itself and was estimated from the typical deviations observed in repeated measurements of RV standard stars over the course of the observing programme. These systematic contributions were added in quadrature to the formal statistical uncertainties to obtain the final RV error estimates and are provided in Table~\ref{tab:rv_accuracy}.

\begin{table}
\centering
\caption{Spectroscopic instrument summary.}
\label{tab:rv_accuracy}
\resizebox{\columnwidth}{!}{
\setstretch{1.1}
\begin{tabular}{lcccc}
\toprule\toprule
Telescope/Instrument  & Set-up & Spec. Count & $\Delta\lambda$  & $\sigma_{v_\mathrm{rad}}$ \\
  &  & & [\AA]&[\kms] \\ 
\midrule
INT/IDS/EEV10 & R1200B  & 1354 &1.0 & 5 \\
SOAR/Goodman & 930 M2 & 1209 &3.2 &  25 \\
NOT/ALFOSC &  Grism 18 & 280 & 2.2 & 15  \\
    NOT/FIES &  low-res & 69 & 25000$^\ast$ & 2  \\
Mercator/HERMES &  HRS & 223 &85000$^\ast$  & 1 \\
NTT/EFOSC2 & Grism 19 & 105 &1.5 & 10 \\
MPG/ESO/FEROS &  - & 547 &48000$^\ast$ & 1 \\
VLT/X-shooter & UVB/VIS  & 385 &9861/18350$^\ast$  & 5 \\
LAMOST DR10  &  LRS/MRS & 120 &3.05/0.7  & 15/5 \\
VLT/UVES & various  & 59 & various & 1--5 \\
\bottomrule
\end{tabular}
}
\tablefoot{Adopted RV uncertainties ($\sigma_{v_\mathrm{rad}}$) per instrument. Where available, values were estimated by adding in quadrature the root mean square of the wavelength calibration solution as well as the RV scatter of standard stars observed during the campaign and represent a lower limit on the total uncertainty. $^\ast$ Stated as resolving power $R = \lambda / \Delta \lambda$. }
\end{table}

\subsection{Photometry}
\label{photometry}
For each target, we searched for time-series photometry from the Transiting Exoplanet Survey Satellite \citep[TESS;][]{Ricker_tess_2015JATIS...1a4003R}, the Zwicky Transient Facility \citep[ZTF;][]{Bellm_ZTF_2019PASP..131a8002B}, and the Asteroid Terrestrial-impact Last Alert System \citep[ATLAS;][]{Tonry_ATLAS_2018PASP..130f4505T}. When suitable data were available, we computed Lomb-Scargle periodograms \citep{Zechmeister_2009AA...496..577Z} and visually inspected the corresponding LCs for evidence of periodic variability, indicative of a close companion (e.g. reflection effects or ellipsoidal modulation). Light curves dominated by noise were discarded. Known pulsators \citep{Uzundag_1_2024AA...684A.118U, Krzesinski_2025AA...700A..71K, Baran_2023A&A...669A..48B, Baran_2024A&A...686A..65B} were flagged so that pulsation signals were ignored in the periodograms and joint RV fitting. No new pulsators were detected. Furthermore, for well-solved binaries, the period range in the TESS, ZTF, and ATLAS periodograms were restricted to a $\pm 10\%$ window around the orbital period determined from the RV analysis to search for small signals induced by binarity. 

For all of our targets, we checked the field of view for TESS full-frame images (FFIs) and the CROWDSAP parameter when shorter 2-minute or 30-second cadence was available (78\% of the sample). Most of our targets returned a CROWDSAP value close to 1, meaning that almost all the flux belongs to the target star. There were several exceptions, including GD~1110 whose LC is dominated by the bright nearby eclipsing binary star HD~219869 \citep{hd219869_2021MNRAS.508.5687H}.  

\subsection{Criterion for binarity}
\label{criterion_for_binarity}
For each star, we tested whether the measured RVs significantly deviate from a constant systemic value (being the inverse-variance weighted mean) to determine if they were RVV or not. The chi-squared statistic was computed and compared to the mean value of all RVs of a given star. From this statistic, we tested the probability that the data agree with a constant RV hypothesis. A threshold of $\log_{10}p < -4$ -- where $p$ is the probability of the observed RV scatter arising by chance under a constant-velocity hypothesis -- was used to securely identify a binary star system \citep[also adopted in][and references therein]{Maxted_2001, Geier_2022}. Given our sample size, we would expect less than one outlier at a $3\sigma$ level. $\log_{10}p$ values between -4 and -2 (0.01 - 1\% probability) are considered as candidates.

\subsection{Determination of orbital solutions}
\label{determination_of_the_orbital_solutions}
For all RV-variable hot subdwarfs with at least five spectra, we performed a systematic period search to determine orbital solutions. With observational baselines spanning up to several years and trial periods as short as 30 minutes, evaluating millions of trial frequencies sequentially becomes computationally expensive. 
We computed a generalised Lomb-Scargle periodogram for each star, sampling uniformly in logarithmic space between frequencies of $48~\mathrm{d}^{-1}$ and $0.01~\mathrm{d}^{-1}$, corresponding to periods of 30 minutes to 100~days. 
The minimum number of frequency samples was set by the ratio of the observational baseline to the shortest trial period, ensuring adequate resolution of even the narrowest spectral peaks. 

At each trial frequency, we fitted the linearised sinusoidal model,
\begin{equation}
  v(t) \;=\; A\,\sin\!\left(\frac{2\pi~t}{P}\right)
         \;+\; B\,\cos\!\left(\frac{2\pi~t}{P}\right)
         \;+\; C,
  \label{eqn:linearised}
\end{equation}

\noindent yielding the best-fit semi-amplitude $K = \sqrt{A^2 + B^2}$, systemic velocity $v_\gamma = C$, and epoch of maximum velocity $t_0 = -(P/2\pi)\arctan(B/A)$ without iterative optimisation.\footnote{This linearisation exploits the  trigonometric identity,  $K\sin(2\pi t/P + \phi_0) \equiv A\sin(2\pi t/P) + B\cos(2\pi t/P)$,  where $A = K\cos\phi_0$ and $B = K\sin\phi_0$. Moving the phase  outside the brackets makes the model linear in $(A, B, C)$, so the globally optimal solution follows directly from the weighted normal equations without iteration. Rather than solving each trial period sequentially, we assembled the $3\times3$ weighted normal-equation system for all trial periods simultaneously and solved them using vectorised matrix operations as implemented in  \textsc{NumPy} \citep{Harris_2020Natur.585..357H}, which uses singular value decomposition internally.}

The resulting power spectrum was then normalised following the approach defined in \citet{Cumming_2004MNRAS.354.1165C}:

\begin{equation}
  z(f) = \frac{N - 3}{2}
         \cdot\frac{\chi^2_{\mathrm{null}} - \chi^2_{\mathrm{model}}(f)}
                   {\chi^2_{\mathrm{model}}(f)},
  \label{eqn:chisq_power}
\end{equation}

\noindent where $\chi^2_{\mathrm{null}}$ is the chi-square of a constant-velocity (flat) model derived from the weighted-mean of the RV measurements and $\chi^2_{\mathrm{model}}$ is the chi-square of the best-fit sinusoid at the trial period. $N$ is the number of epochs, and $3$ is the degrees of freedom of the model. 

To filter unphysical solutions, we applied a constraint based on the binary mass function and physical insight into our hot subdwarf systems.  For each candidate period, we computed the maximum permitted semi-amplitude, $K_{\mathrm{max}}$, assuming an edge-on orbit ($i = 90^\circ$) and a fixed primary mass of $M_{\mathrm{primary}} = 0.15~\mathrm{M}_\odot$, which is a conservative lower limit. For the companion, we note that hot subdwarfs were observed to host massive compact companions, such as massive WDs or even neutron stars \citep[NSs; e.g.][]{Geier_2023AA...677A..11G}. To ensure these rare systems were not filtered out of our analysis, we adopted a maximum companion mass of $M_{\mathrm{comp}} = 3.0~\mathrm{M}_\odot$ \citep[an approximate maximum mass of a NS, depending on the adopted equation of state;][]{Kalogera_1996ApJ...470L..61K}. We used the binary mass function defined as

\begin{equation}
K_{\mathrm{max}}^3 = \frac{2\pi G\, M_{\mathrm{comp}}^3}{P\,(M_{\mathrm{sdB}} + M_{\mathrm{comp}})^2}.
\label{eqn:Kmax}
\end{equation}
Candidate periods yielding fitted $K$ values exceeding $K_{\mathrm{max}}$ were assigned zero power in the periodogram.

For all systems we then assessed the significance of periodogram peaks through the bootstrap false alarm probability (FAP) procedure described in \citet{Munday2025dbl2}. The observed RVs were randomly reassigned 500 times among the original observation timestamps and the power spectrum recomputed at each iteration. This approach searches for randomly injected artificial periodicity from the window function of the observations, providing the null distribution from which FAP thresholds are defined. The 3$\sigma$ and 4$\sigma$ levels, corresponding to the $99.87$ and $99.997$ percentiles of this distribution, were determined from the merged set of resampled periodograms. All unique peaks above 3$\sigma$ from this physically constrained power spectrum were subjected to detailed non-linear fitting using the \texttt{lmfit} Python package \citep{Newville_2014zndo.....11813N} as a second check, whereas those with no peaks above 3$\sigma$ were not considered as real orbital solutions, at times leaving a binary unsolved. Multiple peaks were often detected above the 4$\sigma$ threshold, requiring closer inspection to identify the true orbital period. To address this, we first computed the spectral window function of the observation timestamps for every system and verified that the adopted period does not coincide with its dominant peaks. We discuss these marginal cases individually in Sect.~\ref{highlighted_systems}. 

Where two or more clearly separated aliases are present above the 4$\sigma$ threshold in the power spectrum and are clear of the window function artefacts, we report, alongside the formal uncertainty, the probability that the highest peak is the correct period, computed as the share of the total probability assigned to that peak relative to all peaks in the periodogram. In cases where a single dominant peak is broad and contains numerous closely spaced aliases, we report a period range instead of the formal uncertainty. This range is defined as the 68\% credible interval of this probability distribution within a fixed window around the peak, as indicated by the shaded-blue regions in the zoomed insets of the relevant power spectra.

To identify the most probable orbital period from the available RVs alone and assess parameter uncertainties, we employed the parallel-tempered Markov chain Monte Carlo (MCMC) sampler \texttt{ptemcee} \citep{Vousden_2016MNRAS.455.1919V}. This sampler enables efficient exploration of strongly multimodal period probability distributions. The final posterior was found in some cases to still be multimodal, with aliasing present within seconds of each other due to the large baseline of the observations. In such cases the highest-probability peak was selected, and the period range described above is reported in place of the MCMC uncertainty. For systems with a single, unambiguous peak, we report the 16th and 84th percentiles of the marginal posterior from the MCMC fitting (i.e. the central 68\% credible interval).
For every system, we verified that the MCMC sampler had converged by requiring the production length of the cold chain\footnote{In parallel-tempered MCMC, several chains are evolved simultaneously. The cold chain samples the true probability distribution and provides the final parameter estimates, whereas the hotter chains facilitate the exploration of alternative solutions and prevent convergence to local maxima.} to exceed 50 integrated autocorrelation times \citep[][]{emcee_hammer_2013PASP..125..306F}. We also required efficient exchange between the different temperatures to ensure that the sampler was not confined to a single orbital alias. As a supplementary cross-check, we computed the Gelman-Rubin convergence statistic \citep[$\hat{R}$;][]{Vehtari_rhat_2021BayAn..16..667V} across walkers to confirm consistent sampling of the posterior distribution. For systems in which a range of orbital periods is reported due to residual fine aliasing, we discuss its impact on the derived system parameters -- including the minimum companion mass where relevant -- together with the corresponding convergence diagnostics in Sect.~\ref{highlighted_systems} and Appendix~\ref{app:individual_systems}.

When a clear photometric signal was detected in the \textit{TESS}, \textit{ZTF}, or \textit{ATLAS} LC (high significance; $\mathrm{FAP} \lesssim 10^{-3}$) and affirmed to be of binary origin (e.g. characteristic variations caused by ellipsoidal modulations or reflection effects) by inspection, a photometry-based period prior was included to guide the MCMC towards periods consistent with both the RV and light-curve data. For these systems, we do not provide the RV periodogram as it becomes redundant. For all solved systems using only the RV data, however, we verified the photometry for small photometric signals possibly induced by binarity in a $\pm10\%$ window around the orbital period determined from the RV analysis. If a peak was found, we then fitted the RV data and photometry together to find a consistent solution \citep[also done in][]{Schaffenroth_2023_2}.

We also explored eccentric orbital solutions, allowing the eccentricity $e$ to vary up to $0.3$. While we detected several previously identified eccentric systems (see Appendix~\ref{appendix_eccentricities} for more details), the fits for all other cases tended towards circularity ($e=$~0), likely due to insufficient RV precision for most of our data. We, therefore, adopted circular orbits throughout.

\subsection{Light-curve modelling}
\label{light_curve_modeling}
Of the 34 newly solved systems, five show a reflection effect (one of which also shows a primary and secondary eclipse), three show ellipsoidal modulation, and eight show Doppler boosting. The ellipsoidal modulation and reflection effect systems were modelled with LCURVE \citep[see][for details]{Copperwheat_2010MNRAS.402.1824C}, whereas no new information can be extracted from the Doppler boosting systems. LCURVE has been successfully applied to close hot subdwarf binaries in previous studies \citep[see e.g. ][]{Schaffenroth_2021MNRAS.501.3847S, Schaffenroth_2023_2} and is well suited to our purpose. In this study, we employed a modern C\texttt{++} reimplementaion of LCURVE \citep{Mattig2025}\footnote{publicly available at  \url{https://github.com/Fabmat1/lcurve_re}}. The code represents each stellar component as a mesh of small, flat surface elements, each with a defined area, position, orientation, and brightness. Synthetic LCs were then computed by summing the flux contributions of all visible elements at each orbital phase, accounting for mutual eclipses.

For the hot subdwarf primary, we adopted a quadratic limb-darkening law using the coefficients of \citet{Claret_a_2020A&A...634A..93C}. By contrast, for the MS companion, we adopted those of \citet{Claret_2018AA...618A..20C}, selecting values closest to the atmospheric parameters of each star for the TESS passband. For systems exhibiting a reflection effect, the irradiated hemisphere of the companion was modelled using a blackbody approximation with the absorption factor fixed to unity such that all intercepted flux from the hot subdwarf is used to heat the companion's inner hemisphere. Because the contribution of the unirradiated side of the companion to the total flux is negligible, the companion temperature is poorly constrained by the LC and was therefore fixed to a typical dM value of $3000\pm1000$~K \citep[also adopted in][]{Schaffenroth_2023_2}. The effective temperature and surface gravity of the hot subdwarf were fixed to values from the spectroscopic analysis in Paper~II.

Since most LCs lack eclipses, the component radii and mass ratios cannot be determined from photometry alone. We therefore applied Gaussian priors on the radius and mass of the primary star derived in Paper~II. Together with the spectroscopic mass function from the RV analysis, this allows us to constrain the orbital inclination, mass ratio, and absolute separation for several systems exhibiting either a reflection effect or ellipsoidal modulation, as given in Table~\ref{tab:lc_fitting}. For the reflection-effect systems, trends in the residuals were often seen between the photometry and model. However, this is commonly seen due to shortcomings in modelling the reflection effect, particularly at high inclinations where the effect is strongest. The errors we present on LC solutions hence represent statistical uncertainties and do not reflect errors on the binary modelling.

\subsection{The nature and minimum mass of the companion}
\label{the_nature_of_the_companion}
We determined the minimum companion mass for each system by using Eq.~\ref{eqn:Kmax} in combination with the orbital parameters and the primary masses derived in Paper~II. The nature of the companion, however, is not unambiguously constrained in all cases. Some systems exhibit a pronounced reflection effect, indicative of a cool MS companion, because a compact WD would be too small to produce a detectable irradiation signal. Others show variability phased to the spectroscopic period consistent with Doppler boosting, caused by the extreme orbital motion of the sdB \citep[e.g.][]{Bloemen_2011MNRAS.410.1787B}, or ellipsoidal modulation caused by gravitational deformation. Both signals confirm a WD or an unseen, compact companion. By contrast, a low-mass MS star would produce a dominant irradiation signal. For a subset of binaries, the LCs remain inconclusive. In these cases, sufficiently massive MS companions would likely produce a detectable IR excess in the SED, depending on the brightness of the primary. The absence of such an excess would allow us to exclude an MS companion and identify the companion as a likely WD.

To quantify the photometric detection limits for MS companions on a star-by-star basis, we applied the SED fitting procedure of \citet{Heber2018} used in Paper~II to each target individually, forcing a contribution from a (possibly undetected) MS companion at increasing masses until the composite SED became statistically distinguishable from a single-star fit. These SED fits were performed for fixed companion masses from isolated MS companion models spanning $0.1$--$1.5$~\msun\footnote{The companion flux was modelled using the \textsc{Phoenix} Göttingen spectral library \citep{Husser_2013}.} in increments of 0.05~\msun, with corresponding effective temperatures and radii interpolated from BaSTI zero-age MS tracks \citep{Hidalgo2018ApJ...856..125H}, adopting $\log(Z/Z_\odot) = -0.3$, a mean value for MS companions to hot subdwarfs \citep{Vos_2018}. The atmospheric parameters of the hot subdwarfs were fixed to their spectroscopic values, leaving two free parameters: the angular diameter of the primary and the interstellar colour excess $E(44{-}55)$, modelled using the reddening law of \cite{Fitzpatrick2019}. The radius of the subdwarf was derived from the (free) angular diameter and \textit{Gaia} parallax. For each assumed companion mass, we evaluated the reduced $\chi^2$ and determined a minimum MS mass such that it is excluded at a $3\sigma$ threshold ($p = 0.003$). For 253 stars, fewer than one spurious exclusion is expected. The resulting detection limits depend on the available photometry (especially IR), its quality, and the properties of the primary. Unlike the RV and LC methods, this approach is independent of orbital inclination\footnote{This assumes that the MS companion is not irradiated.} and can reveal an IR excess from a companion even in systems where those diagnostics show no variability.

Two examples are shown in Fig.~\ref{fig:sed_min_det_mass}. The left panel shows EC~01578-1743, a known reflection-effect binary with a dM companion \citep{Schaffenroth_2023_2}. Here, the $\chi^2$ minimum (red arrow) lies above the detection limit at a best-fit companion mass of $0.21^{+0.03}_{-0.04}$~\msun\ ($3\sigma$ uncertainties) broadly consistent with the value of $0.278 \pm 0.004$~\msun\ derived from the precise light-curve modelling of \citet{Schaffenroth_2023_2}. The small discrepancy likely reflects the different sdB stellar parameters adopted in this work relative to the canonical sdB mass assumed by \citet{Schaffenroth_2023_2}. The corresponding IR excess is clearly visible in the SED (right panel of Fig.~\ref{fig:sed_min_det_mass}).
The central panel shows GSC~00141-01628, a newly solved binary with a minimum companion mass of $0.28 \pm 0.02$~\msun\ from the RV curve, whose small photometric variability (see Sect.~\ref{fig:rv_lc_power}. The TESS amplitude $\sim0.02\%$) does not reveal the companion nature. The SED detection limit is $0.15$~\msun\ (vertical dashed line). Since the $\chi^2$ minimum lies below this threshold, no IR excess is detectable, and we conclude the companion is most likely a WD. 

\subsection{Orbital inclination correction and detection efficiency}
\label{injection_recovery}

The observed binary fraction is a lower limit. Close binaries viewed at low orbital inclinations can produce RV amplitudes too small to detect, particularly given that the majority of our spectra are medium resolution ($\delta\lambda \sim 1.0~\AA$).  In a minority of cases, lower resolution observations, a few epochs, or short temporal baselines may reduce sensitivity to low-amplitude or long-period signals. To account for and investigate these biases simultaneously, we employed a Monte Carlo (MC) injection-recovery approach.

For each of the 253 stars in the sample, regardless of whether it was identified as a binary, we estimated the per-star detection efficiency, $\eta_j$, i.e. the probability that our RV data would identify the star as variable if it were in a close binary system similar to other binaries in our sample.
To compute $\eta_j$, we performed an injection-recovery simulation designed to estimate the fraction of binaries our observations would fail to detect. In each iteration, we drew a semi-amplitude ($K$) and period ($P$) from the pool of orbital solutions of our detected binaries ($K = 3.9$ - $377.0$ \kms; $P = 0.049$ - $21.7$ days), propagated through MC sampling to account for parameter uncertainties. We then assigned a random orbital phase, $\phi_0$, drawn uniformly between $0$ and $2\pi$ (equivalently a uniform distribution in cos $i$), and a random orbital inclination, $i$, drawn from the geometric prior for randomly oriented orbits. For an isotropic distribution of orbital poles on the celestial sphere, the probability of observing a given inclination is proportional to $\sin i$ \citep[e.g.][]{Breedt2017, rodriguez2026whitedwarfm}. 

From these parameters, we generated synthetic RVs at the star's actual observation epochs with its per-epoch measurement uncertainties:

\begin{equation}
v_k^{\rm synth} = K \sin i \, \sin\!\left(\frac{2\pi\,t_k}{P} + \phi_0\right)
    + \epsilon_k,
\end{equation}
where $t_k$ is the timestamp of star $j$ at epoch $k$ and $\phi_0$ is Gaussian noise with standard deviation equal to the real RV measurement uncertainty $\sigma_{k}$. We then applied the same $\chi^2$ variability ($\log p$) test used on the real data and recorded whether the signal, or binary, was recovered or not. After ${\sim}45,000$ trials per star ($500$ draws of $K, P$, and $90$ inclinations for each), the detection efficiency is $\eta_j = N_{\rm recovered} / N_{\rm trials}$. 

The corrected short-period binary fraction was then calculated via
\begin{equation}
    f_{\rm corr} = \frac{1}{N_{\rm total}}\left[
        N_{\rm known\text{-}only}
        + \sum_{\mathrm{detections}} \frac{1}{\eta_j} 
    \right],
    \label{eq:fcorr}
\end{equation}
where the sum runs over the 90 originally RV-detected binaries, each weighted by the inverse of its recovery probability, and values $N_{\rm known\text{-}only} = 5$ are the previously known confirmed close binaries. Uncertainties were derived from bootstrapping the full sample of 253 stars with replacement, quoting the 16th and 84th percentiles of the resulting distribution as the $1\sigma$ confidence interval. The resulting corrected binary fractions, $f_{\mathrm{corr}}$, and their uncertainties are reported in Table~\ref{tab:binary_fractions}.

\subsection{Inclination-corrected mass of the companion}
\label{most_probable_mass}

The minimum companion mass $M_{2,\min}$ derived from the binary mass function assumes an edge-on orbit ($i = 90\degr$). To estimate the absolute companion mass distribution for the non-photometrically variable binaries, we used the geometric inclination prior described in Sect.~\ref{injection_recovery}. For each solved binary, we randomly drew a $P$, $K$, and $M_{\mathrm{sdB}}$ 2000 times, assuming Gaussian measurement uncertainties. The mass function was then recomputed, and a random inclination was drawn from the prior a further 300 times in each case ($2000\times300$ draws in total). The companion mass $M_2$ was then predicted by numerically solving
\begin{equation}
    \frac{M_2^3 \sin^3 i}{(M_1 + M_2)^2} = \frac{K_1^3 P}{2\pi G}
\end{equation}

at each trial. These mass arrays were smoothed into a probability density using a Gaussian kernel density estimates (KDE; bandwidth $0.05$~\msun), and the stacked posteriors for each companion class (WD, dM/BD, and undetermined) were determined.

\section{Results}
\label{results}
\subsection{The short-period binary fraction}
\label{close_binary_fraction}
The short-period binary fraction is a key observable for constraining hot subdwarf formation. It directly determines the conditions for unstable mass transfer. Applying the $\log p<-4$ criterion (Sect.~\ref{criterion_for_binarity}), we identify 90 RV variable systems among the 253 non-composite hot subdwarfs in our sample. A further five systems are confirmed close binaries from photometric variability or previous studies, which we did not follow up, bringing the total to 95 and an observed binary fraction of $f_{obs}\sim37.5$\%.

This represents a lower limit because of observational bias. Following the injection-recovery approach detailed in Sect.~\ref{injection_recovery}, we find a corrected short-period binary fraction of $41.4^{+3.6}_{-3.4}\%$ for the 253 non-composite hot subdwarfs, or $34.7^{+2.8}_{-2.9}\%$ when including the 48 wide-orbit composites (identified in Paper~I). The correction increases the observed binary fraction by $\sim4\%$, implying that approximately ten short-period binaries go undetected. These include systems that are face-on, poorly sampled in phase, or host very low-mass companions. This estimate assumes that undetected binaries have orbital properties similar with those already detected and solved.

Breaking the short-period binary fraction down by spectral class (see Table~\ref{tab:binary_fractions}), the sdB and sdO subclasses show the highest corrected binary fractions at $56.0\%$ and $42.5\%$, respectively. The sdOBs, however, have a much lower corrected fraction of $23.1\%$. This has also been seen in \citet{Geier_2022}, who report a lower binary fraction of their 'EHB3' stars, which are analogous to the sdOBs in this paper, and sit at the hot end of the EHB. Further discussion on this finding is given in Sect.~\ref{rvv_classes}.

Of the 253 stars in the sample, 205 ($82.7\%$) have a detection efficiency $> 0.8$, indicating that the majority of our sample has good sensitivity.
Just eight stars without detected companions have $\eta < 0.3$, which should be confirmed with follow-up spectroscopy. 

\begin{table}
\centering
\small
\caption{Observed ($f_{\rm obs}$) and corrected ($f_{\rm corr}$) short-period binary fractions by spectral class and companion types. $N$ is the number of stars in a class, whereas $N_{\rm RV}$ is the number of detected binaries in that class.}
\label{tab:binary_fractions}
\begin{tabular}{lrrrcc}
\toprule\toprule
Class & $N$ & $N_{\rm RV}$  & $f_{\rm obs}$ & $f_{\rm corr}$ \\ [0.1cm]
\midrule 
\multicolumn{6}{c}{\textit{Spectral classes}} \\
\midrule
All    & 253 & 95  & 37.5\% & $41.4^{+3.6}_{-3.4}$\% \\[0.1cm]
sdB    & 143 & 72 & 50.3\% & $56.0^{+5.0}_{-4.7}$\% \\ [0.1cm]
sdOB   &  52 & 11 & 21.2\% & $23.1^{+5.8}_{-6.3}$\% \\ [0.1cm]
sdO    &  25 & 10 & 40.0\% & $42.5^{+12.0}_{-9.0}$\% \\ [0.1cm]
iHe    &  17 &  1 &  5.9\% & $6.1^{+6.1}_{-6.1}$\% \\ [0.1cm]
eHe$^*$    &  16 &  1 &  6.2\% & $6.2^{+6.3}_{-6.2}$\% \\ [0.1cm]
\midrule
below-EHB  & 22 & 9 & 40.9\% & $45.6\substack{+12.2 \\ -11.6}$\%  &  \\
\midrule
Overall$^\dagger$ & 301 & 95 & 31.6\% & $34.7\substack{+2.8 \\ -2.9}$\% &  \\
\midrule
\midrule
\multicolumn{6}{c}{\textit{Companion types}} \\
 Companion &  & $N$ & &  $f_{\rm obs}$ & \\[0.1cm]
\midrule
+WD  &  & 55 &  & $57.9^{+4.9}_{-5.2}$\%  &  \\[0.1cm]
+dM/BD  &  & 19 &  & $20.0^{+4.7}_{-3.5}$\% &  \\[0.1cm]
+WD/MS  &  & 4 &  & $4.2^{+3.1}_{-1.2}$\% &  \\[0.1cm]
+He-sdB$^*$  &  & 1 &  & $1.1^{+2.3}_{-0.3}$\% &  \\[0.1cm]
Unsolved  &  & 16 &  & $16.8^{+4.5}_{-3.2}$\% &  \\[0.1cm]
\bottomrule
\end{tabular}
\tablefoot{All spectral class binary fractions are given as a fraction of the 253 non-composite hot subdwarfs. The overall fraction at the bottom additionally includes the 48 composite systems with identified MS companions from Paper~I, defining the total short-period binary fraction of the sample. Uncertainties for the spectral classes denote 16th--84th percentile bootstrap intervals, and binomial uncertainties are given for the companion types.
$^\ast$PG\,1544+488 \citep{Sener_2014MNRAS.440.2676S} is a known binary system comprising an eHe-sdOB and an He-sdB within 500~pc. It is counted under the spectral class of eHe. 
$^\dagger$ 301 includes hot subdwarf and FGK-type systems from Paper~I.
}
\end{table}

\subsection{The companions to hot subdwarf stars}
\label{companions}
Using the SED-approach detailed in Sect.~\ref{the_nature_of_the_companion}, we classified the companion nature for all systems lacking unambiguous photometric signals. Four systems (LS\,III\,+48\,50, LS\,IV\,$-$13\,2, TYC 4563-2614-1, and EC\,21556$-$5552) have minimum companion masses below the detection limit. They are labelled `$+$MS/WD' in all tables and shown in grey in subsequent figures. Figure~\ref{fig:det_mass} shows the distribution of detection limits for the full sample and has a median of $0.16^{+0.05}_{-0.03}$~\msun, the typical sensitivity through the SED method for our objects.

In the bottom panel of Table~\ref{tab:binary_fractions}, the number of each companion type is given for the 79 solved systems. The 16 detected but unsolved binaries are listed in Table~\ref{table_new_binaries}. We find a higher relative ratio of sd$+$WD to sd$+$dM/BD (2.9:1) compared with previous work \citep{Kupfer_2015, Schaffenroth_2022_1}, which is unsurprising due to the minimised observational biases of this volume-complete study and the photometric signals from dM/BD companions that are easier to detect.

The bottom-right panel of Fig. ~\ref{fig:combined_period_analysis} gives the minimum companion mass distribution, which exhibits two distinct peaks corresponding to sd$+$dM/BD and sd$+$WD companions. There also appears to be a second peak for the sd$+$WD binaries at $\sim0.8$~\msun, which corresponds to orbital periods of less than one day.
The most probable distribution of companion masses (detailed in Sect. \ref{most_probable_mass}) are shown in Fig.~\ref{fig:most_prob_mass}, with median masses and 16th-84th percentile intervals shown in the plot. The $+$dM/BD companion mass distribution is tightly centred at $0.15^{+0.12}_{-0.05}$, whereas the $+$WD systems exhibit a broad distribution with a median of $0.54^{+0.35}_{-0.20}$~\msun. A long tail towards high masses is seen for the $+$WD systems, which is caused by low orbital inclinations combined with several high minimum WD masses seen in the bottom-right panel of Fig. ~\ref{fig:combined_period_analysis} ($\geq0.6$~\msun).

\subsection{The period distribution}
\label{sect:period}

\begin{figure*}[h!]
    \centering
    \begin{subfigure}[t]{0.49\linewidth}
        \centering
        \includegraphics[width=\linewidth]{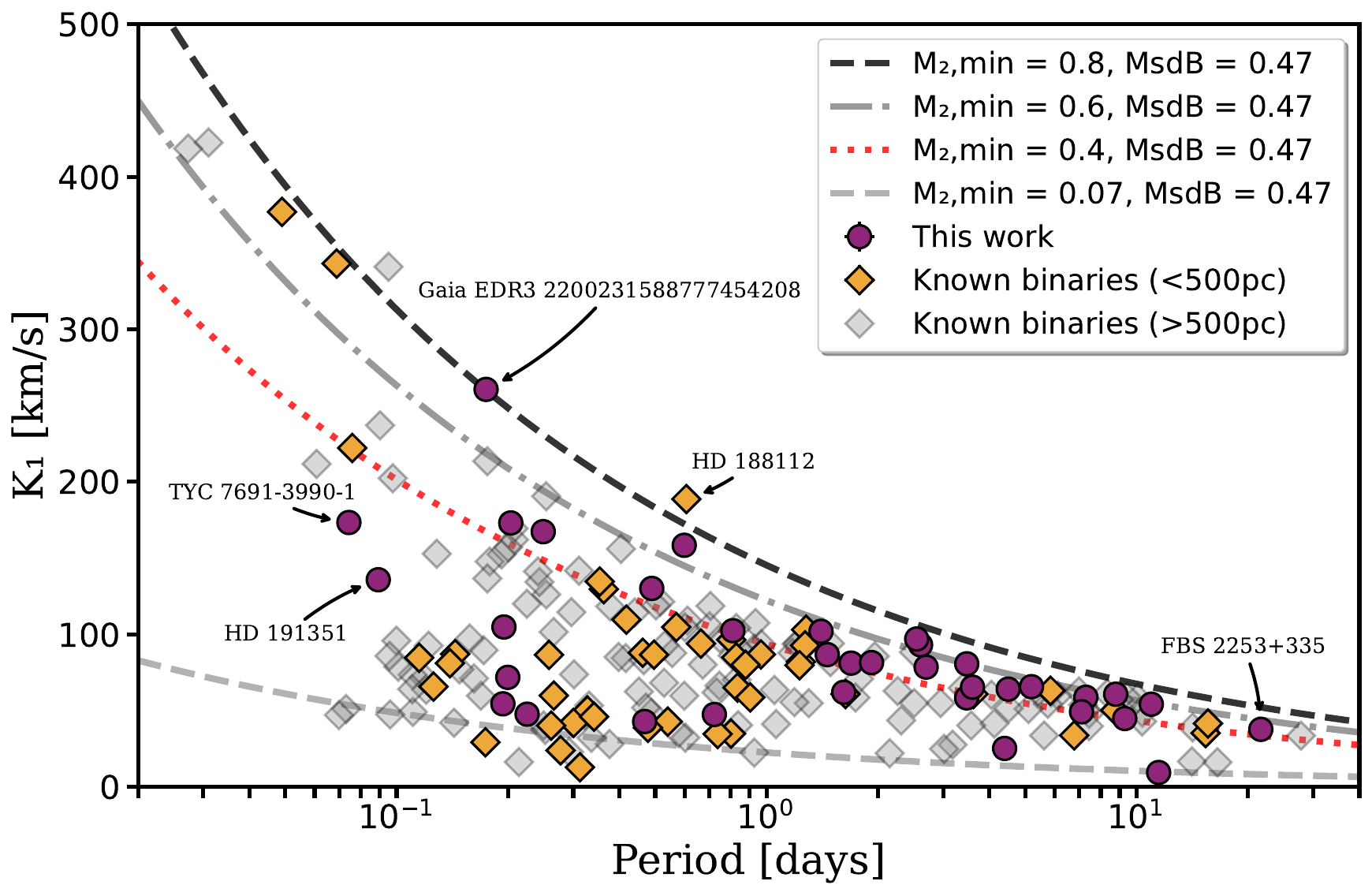}
    \end{subfigure}
    \hfill
    \begin{subfigure}[t]{0.49\linewidth}
        \centering
        \includegraphics[width=\linewidth]{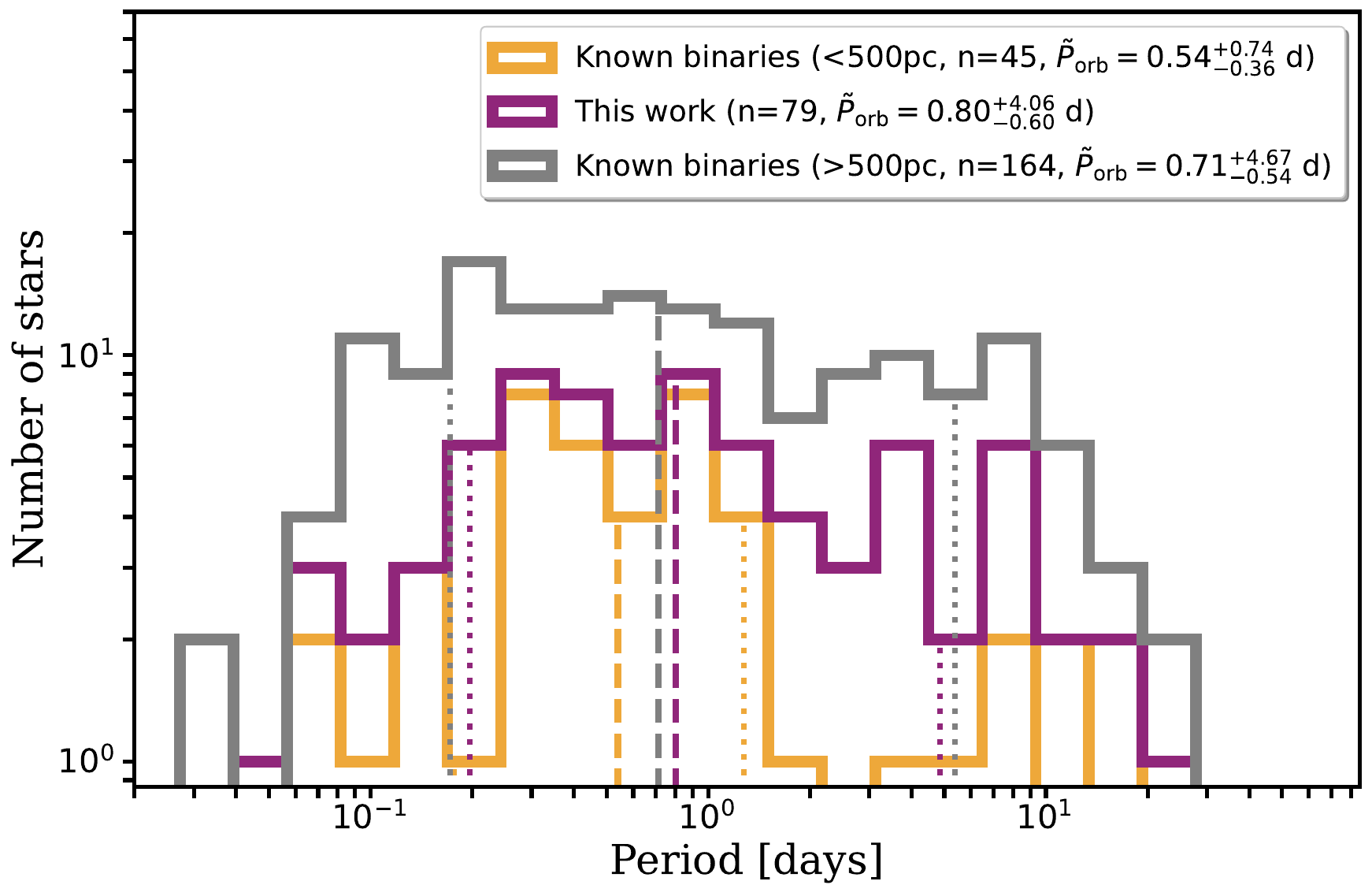}
    \end{subfigure}

    \vspace{0.5em}

    \begin{subfigure}[t]{0.49\linewidth}
        \centering
        \includegraphics[width=\linewidth]{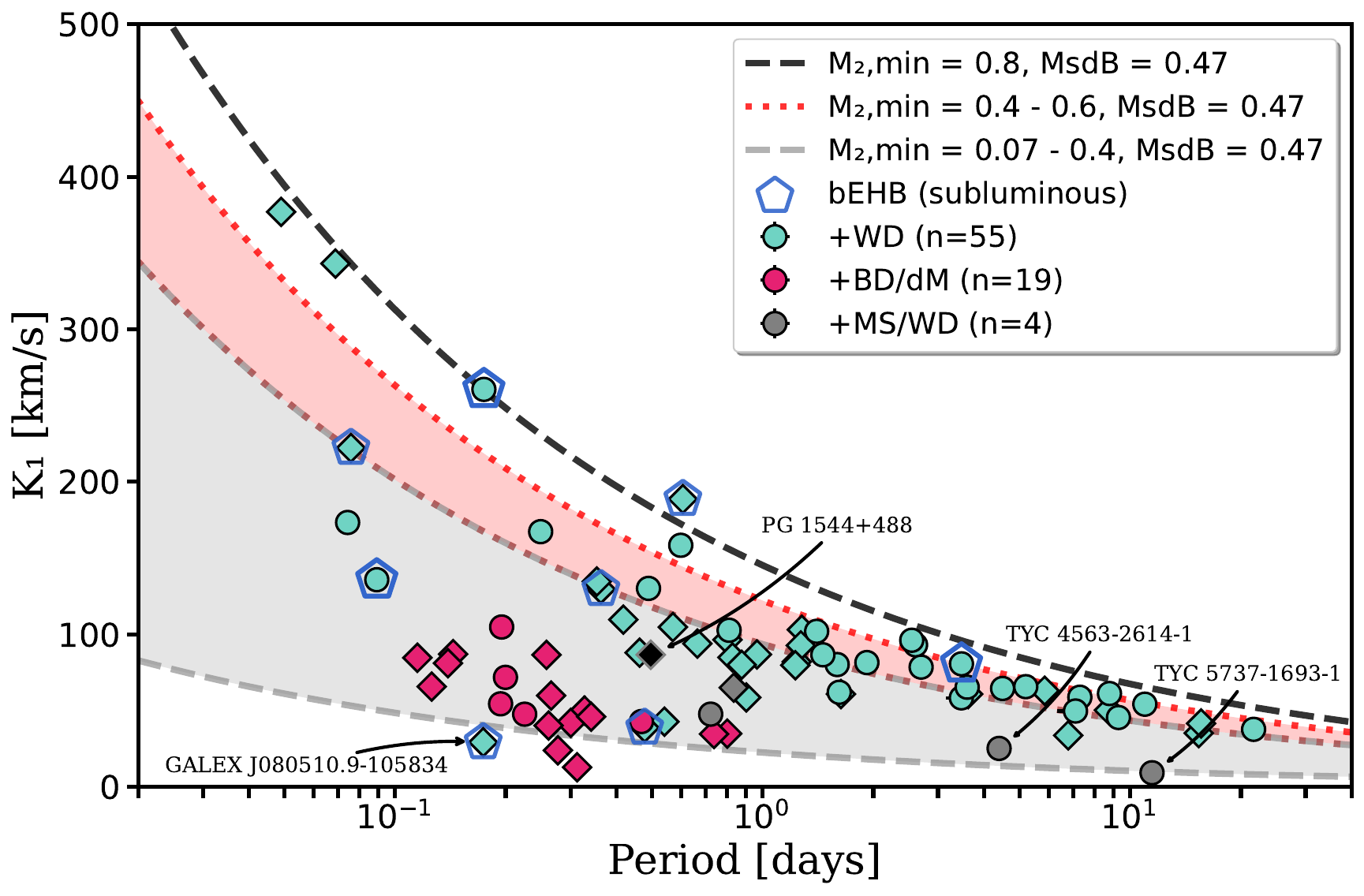}
    \end{subfigure}
    \hfill
    \begin{subfigure}[t]{0.49\linewidth}
        \centering
        \includegraphics[width=\linewidth]{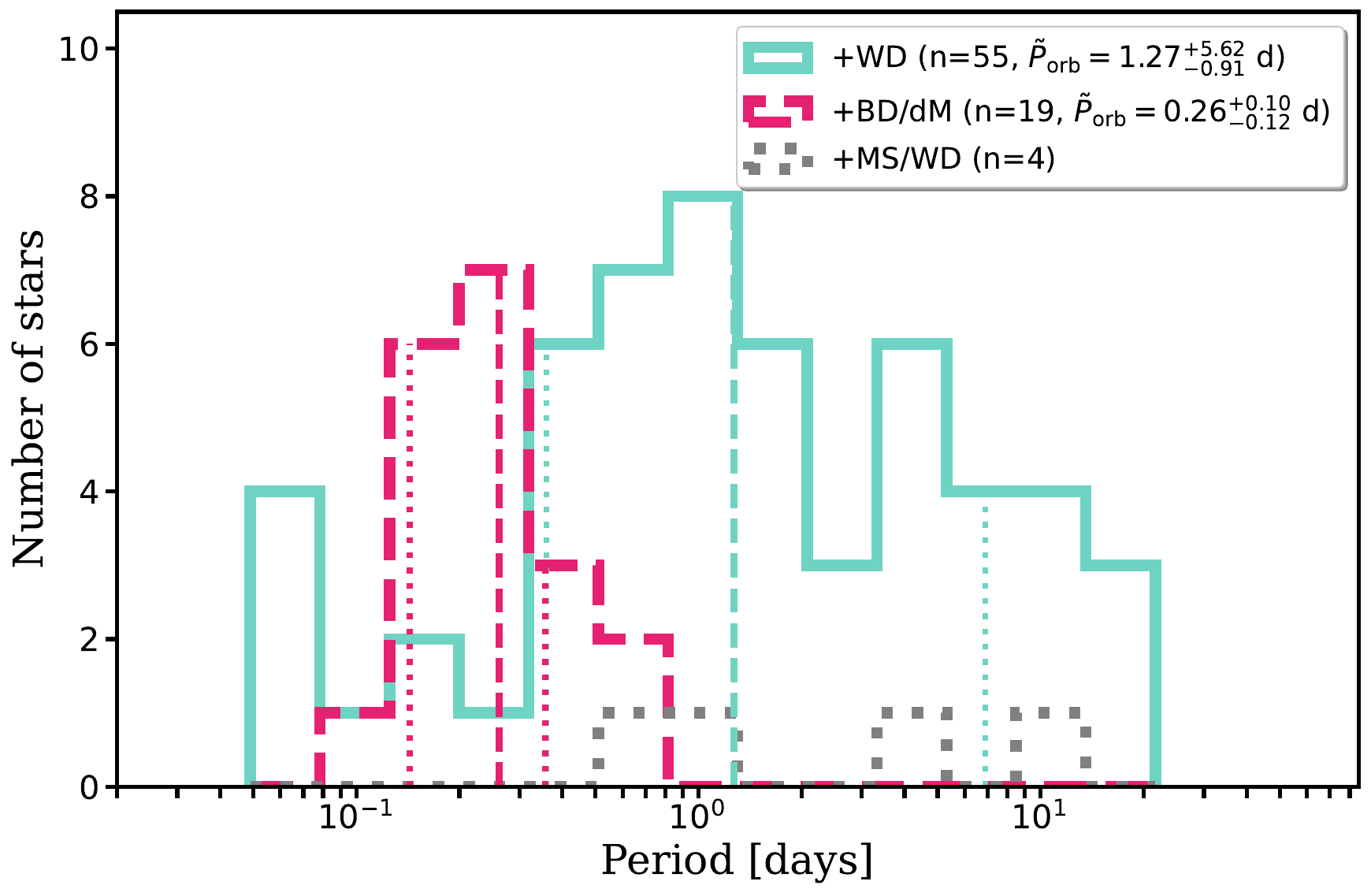}
    \end{subfigure}

    \vspace{0.5em}

    \begin{subfigure}[t]{0.49\linewidth}
        \centering
        \includegraphics[width=\linewidth]{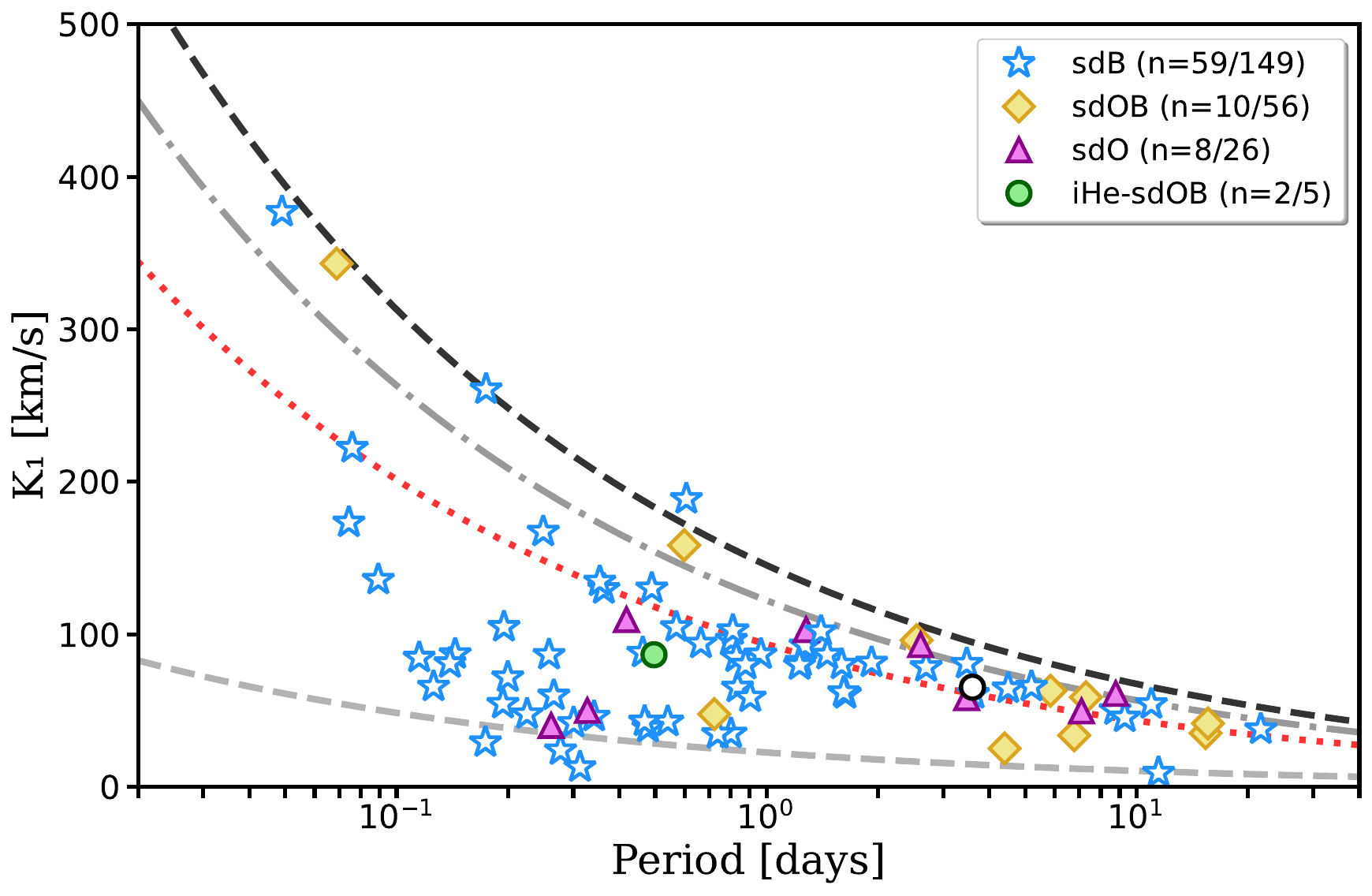}
    \end{subfigure}
    \hfill
    \begin{subfigure}[t]{0.49\linewidth}
        \centering
        \includegraphics[width=\linewidth]{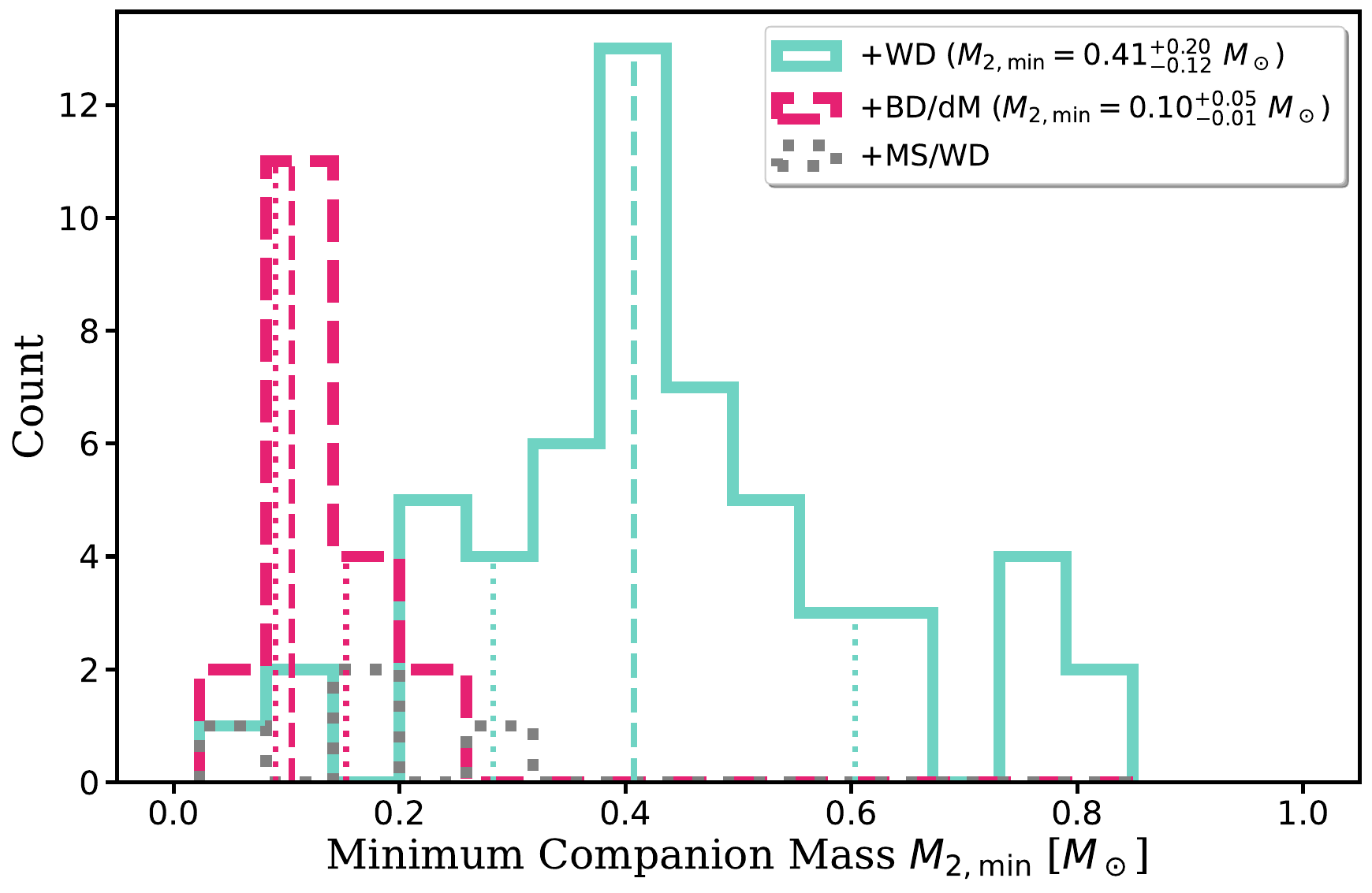}
    \end{subfigure}

    \caption{
        \textit{Top row:} Semi-amplitude vs period (left) and star count vs period as a step histogram (right) for the known (diamonds) and new (circles) binaries. The plotted lines mark minimum companion masses, as defined in the legend.
        \textit{Middle row:} Same as the top row but with the sample divided based on the nature of the companion.
        \textit{Bottom row:} Same as above but divided by spectral class (left) and the companion mass distribution by class (right). The medians of the distributions are provided in the legends and are marked with vertical dashed lines. 
    }
    \label{fig:combined_period_analysis}
\end{figure*}

\begin{figure}
    \centering
        \includegraphics[width=1\linewidth]{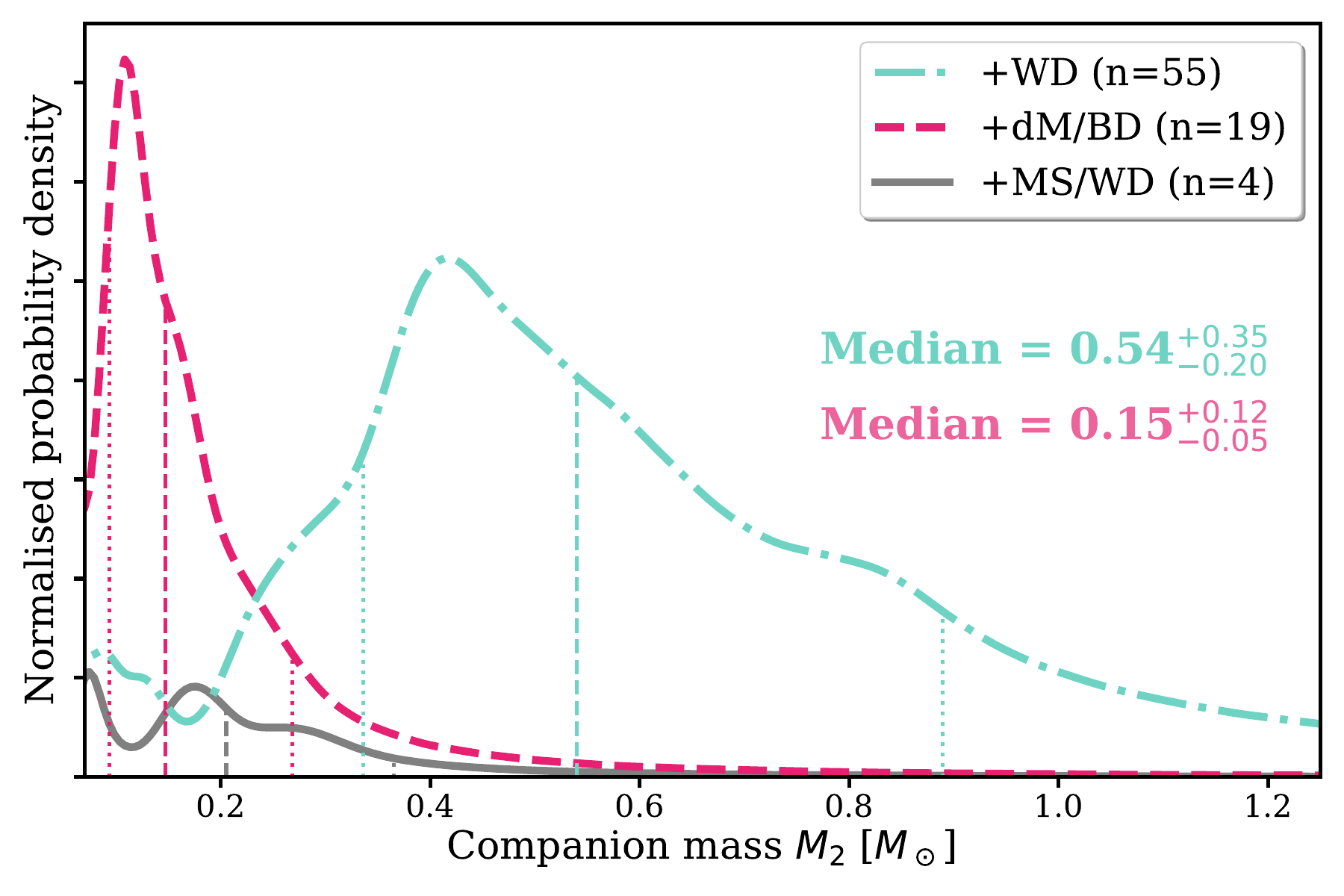}
    \caption{Inclination-corrected mass distributions of the companion types, colour-coded accordingly. The vertical dashed lines represent the median and 16th-84th percentiles of each distribution.}
    \label{fig:most_prob_mass}
\end{figure}

The semi-amplitude versus orbital period distribution for all known close hot subdwarf binaries within 500~pc is shown in the left panels of Fig. ~\ref{fig:combined_period_analysis}. The dashed lines indicate minimum companion masses assuming a canonical hot subdwarf mass of $0.47$~\msun\ and an inclination of $90\degr$. All known binaries beyond 500~pc are shown in grey for comparison. We see fewer objects at orbital periods longer than $\sim1$~day with low ($\lesssim~$50~\kms) semi-major amplitudes compared with the literature compilation, which may point to an intrinsic property of our sample where more massive companions are favoured at longer orbital periods. Several systems of interest are highlighted in these plots, all of which are referred to in the subsequent text. This includes TYC~5737-1693-1 and TYC~4563-2614-1, which are the only two systems with very low semi-amplitudes at orbital periods longer than one day. They may either be oriented at high inclinations or indeed host very low-mass companions, which would be difficult to explain theoretically \citep{Han_2002}. 

In the upper-right and central-right panels of Fig.~\ref{fig:combined_period_analysis}, we observe that the majority of the newly solved binary systems range from 1.5 to 10~days. The comparatively low number of systems identified in this range in previous studies is likely an observational bias, as long-period systems are more difficult to detect.

The systems with identified dM/BD companions are tightly clustered at orbital periods of $\sim6.5\pm3.1$~hours (median, with uncertainties given as the 16th~/~84th percentiles throughout). As shown in the upper-right panel of Fig.~\ref{fig:combined_period_analysis}, the vast majority of these systems lie between 0.1 and 0.4~days, with very few WD companions occupying this period range. No very short-period dM/BD systems (P~$\leq0.1$~days) are present here compared with several known systems from the literature (grey diamonds). Systems with WD companions, by contrast, exhibit a much broader distribution with P$\sim1.25^{+5.59}_{-0.92}$~days and dominate at orbital periods longer than $\sim1$~day. A period peak at $\sim$5--10 days has been reported in previous studies by \citet{Kupfer_2015} and \citet{Schaffenroth_2022_1} (see their Figs.~7 and 17, respectively), but this feature is not clearly evident in our sample.

\subsection{Highlighted individual systems}
\label{highlighted_systems}

\begin{figure*}
    \centering
    \begin{subfigure}[t]{0.32\linewidth}
        \includegraphics[width=\linewidth]{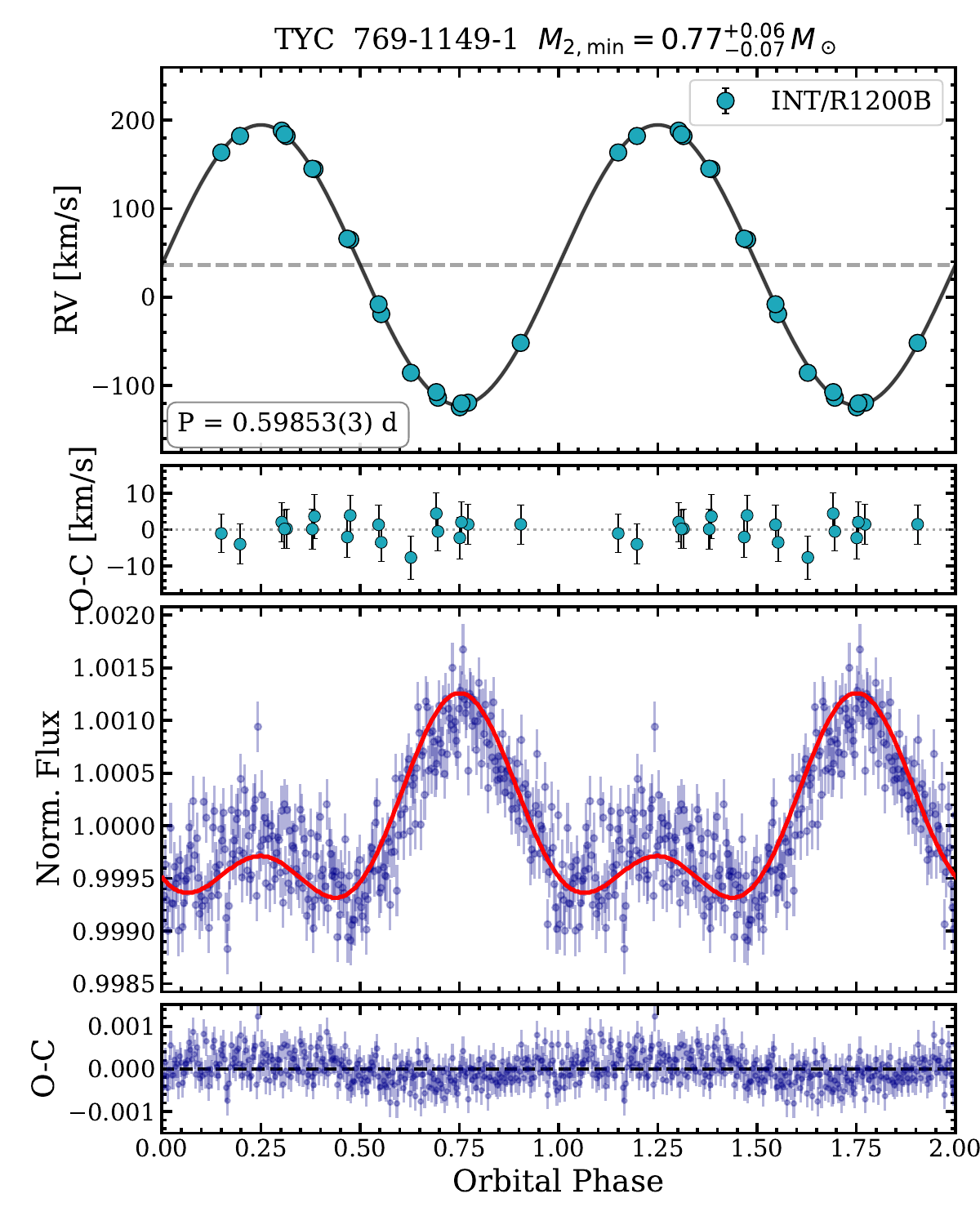}
    \end{subfigure}\hfill
    \begin{subfigure}[t]{0.32\linewidth}
        \includegraphics[width=\linewidth]{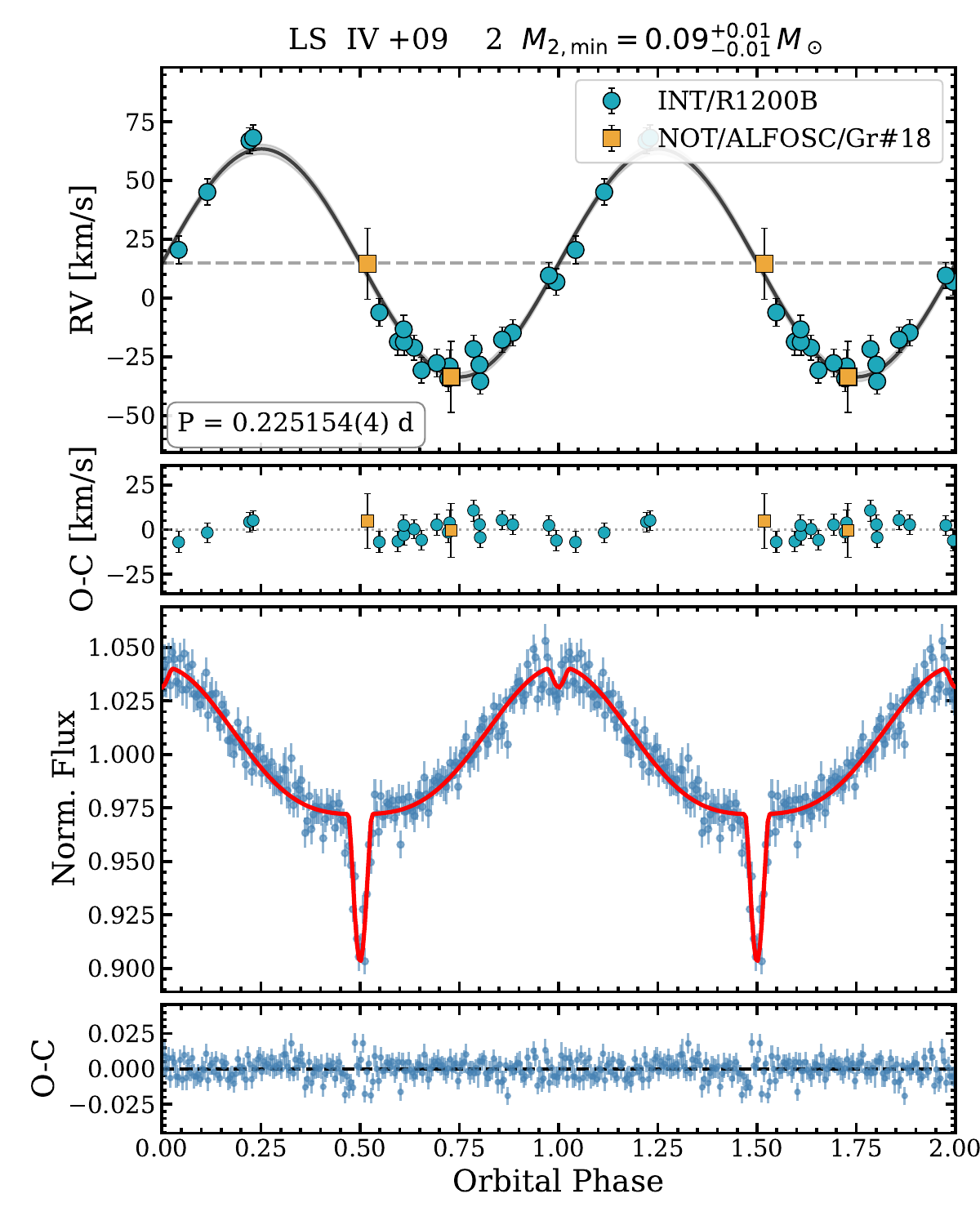}
    \end{subfigure}\hfill
    \begin{subfigure}[t]{0.32\linewidth}
        \includegraphics[width=\linewidth]{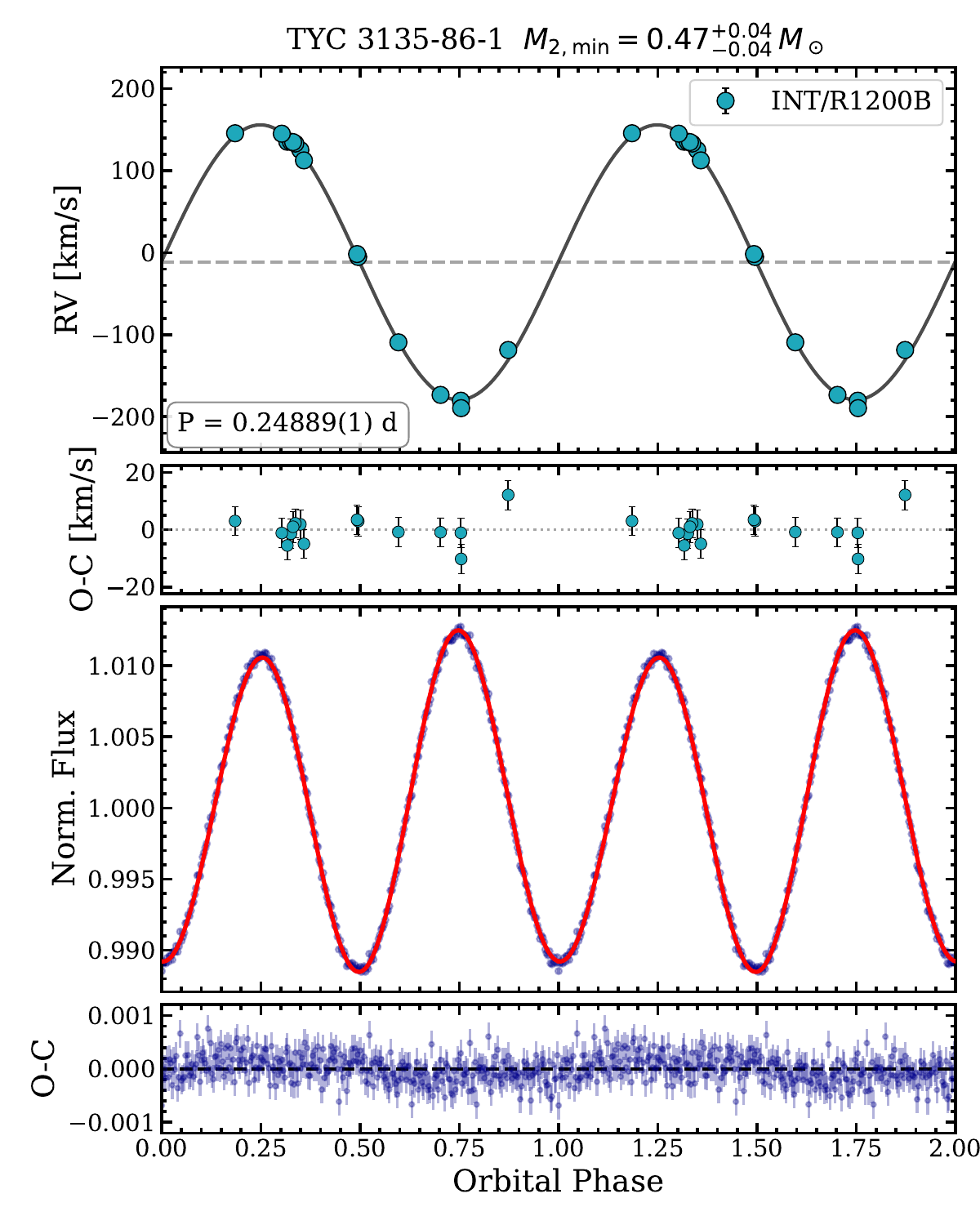}
    \end{subfigure}

    \begin{subfigure}[t]{0.32\linewidth}
        \includegraphics[width=\linewidth]{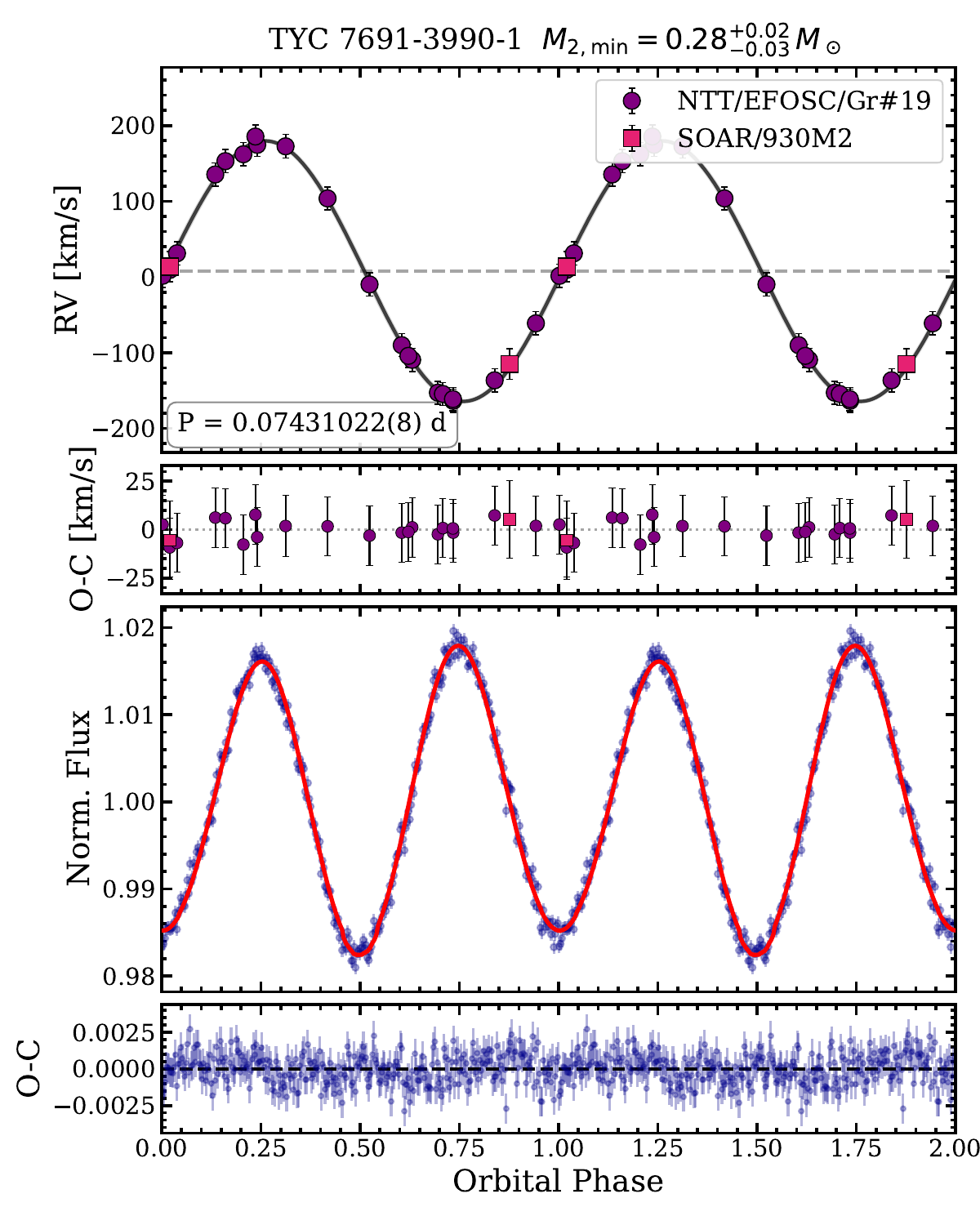}
    \end{subfigure}\hfill
    \begin{subfigure}[t]{0.32\linewidth}
        \includegraphics[width=\linewidth]{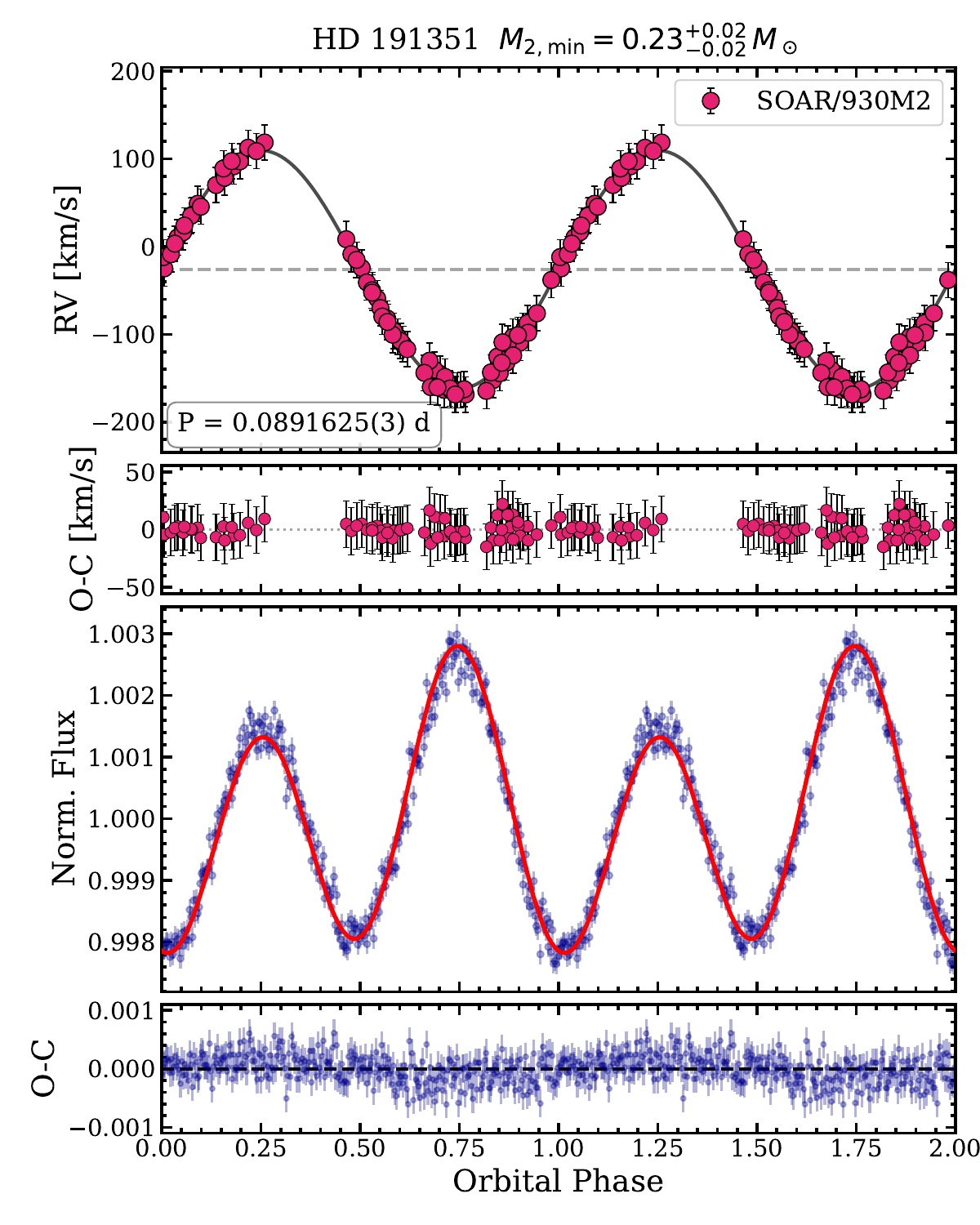}
    \end{subfigure}\hfill
    \begin{subfigure}[t]{0.32\linewidth}
        \includegraphics[width=\linewidth]{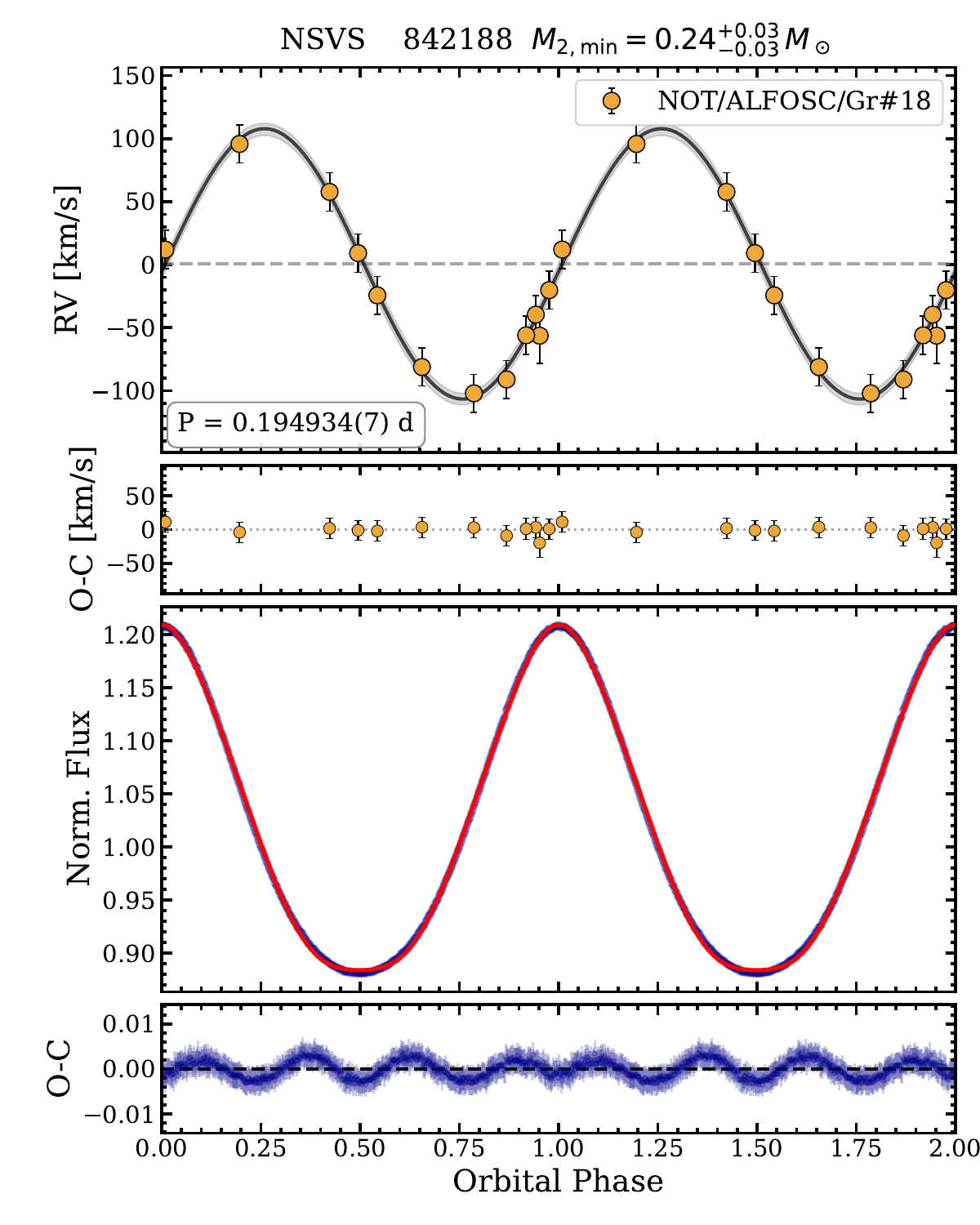}
    \end{subfigure}\hfill    

    \begin{subfigure}[t]{0.32\linewidth}
        \includegraphics[width=\linewidth]{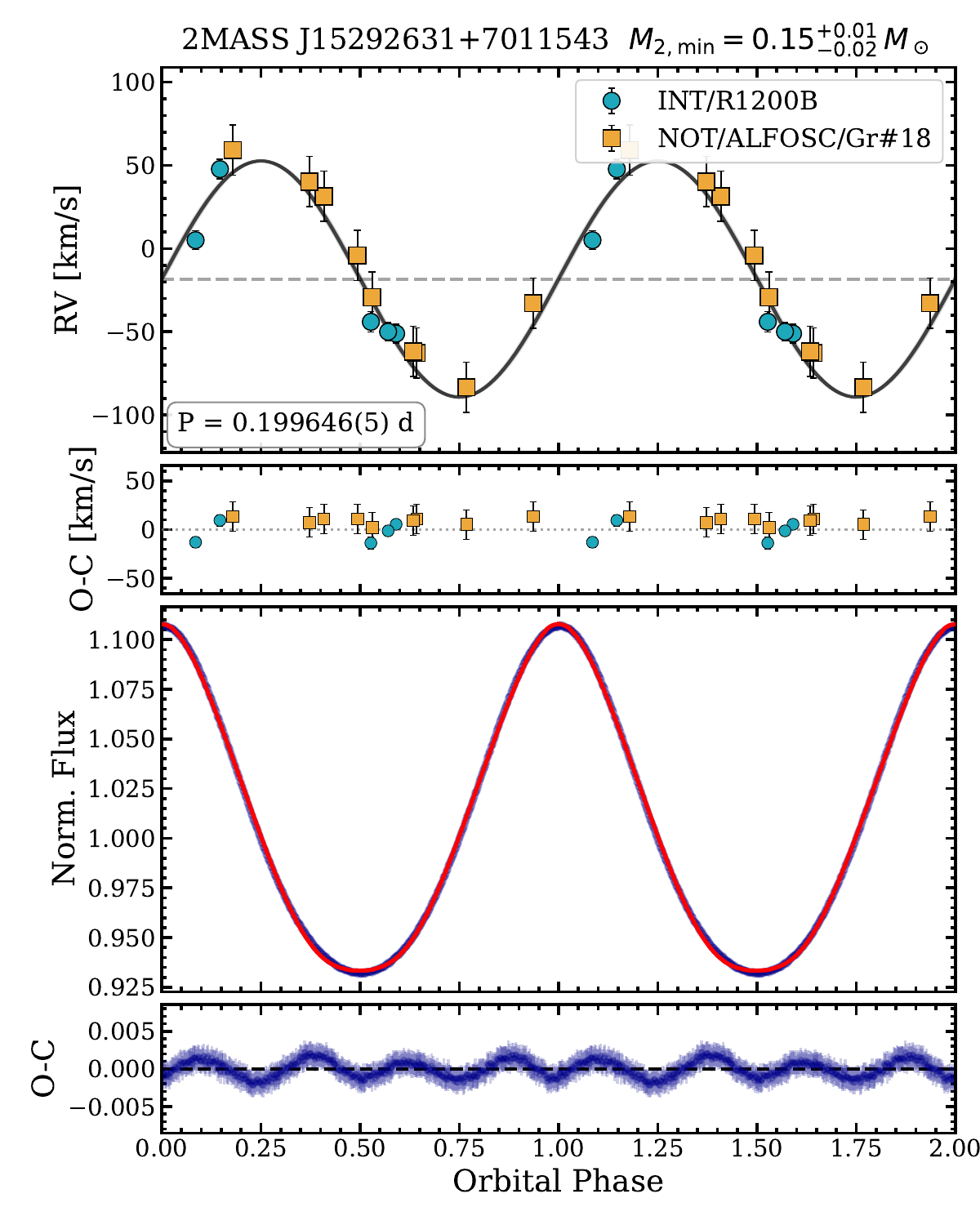}\\
        \end{subfigure}\hfill
    \begin{subfigure}[t]{0.32\linewidth}
       \includegraphics[width=\linewidth]{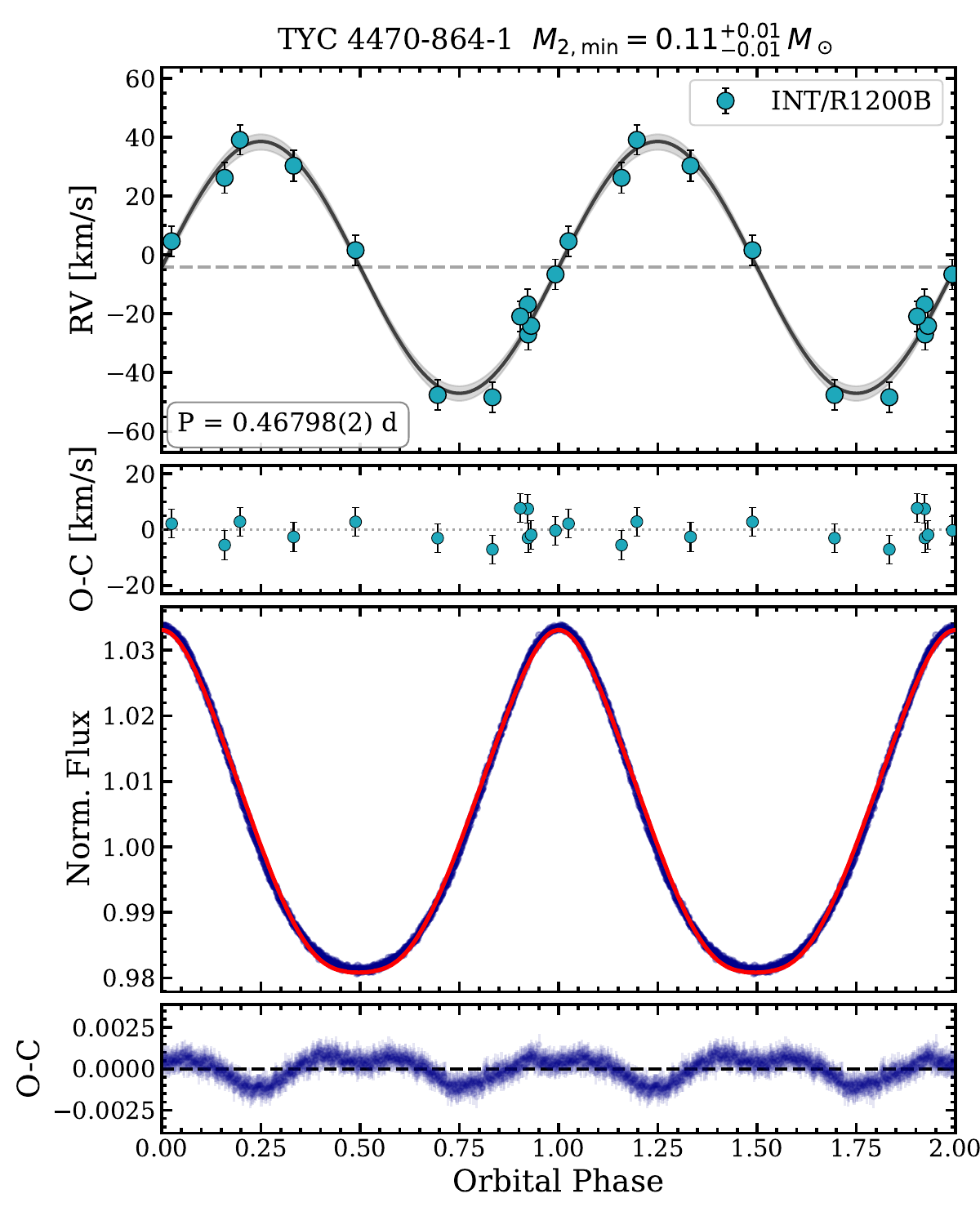}\\
        \end{subfigure}\hfill
    \begin{subfigure}[t]{0.32\linewidth}
        \includegraphics[width=\linewidth]{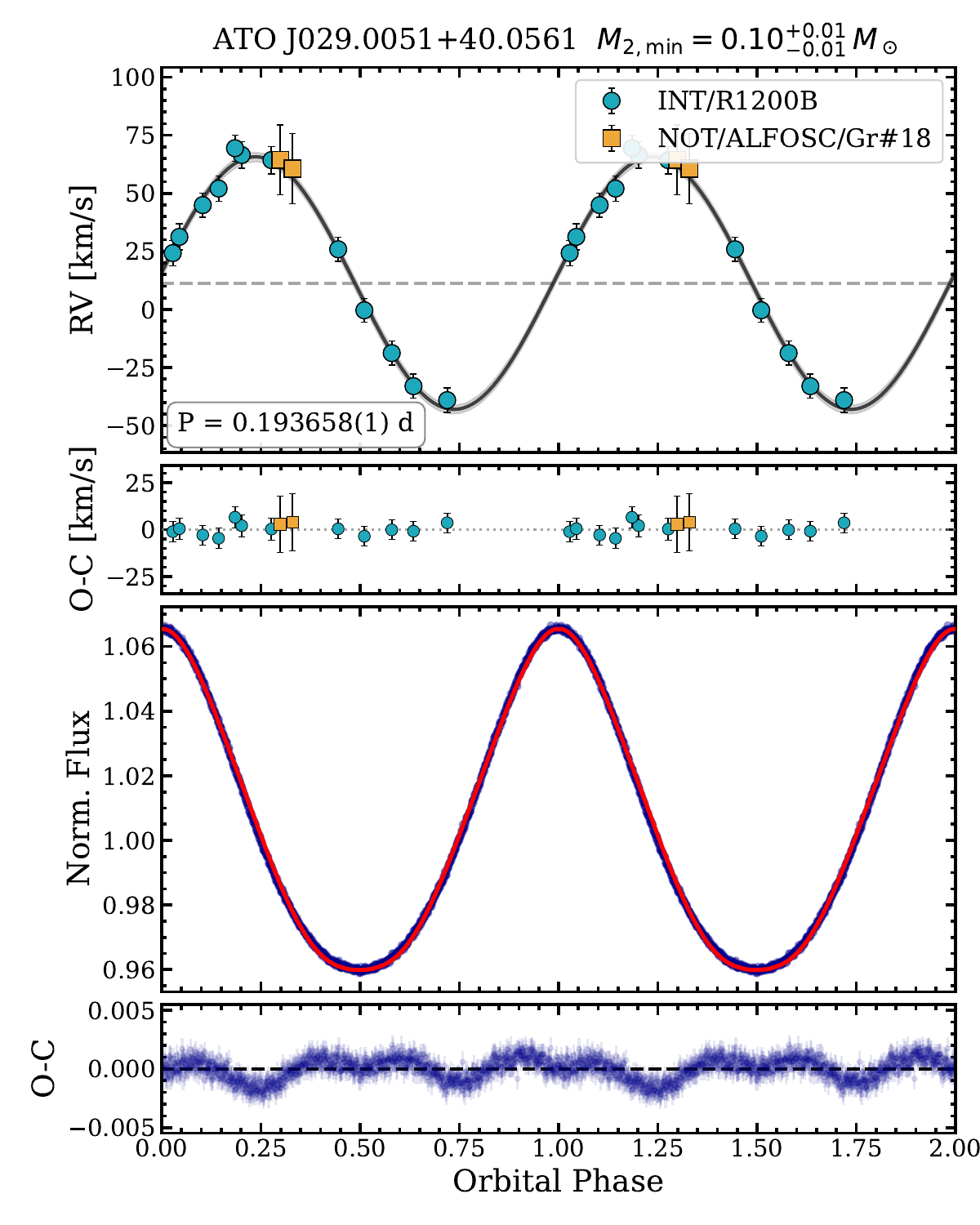}\\
    \end{subfigure}
    \caption{RV curves (top) and LCs (bottom) of nine new highlighted binary systems discussed in Sect.~\ref{highlighted_systems}. The RV and photometric data were jointly fitted, as described in Sect.~\ref{determination_of_the_orbital_solutions}. TESS data are shown in dark blue, whereas the ATLAS data utilised for LS~IV$+092$ are shown in steel blue. The residuals (observed - computed) are shown under each RV curve and LC. The residuals seen for the reflection-effect systems are likely due to shortcomings in the models and are seen in previous work.}
    \label{fig:highlighted_systems}
\end{figure*}

\begin{figure*}
    \centering
    \begin{subfigure}[t]{0.32\linewidth}
        \includegraphics[width=1\linewidth]{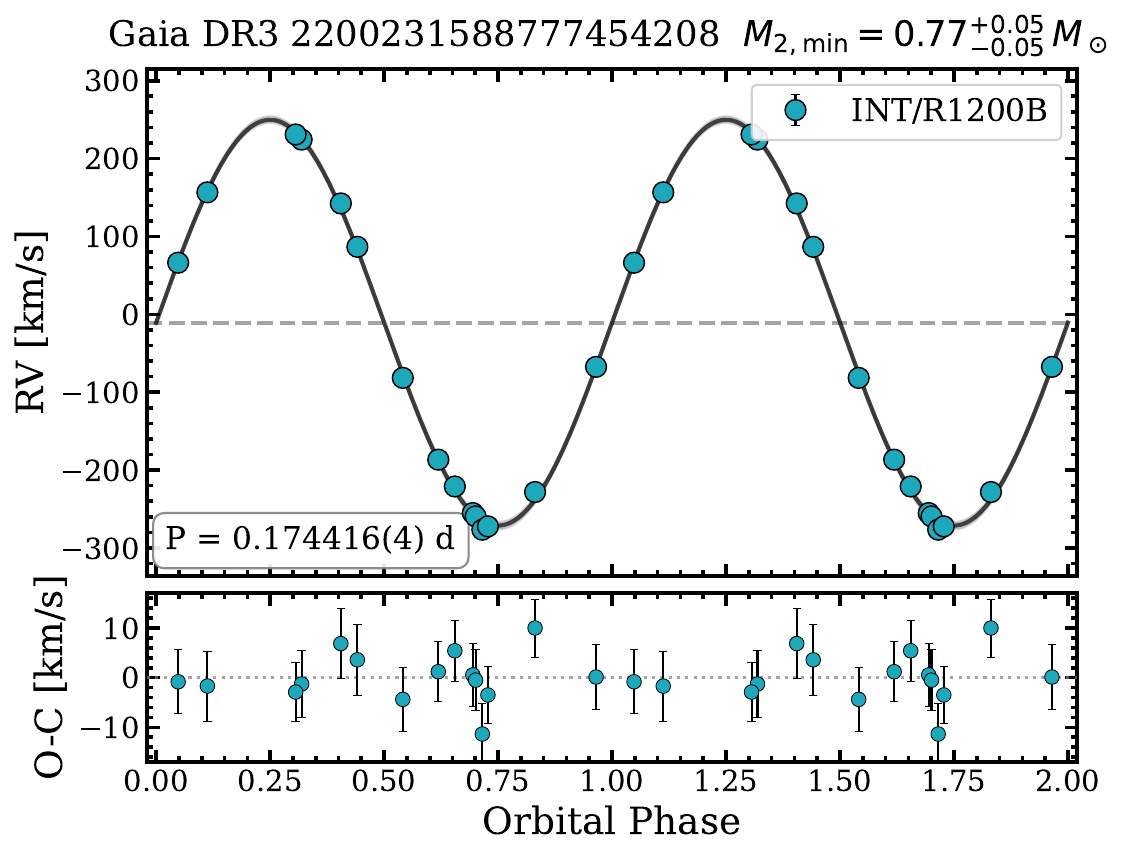}\\
        \includegraphics[width=1\linewidth]{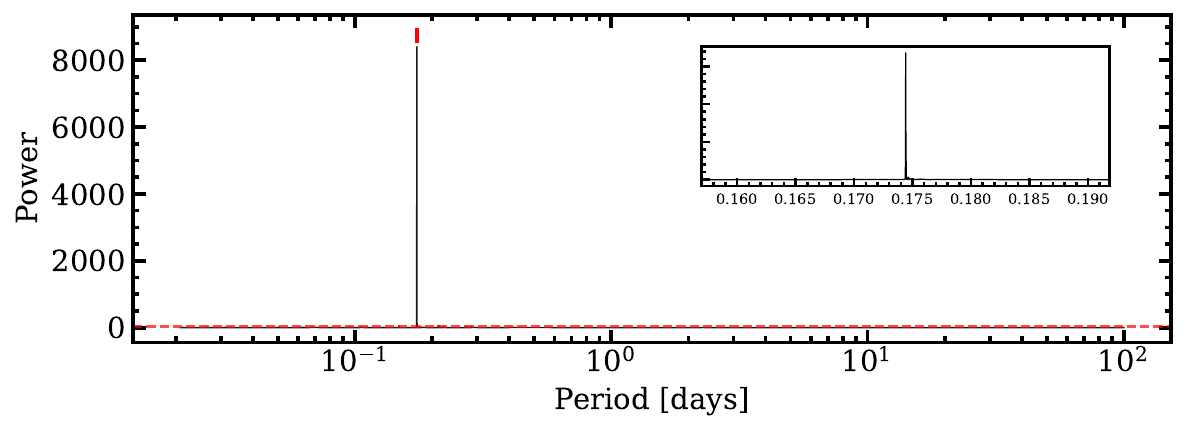}
        \end{subfigure}\hfill
    \begin{subfigure}[t]{0.32\linewidth}
        \includegraphics[width=1\linewidth]{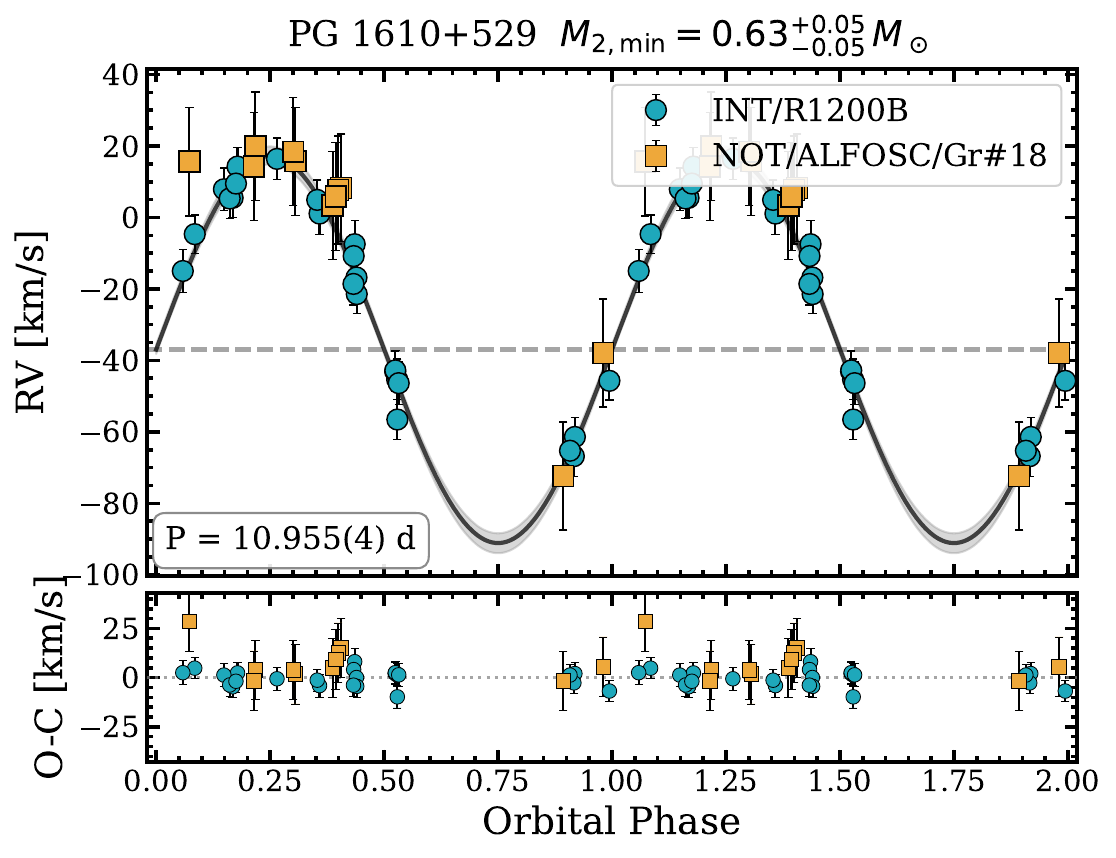}\\
        \includegraphics[width=1\linewidth]{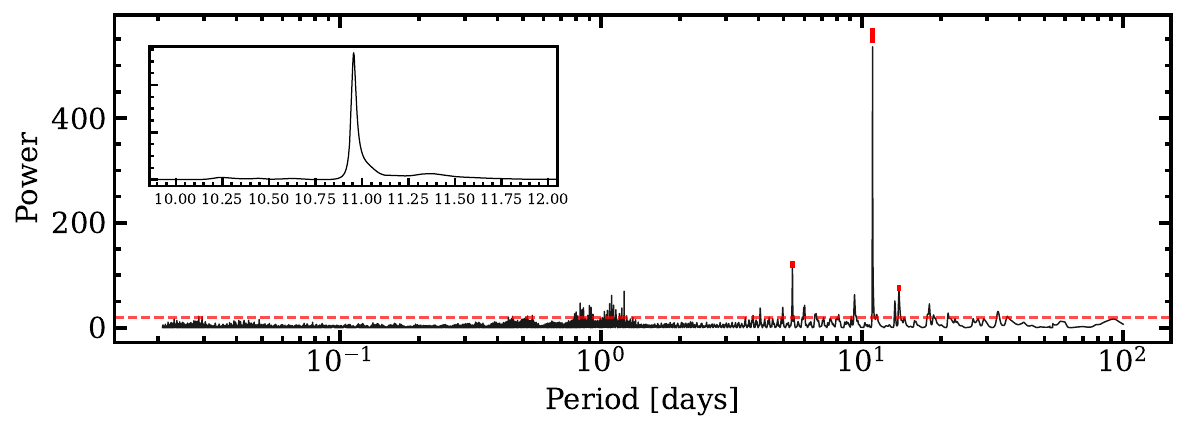}
        \end{subfigure}\hfill
    \begin{subfigure}[t]{0.32\linewidth}
        \includegraphics[width=1\linewidth]{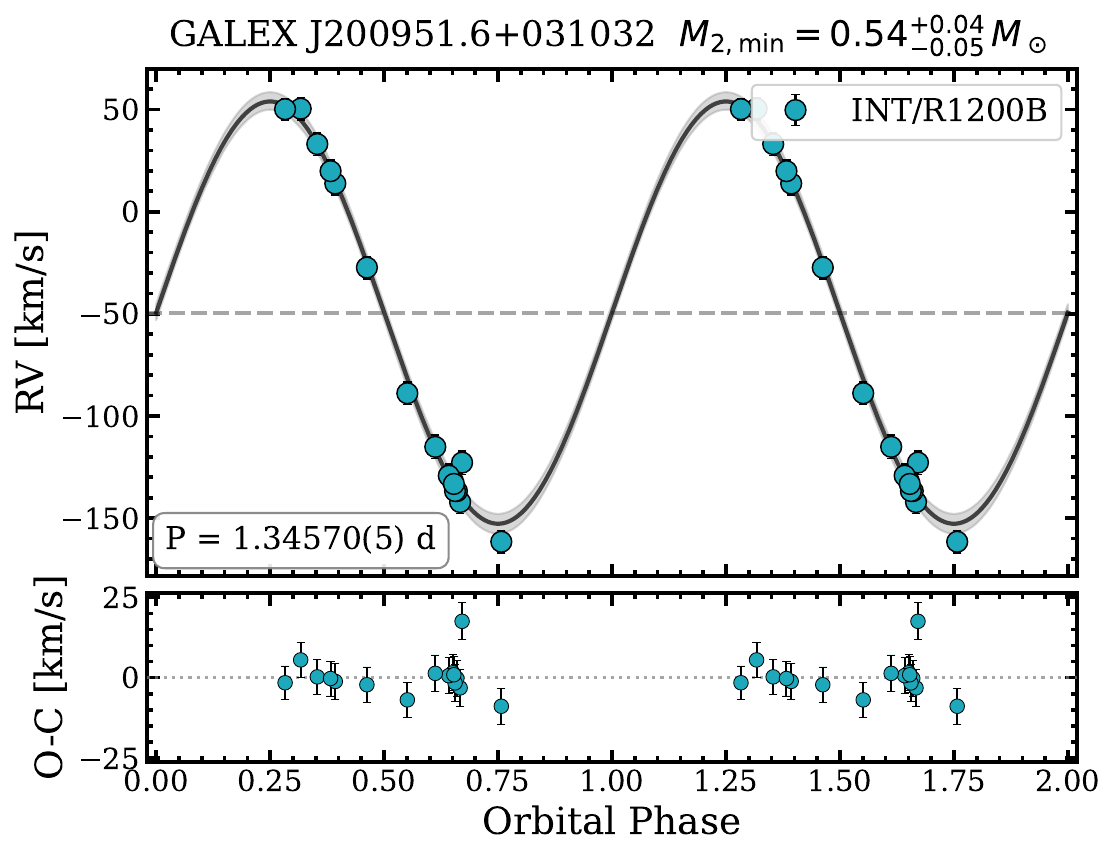}
        \includegraphics[width=1\linewidth]{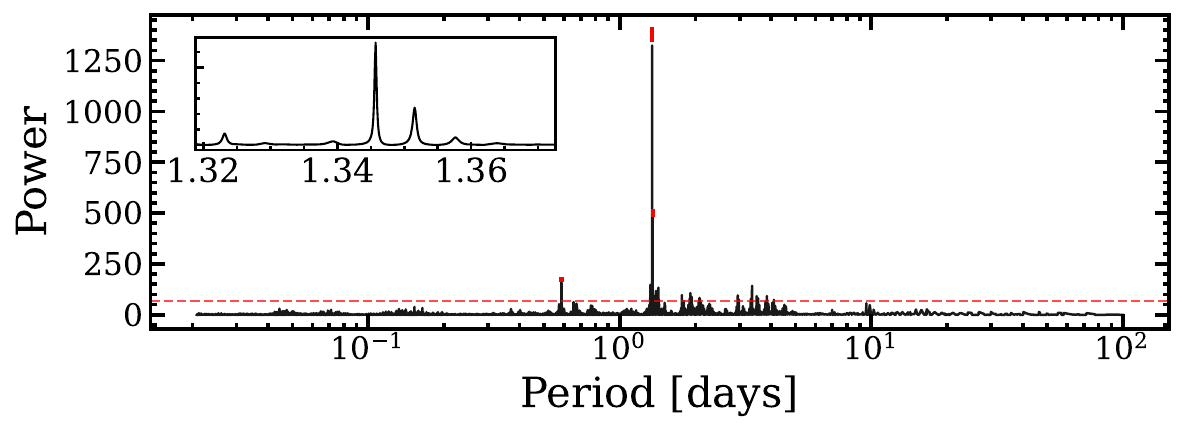}
    \end{subfigure}
    \caption{RV curves and power spectra for the three systems discussed in Sect.~\ref{highlighted_systems}. No time-series photometry is available for these stars. }
    \label{fig:power_rv_examples}
\end{figure*}

Radial velocities and LCs for a selection of noteworthy systems from the sample are presented in Figs.~\ref{fig:highlighted_systems} and \ref{fig:power_rv_examples}, where all are commented on in more detail below. All remaining newly solved systems can be seen in Figs.~\ref{fig:rv_lc_power} and \ref{fig:rv_power} in the appendix. In Fig.~\ref{fig:highlighted_systems} we highlight nine systems that are representative of key findings in this work, including LC constrained solutions, high companion masses, and evolutionarily significant configurations. Figure~\ref{fig:power_rv_examples} shows three example systems with only RV curves and their corresponding power spectra as photometry was not available.

\paragraph{TYC\,769-1149-1} (Fig.~\ref{fig:highlighted_systems}, upper-left panel) is an sdOB$+$WD and has an orbital period of $P=0.5985526(2)$ days. A clear Doppler boosting signal is displayed in the TESS LC, with an amplitude of $0.06$\% and was jointly fitted with the RV data. Using the SED-derived sdB mass of $0.61\pm0.12$~\msun, the minimum companion mass is $0.77\pm0.06$ \msun\, bringing the total minimum system mass to $1.38$~\msun. The most probable mass accounting for orbital inclination pushes this towards the Chandrasekhar limit, making TYC~769-1149-1 a type~Ia progenitor. With a period of just over 14 hours, it is a massive double WD binary progenitor in a relatively wide orbit. Assuming the two components evolve in isolation, we adopted the merging time equation from \citet{Peters_1964PhRv..136.1224P} and find that the system will merge due to angular momentum loss through gravitational wave radiation in $28.73^{+6.3}_{-4.4}$~Gyr and possibly detonate as a SN~Ia. Nevertheless, detailed modelling is needed to confirm this. A nearby (49~pc) double WD binary set to detonate in 23~Gyr was recently discovered \citep{Munday_2024MNRAS.532.2534M, Munday_2025NatAs...9..872M} with a strikingly similar orbital period of 14.23~hours. Although TYC\,769-1149-1 does not belong to the population of hot subdwarfs sitting below the EHB (see Sect.~\ref{intermediate_mass}), its high mass strongly suggests descent from a binary system comprising two early B-type MS stars \citep[see figure 4 in][]{Arancibia_Rojas_2024MNRAS.52711184A}. It did not therefore ignite its helium core degenerately as is often assumed. Its position high above the terminal-age extreme horizontal branch (TAEHB) in the HRD (see the annotation in the right panel of Fig.~\ref{fig:mass_m2}) is consistent with the high mass of the hot subdwarf and is also comparable to the recently discovered massive hot subdwarf binary LAMOST~J065816.72$+$094343.1 \citep{Mattig_2026A&A...706L...2M}.

\vspace{-10pt}
\paragraph{LS\,IV$+$09\,2} (Fig.~\ref{fig:highlighted_systems}, upper-central panel) was identified as a HW~Vir (eclipses and reflection effect; sdB$+$dM) in \citet{Ranaivomanana_2025AA...693A.268R} and has an orbital period of $0.225154(4)$~days. We followed up and solved for the RV curve in this work and modelled its ATLAS LC (steel-blue data points). We constrained its orbital inclination to $76.3\pm1.3\degr$, giving a fairly low companion mass of $0.098\pm0.022$~\msun, which is close to the H burning cut-off mass. 

\vspace{-10pt}
\paragraph{TYC\,3135-86-1} (Fig.~\ref{fig:highlighted_systems}, upper-right panel) was first spectroscopically confirmed as an sdB in Paper~I. 
A more detailed analysis in Paper~II reveals the mass of the sdB to be $0.47^{+0.08}_{-0.07}$~\msun\ but with a rather large radius of $0.355^{+0.005}_{-0.005}$~\rsun. Its spectroscopic parameters \teff$\approx21~000\,{\rm K}$, \logg$\approx5.0$, and \logy$\approx-1.95$ also position it high on the EHB in the Kiel and HRD (see Fig.~\ref{fig:mass_m2}). It is possible, therefore, that this sdB is in a bloated state shortly after its formation and is currently evolving towards the EHB, though a post-EHB evolutionary state is also possible. Clear ellipsoidal modulation and Doppler boosting is seen in the TESS LC when fitted together with the RV data (0.2489(1)~days) revealing the nature of the companion to be a WD of minimum mass $0.48\pm0.04$~\msun. The LC analysis constrains the orbital inclination to $60.6\pm12.3\degr$, which gives the WD a mass of $0.57\pm0.07$~\msun\,, bringing the total system mass to $\sim1.04$~\msun\ and qualifying it as another massive double WD progenitor. We find that the system will merge in $4.4\pm0.3$~Gyr and possibly produce a peculiar thermonuclear transient.

\vspace{-10pt}
\paragraph{TYC\,7691-3990-1} (Fig.~\ref{fig:highlighted_systems}, central-left panel) was initially selected by \citet{Geier_2019} and later reported as a pulsator based on TESS FFI data by \citet{Sahoo_I_2020MNRAS.499.5508S}. However, it was first spectroscopically classified as an sdB in Paper~I.
\citet{Ranaivomanana_2025AA...693A.268R} recently recorded it as having a $0.891169$ day orbital period from the \textit{Gaia} LC with a sinusoidal light variation. However, our spectroscopic follow-up finds a very short orbital period of $0.0743(2)$~days (1.78~hours), fitted together with the TESS LC, which shows clear ellipsoidal deformation. This object has the shortest orbital period among the newly discovered systems and the third shortest in the full 500~pc sample (after CD-30~11223 and HD~265435). With a semi-amplitude of $K = 175\pm3$~\kms, it occupies an isolated region of the $K$-period diagram (annotated in the central-left panel of Fig.~\ref{fig:combined_period_analysis}), lying between the loci of high- and low-mass companions. The minimum mass of the WD companion is $0.28\pm0.02$ \msun, which increases only slightly to $0.285\pm0.008$ \msun\ given our inclination of $87.6\pm14.7$ degrees constrained by the TESS LC fit. This means that the companion is likely a He-WD. The system will also merge imminently in $320\pm4$~Myr, likely soon after the sdB evolves into a WD. If, however, it merges during the sdB lifetime, this may result in the merger product becoming a rejuvenated shell-helium-burning \citep{Justham_2010} hot subdwarf. 

\vspace{-10pt}
\paragraph{HD\,191351} (Fig.~\ref{fig:highlighted_systems}, central panel) is the brightest and nearest of the newly solved systems presented in this work, at a distance of approximately $170$~pc. The star is also below the EHB (Sect.~\ref{intermediate_mass}; $\log L/L_{\odot} = 0.77\pm0.02$), qualifying it as a likely intermediate-mass MS star progeny. \citet{He_2025MNRAS.543.2243H} recently identified it as an ellipsoidally deformed system with an orbital period of $\sim$2.14~hours, in agreement with our findings. We refine its orbital period to $0.089162439(6)$~days. Similar to TYC\,7691-3990-1 above, the derived minimum mass of the WD companion is fairly low at $0.23\pm0.02$~\msun, positioning it in the sparsely populated region in the $K$-period diagram (annotated in the central-left panel of Fig.~\ref{fig:combined_period_analysis}). Fitted together with the TESS LC, we find the orbital inclination to be $31.7\pm2.9$\degr, resulting in a WD companion mass of $0.562\pm0.061$~\msun\, and moving the system up in Fig.~\ref{fig:combined_period_analysis} to the shaded-red region. With an sdB mass of $0.52\pm0.07$~\msun\ from the SED, the system will likely merge in $295\pm40$~Myr.

\vspace{-10pt}
\paragraph{NSVS~842188} (Fig.~\ref{fig:highlighted_systems}, central-right panel) was first identified as a blue object in the First Byurakan Spectral Sky Survey \citep[FBS;][]{Abramyan_1994Ap.....37..224A}. It was later flagged as variable in \citet{Chen_2020ApJS..249...18C} and \citet{Ranaivomanana_2025AA...693A.268R}, but was only recently classified as an sdB \citep{Barlow_2022ApJ...928...20B}. We find an orbital period of $0.19493418(2)$~days, fitting the RV data and TESS reflection effect signal jointly. Its orbital inclination is $70.5\pm4.8\degr$, and the dM companion mass is $0.266\pm0.031$~\msun.

\vspace{-10pt}
\paragraph{2MASS~J15292631+7011543} (Fig.~\ref{fig:highlighted_systems}, lower-left panel) is also first identified as a blue object in the FBS \citep{Abramyan_1994Ap.....37..224A}. The object, which displays a clear reflection effect in its TESS LC, is further identified in \citet{Barlow_2022ApJ...928...20B} as an sdBV in the instability strip with a dozen further signals revealed after pre-whitening of the reflection effect signal. These are interpreted as g-mode pulsations. We find an orbital period of $0.19964386(3)$~days fitted jointly with the TESS LC. The orbital inclination is constrained to $71.2\pm4.8\degr$, and the dM companion mass is $0.164\pm0.024$~\msun.

\vspace{-10pt}
\paragraph{TYC\,4470-864-1} (Fig.~\ref{fig:highlighted_systems}, lower-central panel) was first found to show photometric variability originating from both binarity (reflection-effect) and pulsations \citep{Krzesinski_2022AA...663A..45K, Krzesinski_2022MNRAS.516.1509K, Uzundag_1_2024AA...684A.118U}. A joint fit of RVs and TESS photometry gives an orbital period of $0.46798033(1)\,$days. The orbital inclination and dM companion mass are $80.2\pm4.8\degr$ and $0.118\pm0.014\,$\msun, respectively.

\vspace{-10pt}
\paragraph{ATO~J029.0051+40.0561} (Fig.~\ref{fig:highlighted_systems}, lower-right panel) is flagged as a variable object in ATLAS \citep{Heinze_2018AJ....156..241H} and confirmed spectroscopically as an sdB in \citet{Lei_2023ApJ...942..109L_spectra}. A reflection effect was later found by \citet{He_2025AA...693A.121H}. 
Our RV data jointly with TESS give a period of $0.19365852(4)$~days, an orbital inclination of $73.8\pm4.0\degr$, and a dM companion mass of $0.119\pm0.014$~\msun. 

\vspace{-10pt}
\paragraph{Gaia~DR3~2200231588777454208} (Fig.~\ref{fig:power_rv_examples}, left panel) is a newly identified object with an orbital period of $0.17441(3)$~days. It hosts a WD companion with a minimum mass of $0.77\pm0.05$~\msun, bringing the total mass to 1.19~\msun\, and qualifying it as a progenitor to a massive double WD binary. The merger time for this object is $1.5\pm0.2$~Gyr. The system is annotated in the upper-left panel of Fig.~\ref{fig:combined_period_analysis}. No photometric signals are detected at $\pm10$\% around its orbital period in TESS, ZTF, and ATLAS to constrain the orbital inclination and thus the companion mass. The sdB primary is below the EHB with a $\log L/L_{\odot} = 0.73$ (annotated in the HRD in Fig.~\ref{fig:mass_m2}), meaning it may descend from two B-type MS stars.

\vspace{-10pt}
\paragraph{PG~1610$+$529} (Fig.~\ref{fig:power_rv_examples}, central panel) was identified early as an sdB in the Palomar--Green survey \citep{Green_1986} and is included in the first catalogue of spectroscopically identified hot subdwarf stars \citep{Kilkenny_1988}. We followed up and solved this binary for the first time using the INT and the NOT, acquiring 35 epochs over a baseline of 789~days. We find a best-fit orbital period of $10.955(4)$~days from the RV curve. Two further small peaks appear above the $4\sigma$ threshold (see the periodogram in the central panel of Fig.~\ref{fig:power_rv_examples}), but the RV curves folded on these periods are a very poor match to a sinusoidal solution. Inspection of the window function confirms both to be sampling aliases, and the MCMC sampler converges on the highest peak with an integrated autocorrelation time of $<1.7$~steps. This system has the second-longest orbital period among the newly solved binaries, and the WD companion has a fairly massive minimum mass of $0.63\pm0.05$~\msun.

\begin{figure}
    \centering
        \includegraphics[width=1\linewidth]{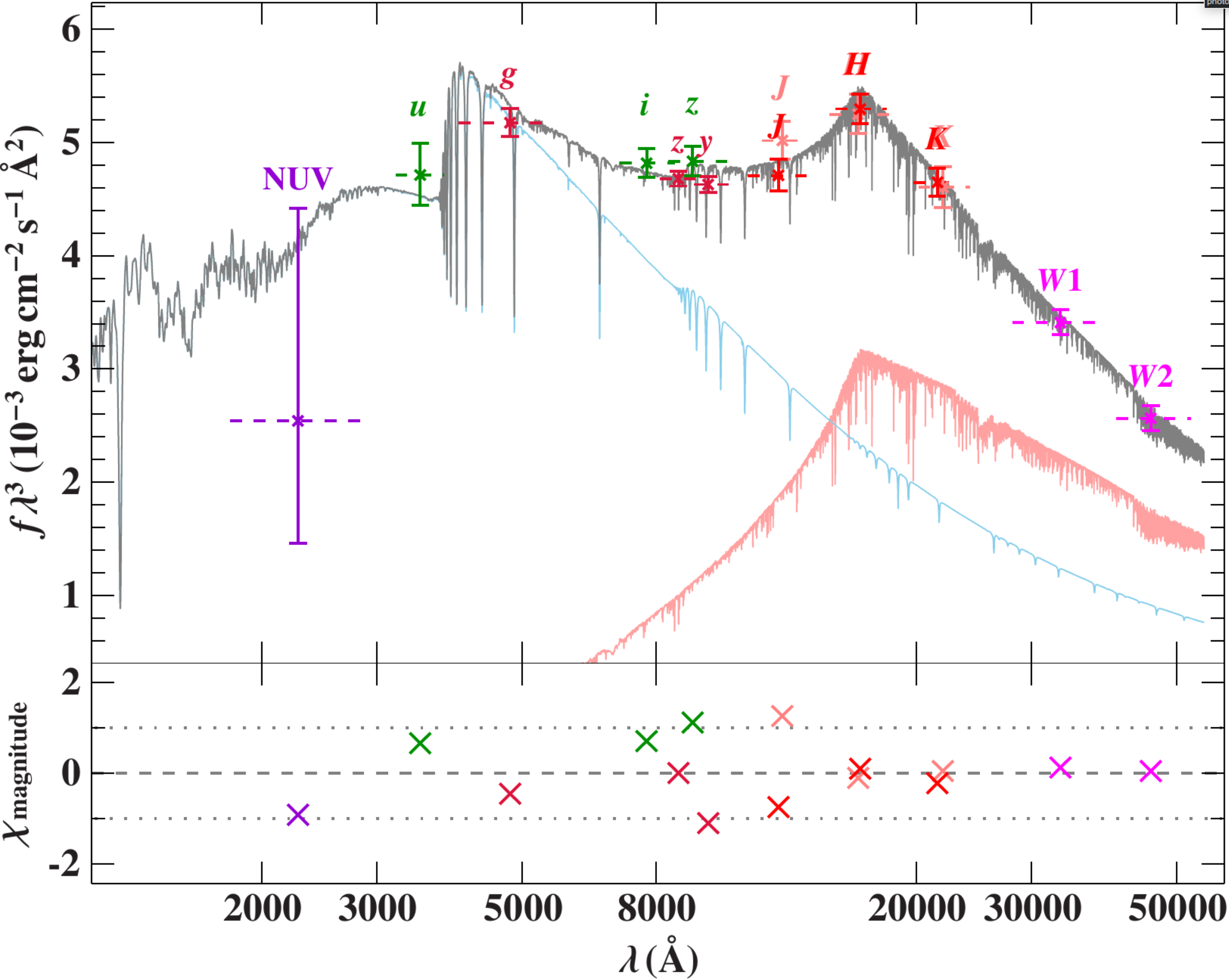}
    \caption{SED of GALEX~J200951.6$+$031032. Top panel: Filter-averaged fluxes converted from observed magnitudes from different photometric surveys. To ease the slope of the distribution, the flux $f_\lambda$ is multiplied by the wavelength to the power of three. Photometric fluxes are displayed as coloured data points with their respective uncertainties and filter widths (dashed lines). The best-fit composite model is shown in grey, which combines the light from the hot subdwarf (light blue) and MS (light red) components. The uncertainty-weighted residual $\chi^{2}$ are shown in the lower panel, demonstrating the quality of the fit.}
    \label{fig:new_k}
\end{figure}

\vspace{-10pt}
\paragraph{GALEX~J200951.6$+$031032} (Fig.~\ref{fig:power_rv_examples}, right panel) was classified in Paper~I as an sdB+K3V owing to the strong IR excess visible in its SED fit (see Fig.~\ref{fig:new_k}). No absorption features originating from the K-type companion are visible in the acquired INT/R1200B optical spectrum ($\sim3700$--$5200$~\AA) as such a companion contributes only 1-2\% of the total system flux at $\sim5000$~\AA. Unlike most composite hot subdwarfs, however, this system exhibits large RV variations, and subsequent follow-up observations yield an orbital solution with $P=1.3457(5)$~days. Combining the hot subdwarf mass of $0.45\pm0.12\,$\msun\ with this orbital solution gives a minimum companion mass of $0.54^{+0.04}_{-0.05}\,$\msun\,, in good agreement with the SED-derived companion mass of $0.49\pm0.15\,$\msun. Low-frequency signals are present in the TESS periodogram of GALEX~J200951.6$+$031032. At these orbital and stellar parameters a strong reflection effect would be expected even at low orbital inclination (as seen in GALEX~J2205-3141; \citealt{Schaffenroth_2023_2}), yet none is observed. The TESS signals are more likely attributable to stellar spots on the K-type companion. We therefore conclude that the RVV originates from a compact WD in the short-period orbit and that the agreement between the minimum and SED-derived companion masses is coincidental.

The SED-derived parameters of the K-type companion ($M=0.49\pm0.15$~\msun, $R=0.65\pm0.05$~\rsun) are consistent with MS evolutionary tracks \citep{Baraffe_2015A&A...577A..42B}. Furthermore, the sdB and K-type are therefore consistent with being at the same distance, supporting the interpretation of this system as a triple, with the K-type as a wide tertiary. The Gaia RUWE of 2.33 is somewhat inflated, as is typical for composite hot subdwarf systems (Paper~I), though the companion is not resolved as a separate source. Long-term IR spectroscopic monitoring is required to confirm the triple nature of this system and potentially constrain the tertiary orbit. 
This system is similar to the triple candidate GD~319A, which is in the 500~pc sample with a spectral class of sdB$+$K3V. A short-period orbital solution of $\sim0.6$~days was derived by \citep{Maxted_2000MNRAS.311..877M} and gives a minimum companion mass of $0.09$~\msun. Several epochs taken during our programme refine the orbital solution to be $P=0.6027511(7)$~days. Because no TESS photometric signal is seen in its LC, this system is also a likely a potential triple (see Geier et al. in prep.), and therefore the second in the 500~pc sample with interestingly similar characteristics. As composite hot subdwarfs, GALEX~J200951.6$+$031032 and GD~319A are not included in the binary statistics in Table~\ref{tab:binary_fractions} but discussed here for completeness and future reference. 

\section{Binary population properties and discussion}
\label{discussion}

\begin{figure*}[h!]
    \centering
    \begin{subfigure}[t]{0.48\linewidth}
        \centering
        \includegraphics[width=\linewidth]{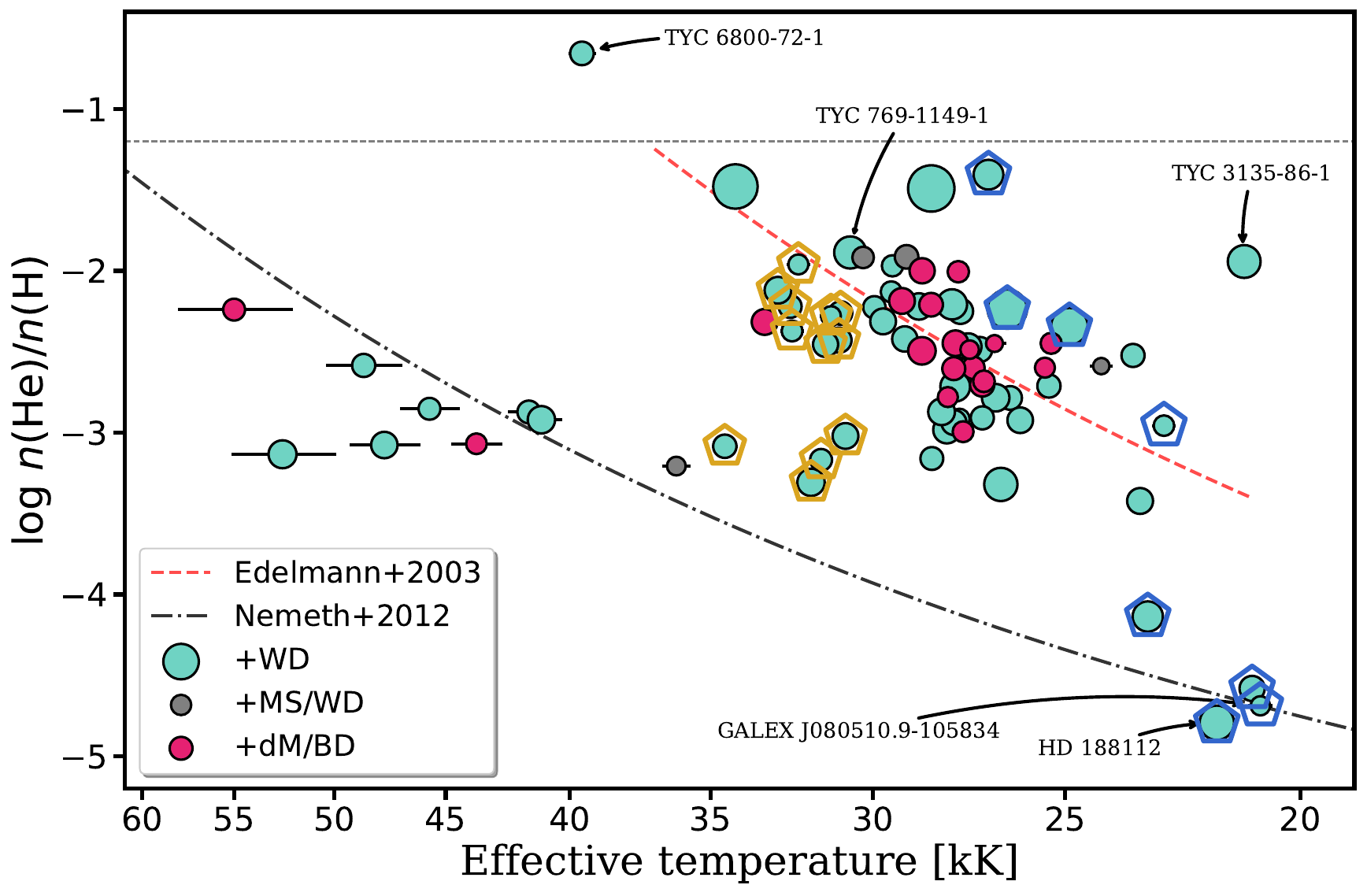}
    \end{subfigure}
    \hfill
    \begin{subfigure}[t]{0.48\linewidth}
        \centering
        \includegraphics[width=\linewidth]{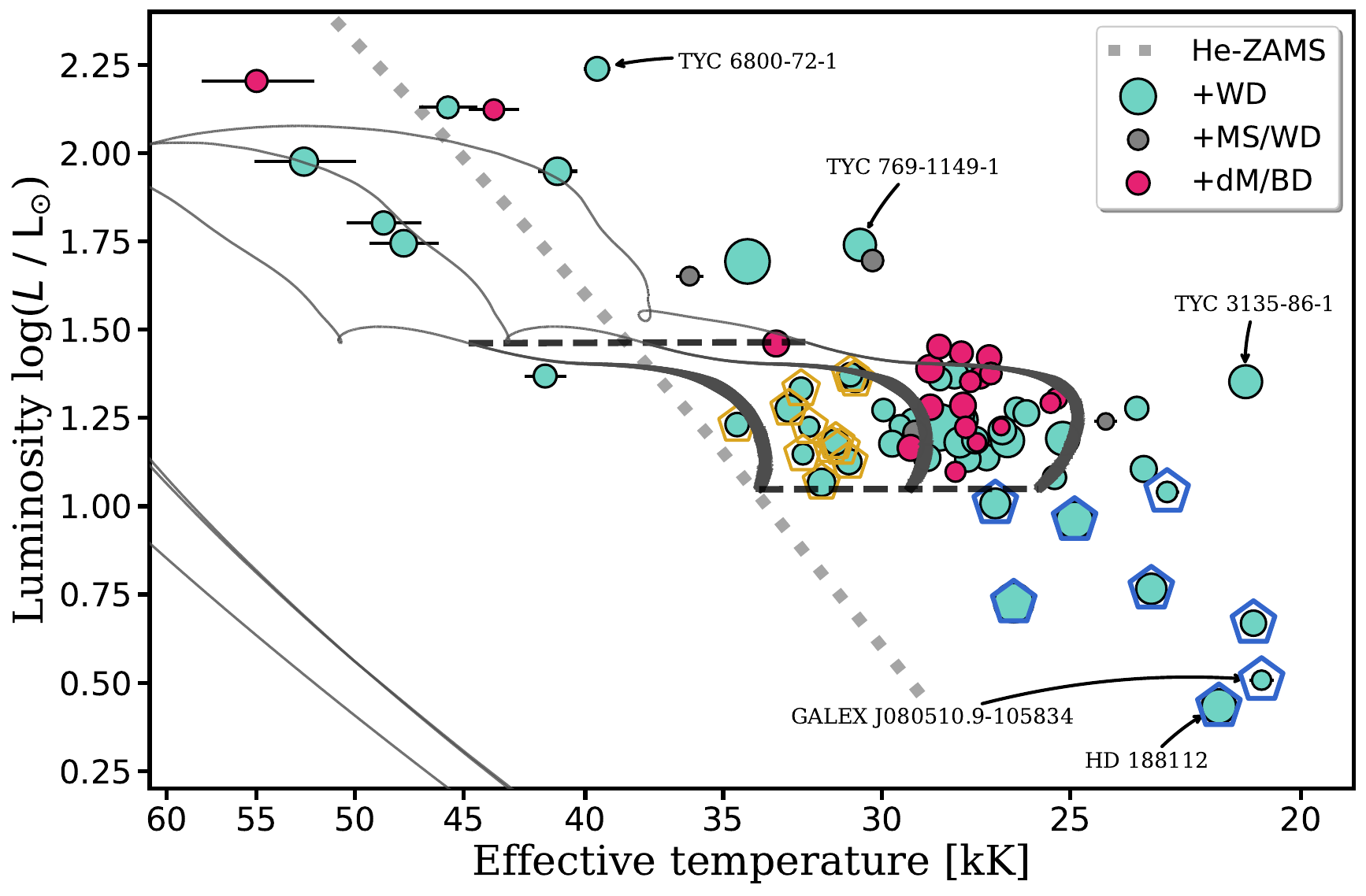}
    \end{subfigure}
    \vspace{0.5em}
    \caption{\textit{Left:} Helium abundance - $T_{\mathrm{eff}}$ diagram for the 500~pc sample binaries, coloured by companion type and sizes scaled by RV semi-amplitudes. The helium sequences identified in earlier studies \citep{Edelmann_2003AA...400..939E, Nemeth_2012MNRAS.427.2180N, Luo_2016, Lei_2018ApJ...868...70L} are shown. \textit{Right:} HRD of the same binaries with evolutionary tracks from \citet{Han_2002} for comparison. The blue and gold pentagons highlight subluminous (below EHB) and hot-EHB stars, respectively.}
    \label{fig:mass_m2}
\end{figure*}

\begin{figure}
    \centering
        \includegraphics[width=1\linewidth]{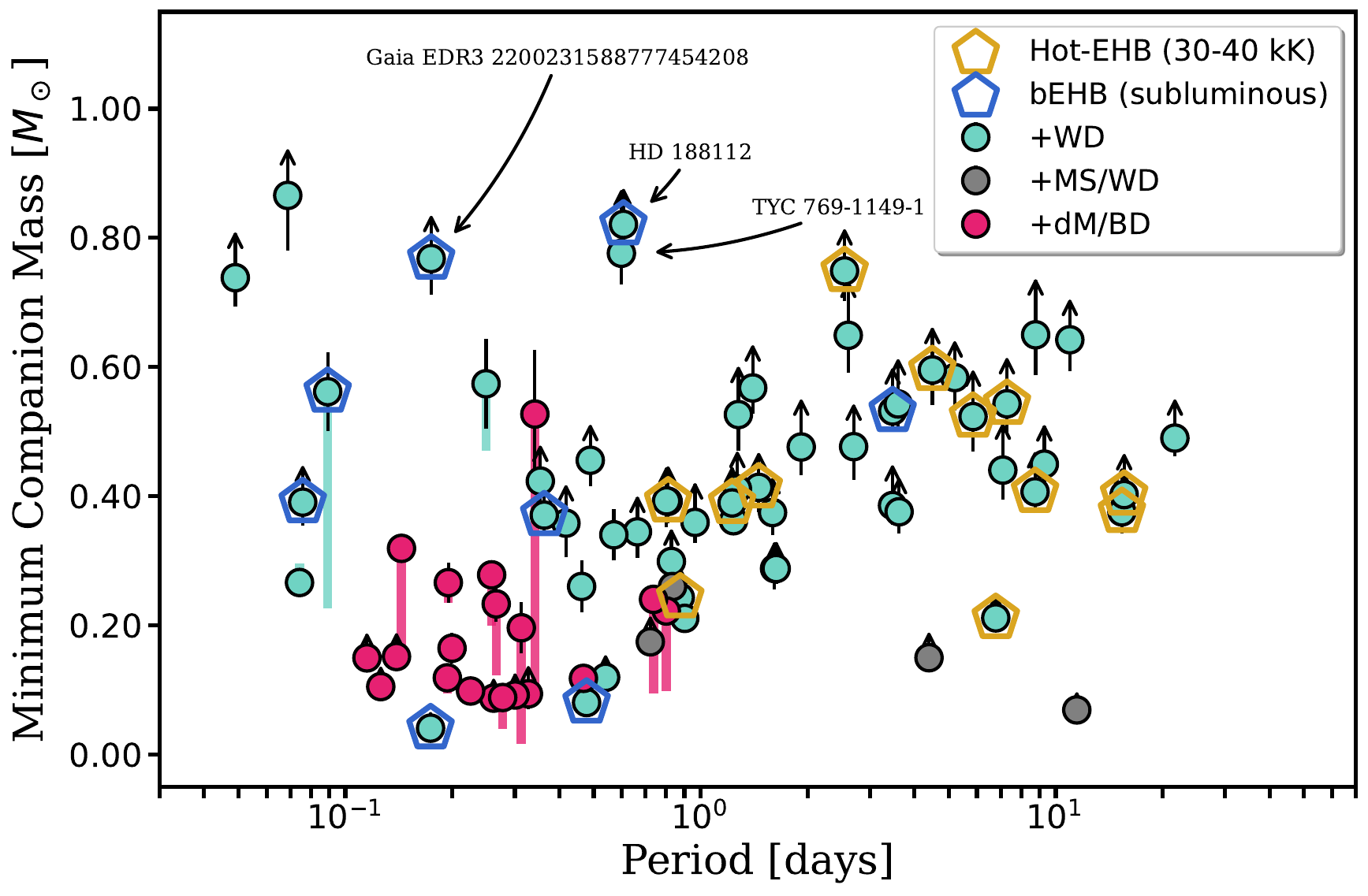}
        \includegraphics[width=1\linewidth]{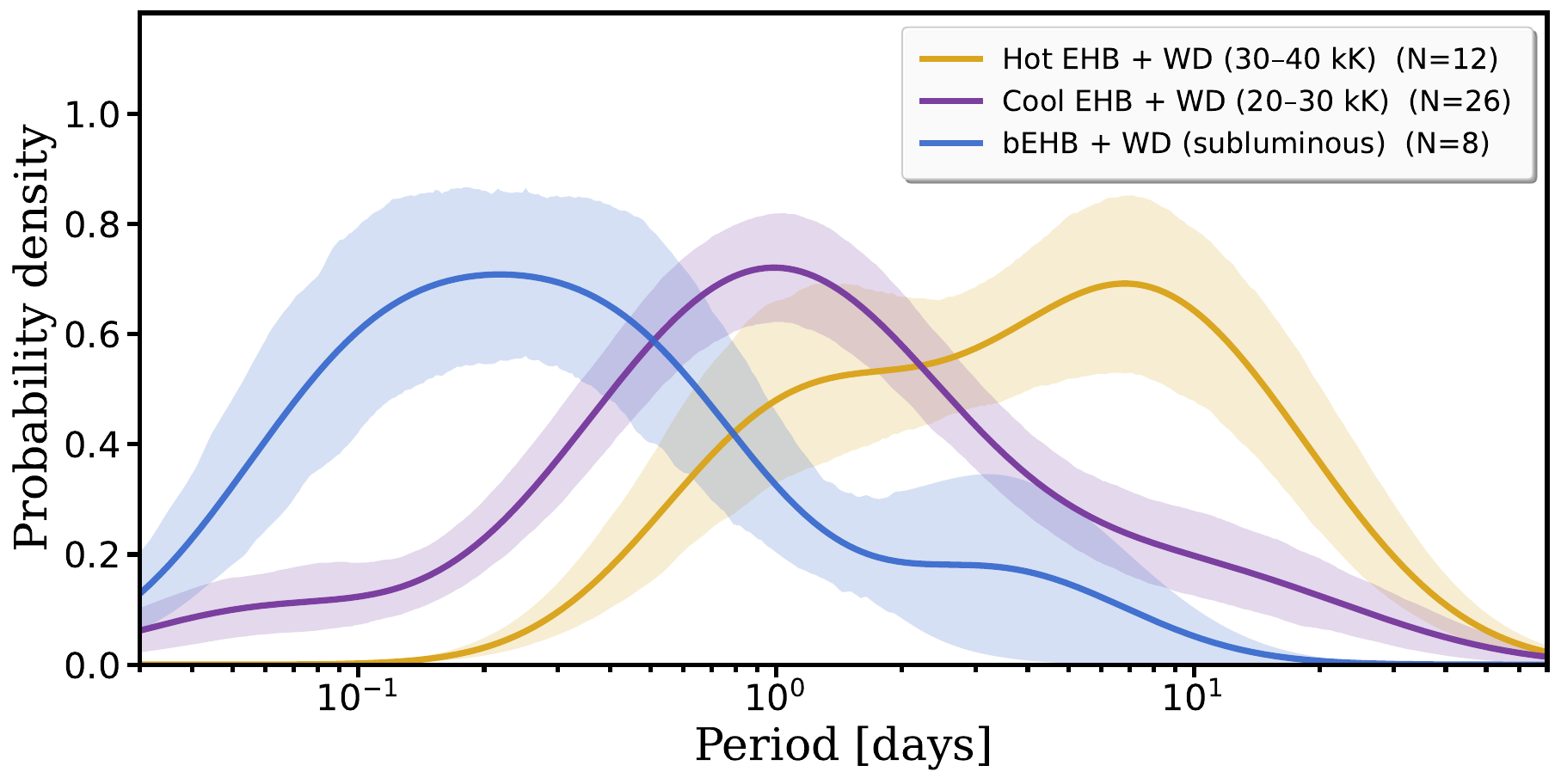}
    \caption{\emph{Top: }Minimum companion mass vs orbital period for all close binaries in the 500~pc sample. The upward arrows indicate systems where only a lower mass limit is available due to an unconstrained orbital inclination. The vertical lines connect the minimum mass to the true companion mass for systems with a known inclination from LC modelling (Table~\ref{tab:lc_fitting} or from \citealt{Schaffenroth_2023_2}). \emph{Bottom: } Normalised probability density period distributions for the hot-EHB, cool-EHB, and bEHB systems hosting WDs in our sample. }
    \label{fig:mass_period}
\end{figure}

The HRD for the short-period binary population is shown in the right panel of Fig.~\ref{fig:mass_m2}, where the luminosities and effective temperatures are adopted from Paper~II. The dotted grey line marks the helium MS from \citet{Paczynski1971}. The horizontal dashed black lines indicate the solar metallicity zero-age extreme horizontal branch (ZAEHB; lower) at $\log L/L_{\odot} = 1.05$ corresponding to the start of the 0.45~\msun\ evolutionary tracks from \citet{Han_2002}, and the TAEHB (upper) at the point of core helium depletion. Three (post-)EHB evolutionary tracks for envelope masses of 0.00, 0.001, and 0.005~\msun\ (left to right; $M_\textrm{total} = 0.45$ \msun) from \citet{Han_2003} are shown as solid black lines where the thickness of the lines are scaled in proportion to the evolutionary lifetime. The size of the markers are scaled in size with their semi-amplitudes. The HRD distribution of our binary sample is inhomogeneous, in agreement with previous studies \citep{Geier_2022, He_2025AA...693A.121H}. We discuss several of the observed properties below.

\subsection{The hot-EHB stars}
\label{rvv_classes}

In Sect. \ref{close_binary_fraction} we noted that the binary fraction of the sdOB spectral class is much lower than the sdB and sdO hot subdwarfs ($23.1^{+5.8}_{-6.3}$\%; Table \ref{tab:binary_fractions}). In the HRD in Fig.~\ref{fig:mass_m2}, the sdOBs are located in the region between the ZAEHB and TAEHB and are typically hotter than $\sim$30~000~K. 
\citet{Geier_2022} report a lower fraction of their EHB3 stars, which were selected based on their position in the Kiel diagram at the hot end of the EHB. For the purpose of comparing this region in the HRD to that of \citet{Geier_2022} and others and to reduce the dependency on spectral classification by visual inspection (Paper~I), we selected the stars in our sample located in the 'hot-EHB' of the HRD. These are defined as having $30~000~\mathrm{K}\leq$ \teff\ $\leq40~000~\mathrm{K}$ and $1.05\leq$ $\log L/L_{\odot}$ $\leq1.4$. This selection returns 16 detected binaries, which corresponds to a corrected binary fraction of $f_{\mathrm{corr}}=28.4^{+5.2}_{-6.6}\%$. The 12 solved hot-EHB binaries are indicated with gold pentagons in both panels of Figs.~\ref{fig:mass_m2} and~\ref{fig:mass_period}.

We also find that the hot-EHB stars exclusively host WD companions. Interestingly, most hot-EHB stars have orbital periods longer than one day, with a median of $\sim$5.1 days. This is best seen in the lower panel of Fig.~\ref{fig:mass_period}, where we show smoothed, normalised KDEs for a clear comparison between the hot-EHBs (gold), the cool-EHBs (purple; $20~000~\mathrm{K}\leq$ \teff\
$\leq30~000~\mathrm{K}$), and the below-EHB (bEHB) objects (blue). The KDEs were constructed by summing a Gaussian kernel centred on each system, computed in $\log_{10}P$ with a smoothing bandwidth of $0.3$~dex. The shaded regions show the 16th and 84th percentiles obtained by resampling each sub-class 2000 times. Systems hosting $+$dM/BD companions are not included.
Since the majority of binaries in the sample have periods shorter than one day, this over-density of longer-period hot-EHB systems is unlikely to be a statistical effect and instead appears to be an intrinsic property of the population. We compare this to the \emph{cool-EHB} objects, which occupy the 20~000 and 30~000~kK range on the EHB and exhibit a much tighter distribution with a median of $\sim1$~day. A similar deficit of short-period binaries at the hot end of the EHB is visible in Fig. 14 of \citet{Schaffenroth_2022_1}. Both \citet{Geier_2022} and \citet{He_2025AA...693A.121H} report a lower binary fraction for this region in the HRD. This is unsurprising given that longer-period systems have intrinsically lower semi-amplitudes and are therefore harder to detect and solve. Furthermore, the absence of dM/BD companions to sdOBs in this work is consistent with the findings of \citet{Schaffenroth_2023_2}. Their photometry-based survey would be sensitive to such systems via their reflection effect signals, suggesting such systems are intrinsically rare rather than affected by a detection bias. Although hot-EHB stars in binaries seem to have longer orbital periods, we are unlikely to miss them in our survey unless the companions are of very low mass ($\lesssim0.2$~\msun) or their orbital periods are longer than $\sim$30 days. This last point may hold given that \citet{Geier_2023AA...677A..11G} recently discovered the first hot subdwarf (spectroscopically classed as sdOB with \teff$\sim$30~kK) hosting a massive ($1.5 ^{+0.37}_{-0.50}$~\msun), compact companion in a wide (P $=$ 892.5 $\pm$ 60.2 d) orbit. Non-RV-variable iHe-sdOBs occupy a similar HRD region, but our classification scheme and high quality data allow us to exclude them as a source of contamination, which was not possible in \citet{Geier_2022}. However, non-RV-variable iHe-sdOBs might be at least partially linked to the hot-EHB population through a distinct evolutionary pathway or represent a different evolutionary stage of the same pathway. \citet{Marcello_2008AA...491..253M} suggested that some hot-EHB stars evolve from non-RV-variable iHe-sdOBs, meaning the hot-EHB population may draw from both this channel and the canonical RV-variable sdB formation pathways.
The resulting lower binary fraction and longer orbital periods of iHe-sdOBs may therefore reflect this mixed origin, though larger samples and detailed modelling will be needed to disentangle the relative contributions of each channel.

The left panel of Fig.~\ref{fig:mass_m2} shows the helium abundance--\teff\ diagram. The helium sequences identified by \citet{Edelmann_2003AA...400..939E} and \citet{Nemeth_2012MNRAS.427.2180N} are marked, along with the horizontal dashed line at \logy\ $=-1.2$, which distinguishes the helium-enriched objects above it. 
A single object lies above this line, TYC\,6800-72-1. This object is a newly identified iHe-sdOB with \logy\ = $-0.74$. It has an orbital period of 3.6001(6)~days with a WD companion of minimum mass $0.55^{+0.05}_{-0.06}$~\msun. With the exception of PG\,1544$+$488 -- a known eHe-sdOB$+$He-sdB double hot subdwarf system \citep[][ annotated in the central-left panel of Fig. \ref{fig:combined_period_analysis}]{Jeffery_2014MNRAS.440.2676S} -- which is omitted to preserve the dynamic range of the figure, none of the extremely helium-rich objects shows indications of RVV. Thus, they are likely merger products \citep{Han_2002}. 

\subsection{Intermediate-mass progenitors}
\label{intermediate_mass}
It was found in Paper~II that 10\% of sdBs are located below the $0.45$~\msun\ ZAEHB (bEHB) in both the Kiel and HR diagrams (also named bEHB by \citealt{Geier_2022}) and that those objects, on average, have lower masses than the bulk of the hot subdwarf population. This may suggest that these objects are the progenies of intermediate-mass stars, which ignited helium either non-degenerately or under partly degenerate conditions. These objects are subluminous compared to the bulk of the sdB population. They were selected in Paper II as having $\log L/L_{\odot} \leq 1.05$, which reflects the zero-age EHB of the $0.45$~\msun\ evolutionary tracks from \citet{Han_2002}. In the central-left panel of Figs.~\ref{fig:combined_period_analysis}, ~\ref{fig:mass_m2}, and \ref{fig:mass_period}, these objects are indicated with blue pentagons. 

We find that all objects located below the ZAEHB appear to have WD companions, suggesting a distinct evolutionary pathway. In Fig.~\ref{fig:mass_m2}, marker sizes scale with RV semi-amplitude, revealing that bEHB-sdBs exhibit systematically larger RVV compared to their EHB counterparts (see also the central-left panel of Fig.~\ref{fig:combined_period_analysis}). This is consistent with that observed in \citet{He_2025AA...693A.121H}. Their WD companions are also more massive on average, particularly when excluding GALEX~J080510.9-105834, which has an unusually low minimum companion mass of $\sim 0.04$~\msun\ (annotated in Fig.~\ref{fig:combined_period_analysis}). 

With the exception of UCAC4~575-030949, all bEHB-sdBs in our sample have orbital periods shorter than one day with a median of $\sim0.25$~days (see the lower panel of Fig. \ref{fig:mass_period}), in stark contrast to the hot-EHB stars. Two systems (Gaia~DR3~2200231588777454208 and HD~188112) host companions with minimum masses exceeding 0.75~\msun, which have parameters similar to the double detonation progenitor PTF1~J2238$+$7430 modelled in detail by \citet{Hernandez_2026A&A...708A.108H}. Gaia~DR3~2200231588777454208 in particular, is newly identified in this work with an orbital period of 0.174416(4)~days and has a combined mass of 1.19\,\msun\,, qualifying it as a progenitor to a massive double WD binary (Sect.~\ref{highlighted_systems}). 

\section{Birthrates}
\label{birthrates}
Leveraging the binary population presented in this paper, this section provides birthrate calculations for various subsets of our sample. To derive a birthrate, we require a space density $\rho$ and an estimated lifetime $\tau$. We assigned a local space density to each object following the method described in section 7.5 of Paper~II, normalised to the number of stars in each subsample. This $z-$weighted density model accounts for the declining space density away from the Galactic mid-plane, implying a position-dependent contribution to each birthrate. The scale heights adopted for the thin and thick discs are $z_{h,\mathrm{thin}} = 300$~pc, $z_{h,\mathrm{thick}} = 900$~pc \citep{McMillan_2017MNRAS.465...76M}, and the mid-plane space density taken from Paper~I is $\rho_{0} = 5.17 \pm 0.33 \times 10^{-7}$ stars pc$^{-3}$. The total lifetime, $\tau$, of each hot subdwarf was estimated by interpolating within the theoretical evolutionary tracks of \citet{Han_2002} in the HRD unless otherwise stated, and the birthrate was then obtained by dividing the local space density by the estimated lifetime. Uncertainties were carried through the calculation using MC arrays and Poisson uncertainties were added in quadrature to account for low-number statistics. 

We also scaled our results to obtain rates for the Galaxy. This was performed using the ratio of integrated stellar mass within 500~pc to the total stellar Galactic mass, adopting the thin and thick disc density profiles, as well as the Galactic bulge density profile of \citet{McMillan_2017MNRAS.465...76M}. The gas discs and dark matter halo were excluded, which is appropriate for our sample.

\subsection{Birthrates by companion type}
Of the solved binaries in our sample, 19 host $+$dM/BD companions (see Table~\ref{tab:binary_fractions}), are detected primarily through their LC variations. The unsolved systems (Table~\ref{table_new_binaries}) can be safely excluded as $+$dM/BD candidates, given their lack of LC variability. 
By dividing by the estimated lifetimes for each sdB$+$dM/BD system (derived from track interpolation), we estimate the local birthrate of these systems to be $1.5 \pm 0.4 \times 10^{-16}$ stars pc$^{-3}$ yr$^{-1}$. Scaled by the mass ratio of the Galaxy to our 500~pc sample, we find an inferred Galactic birthrate of $2.2^{+0.5}_{-0.5} \times 10^{-4}$ yr$^{-1}$ for sdB$+$dM/BD binaries. 

We identify 55 systems in our sample hosting WD companions. The birthrate of sdB$+$WD systems is $4.2 \pm 0.6 \times 10^{-16}$ pc$^{-3}$ yr$^{-1}$. This value rises to $5.4 \pm 0.4 \times 10^{-16}$ pc$^{-3}$ yr$^{-1}$ if we assume the 16 unsolved systems host WD companions, which is likely. These correspond to Galactic birthrates of $6.1^{+0.8}_{-0.8}  \times 10^{-4}$  yr$^{-1}$ and $7.9^{+1.2}_{-1.2}\times 10^{-4}$ yr$^{-1}$, respectively, which is a ratio of about 2.8-3.6:1 compared to the birthrate of the $+$dM/BD systems.

\subsection{SN~Ia occurrence rates}
\label{transients}
There are two systems in the 500~pc sample hosting WD companions, CD$-$30~11223 and HD~265435, which will possibly undergo a SN~Ia. CD$-$30~11223 \citep{Geier_2013AA...551L...4G, Deshmukh_2024MNRAS.527.2072D} is predicted to produce a SN~Ia via the double-detonation channel during the hot subdwarf lifetime in $\sim$55 Myr \citep{Geier_2013AA...551L...4G, Deshmukh_2024MNRAS.527.2072D}, whereas HD~265435 is expected to do so via a double-degenerate dynamical merger channel in $\sim$70 Myr after the sdB evolves into a WD \citep{Pelisoli_2021NatAs...5.1052P}. Assuming CD$-$30~11223 is the only system in the sample that will undergo a SN in 55~Myr \citep{Deshmukh_2024MNRAS.527.2072D} through the single degenerate channel, we calculate an inferred Galactic occurrence rate of $3.3^{+0.5}_{-0.4} \times 10^{-5}$ yr$^{-1}$. This is $1.2^{+0.4}_{-0.3}$\% of the predicted Milky Way SN~Ia rate \citep[$2.8\pm0.6 \times 10^{-3}$ yr$^{-1}$;][]{Liu_article}. 
Our estimate is consistent with the recent binary population synthesis study of \citet{abinaya_2025arXiv251111998R}, who report that hot subdwarf (helium donor) binaries with CO WD companions undergoing sub-Chandrasekhar double detonations contribute approximately 1\% of the observed SN~Ia rate.

Besides HD~265435, several other binaries in the sample have total combined masses close to Chandrasekhar mass, which lies between 1.35--1.38~\msun\ \citep{Jordan_2012ApJ...759...53J, Rotondo_2011PhRvD..84h4007R, Leung_2018ApJ...861..143L, Althaus_2022A&A...668A..58A}. For our calculations we therefore adopted a value of $M_{\mathrm{Ch}}=1.365$~\msun, above which we assume the system will explode when the total mass exceeds this. Carrying through uncertainties with MC arrays on the hot subdwarf mass, WD mass, Poisson errors, and inversely summing the merger timescales, we estimate the Galactic type Ia occurrence rate through the double degenerate channel to be $3.7^{+2.6}_{-2.1} \times 10^{-5}$ yr$^{-1}$. This corresponds about $1.3^{+0.4}_{-0.2}$\% of the Milky Way SN~Ia rate \citep{Liu_article}. We excluded the 16 unsolved binaries as candidate type~Ia progenitors from the calculated rate based on the permitted parameter range of RV and LC solutions. 

Combining both CD-30~11223 and HD~265435, this implies a total contribution of $2.5^{+0.7}_{-0.5}$\% towards the Galactic SN~Ia rate, inferred from our 500~pc sample. As also recognised in previous observational studies \citep[e.g.][]{Geier_2013AA...551L...4G, Pelisoli_2021NatAs...5.1052P}, our first, volume-complete constraint on the SN~Ia rate through these channels confirms that sdB$+$WD systems contribute a minor, yet still significant, fraction of Galactic SNe~Ia.

\subsection{Merger rates}
\label{mergers}
The systems where the WD does not explode will likely evolve into a single massive C/O WD. Here, we performed the same calculation as in Sect. \ref{transients} to estimate the merger rate of sdB$+$WD by summing up the merger times for those with a total mass $M_\mathrm{tot}$\footnote{The total system mass was taken as $M_\mathrm{tot} = M_\mathrm{sdB} + M_\mathrm{WD,min}$ in case the inclination of the system was unknown.} below the Chandrasekhar limit $M_\mathrm{Ch}$ given their uncertainties. Here, we exclude CD$-$30~11223 because it has been modelled to explode during the sdB lifetime \citep{Deshmukh_2024MNRAS.527.2072D}. We estimate this merger rate to be $2.8 ^{+2.0}_{-1.5} \times 10^{-5}$ yr$^{-1}$ for the Galaxy. 

While none of these systems will merge during the hot subdwarf lifetime, they may contribute to the hybrid CO$+$He-WD merger channels proposed by \citet{Justham_2010} and \citet{MillerBertolami_2022MNRAS.511L..60M} to form specific types of He-sdO stars, or to other types of transients resulting from WD$+$WD mergers. The Galactic birthrate of the 16 eHe-sdOs in our sample is $2.3 \pm 0.8 \times 10^{-4}$ stars yr$^{-1}$. Therefore, the hybrid C/O + WD merger rate above could not explain more than $\sim12\%$ of the eHe-sdO population, even if all WD companions are assumed to be HeWDs.\footnote{This assumes that the current sdB$+$WD population is representative of the progenitor population that formed the eHe-sdOs. However, given the long merger delay times associated with the merger channels, this may have occurred several gigayears ago.} 

Finally, for the $+$dM/BD systems, we do not find any with $P<0.1$\,d, and indeed none are predicted to merge within the hot subdwarf lifetime. Due to gravitational wave emission the systems orbit will eventually shrink and the system will come into contact as a WD$+$dM/BD, possibly resulting in a cataclysmic variable. Therefore, by summing the inverse of their merger times\footnote{This assumes a steady state for the hot subdwarf binary formation rate. }, we can estimate an occurrence rate for these systems of $7.0 \pm 2.0 \times 10^{-6}$ yr$^{-1}$ for the Galaxy. This may contribute towards the CV population.

\section{Conclusions}
We presented over 5000 RV measurements of 253 non-composite hot subdwarf stars within a 500~pc volume-limited sample, detecting 50 new short-period binary systems and solving for the orbital solutions of 34 followed-up systems. Among these, 159 hot subdwarfs appear to be RV-stable within our uncertainties, where $\sim57$\% have been covered with at least three medium- to high-resolution spectra (Mercator, FIES, X-shooter, UVES, FEROS, or FOCES).
Most of the newly solved binaries are in the orbital period range of 1--20 days, a parameter space previously under-represented in the literature due to observational biases. Following a detailed analysis of the short-period binaries presented in this paper, the properties of the population can be summarised as follows:
\begin{itemize} 
    \item We find an overall short-period binary fraction of $34.7^{+2.8}_{-2.9}\%$ for the full sample of 301 hot subdwarfs when accounting for inclination bias and detection efficiency. 
    Excluding the 48 composites, as done in some previous studies, the short-period binary fraction is $41.4^{+3.6}_{-3.4}\%$. Both numbers are in line with previous estimates when broadly accounting for selection effects. The majority of companions are WDs (55), with only 19 dM/BD companions. The remaining 16 systems likely host WD companions. If it were a MS companion, it would be detectable via the SED or LC, and no such signatures are observed.  
    
    \item A far lower binary fraction of  $23.1^{+5.8}_{-6.3}\%$ is found for the sdOB spectral class compared with the sdB and sdO spectral class. This suggests a different evolutionary pathway or evolutionary stage. When selecting the hot-EHB (see Sect.\ \ref{rvv_classes}) to compare with previous work, we find that this population exclusively hosts WD companions on orbital periods longer than one day with a median of about $5.1$~days. The remaining objects on the EHB (cool-EHB; $20~000~\mathrm{K}\leq$ \teff\ $\leq30~000~\mathrm{K}$) have a median period of $\sim1$~day. 
    
    \item We find that all bEHB ($\log L/L_{\odot} \leq 1.05$) objects in our sample host WD companions. All but one of these exhibit orbital periods of less than one day, which starkly contrasts with the hot-EHBs. The WDs in these systems are, on average, more massive than the other WD companions. This is consistent with the observational findings of \citet{Schaffenroth_2022_1} as well as with the predictions by binary population synthesis \citep{Han_2002, Han_2003}. 
    
    \item For the first time, we derived volume-complete formation rates for hot subdwarf binaries within 500~pc, separately for the sdB/sdO$+$WD ($5.4 \pm 0.4 \times 10^{-16}$ pc$^{-3}$ yr$^{-1}$) and sdB/sdO$+$dM/BD ($1.5 \pm 0.4 \times 10^{-16}$ stars pc$^{-3}$ yr$^{-1}$) systems. Scaled to the full Galactic disc using the stellar mass distribution of \citet{McMillan_2017MNRAS.465...76M}, this corresponds to $7.9^{+1.2}_{-1.2} \times 10^{-4}$ yr$^{-1}$ and $2.2^{+0.5}_{-0.5}\times 10^{-4}$ yr$^{-1}$, respectively. For the sdB$+$WD population, we derived a Galactic merger rate and compared it with the observed formation rate of extreme helium subdwarfs, for which we found a maximum contribution of $12\%$. For channels involving a hot subdwarf phase, we estimate Galactic occurrence rates of $3.3 \pm 0.5 \times 10^{-5}$ yr$^{-1}$ (based on CD$-$30~11223) for single degenerate SN and double degenerate SN of $3.7 \pm 2.5 \times 10^{-5}$ yr$^{-1}$ (based on HD~265435). This is consistent with the results of \citet{Pelisoli_2021NatAs...5.1052P}. Our calculations also represent a contribution rate of $\sim$$2.5^{+0.7}_{-0.5}\%$ towards the total Galactic SN~Ia rate for channels involving a hot subdwarf stage. 

\end{itemize}
\noindent
We provided a detailed characterisation of the short-period binary population within 500 pc, offering key constraints on the formation channels of hot subdwarf stars for the next generation of binary population synthesis studies. 
The 48 wide systems with identified A/F/G-type companions from Paper~I, two of which we identify as triple candidates in this work (Geier et al. in prep.), will be the subject of a future paper completing the characterisation of the 500~pc sample.

\section{Data availability}
\label{sect:data_availability}
All RV measurements are provided together with those calculated in this paper as an online table accessible via CDS services via anonymous ftp to \url{cdsarc.u-strasbg.fr (130.79.128.5)} or via \url{http://cdsweb.u-strasbg.fr/cgi-bin/qcat?J/A+A/}.


\begin{acknowledgements}

H.~D. was supported by the Deutsche Forschungsgemeinschaft (DFG) through grants GE2506/17-1 and GE2506/9-2.

M.~D. was supported by the Deutsches Zentrum für Luft- und Raumfahrt (DLR) through grant 50-OR-2304. 

V.~S. and M.~P. received funding by the Deutsche Forschungsgemeinschaft (DFG) through grants GE2506/9-1 and GE2506/12-1.

D.~S. acknowledges funding by DFG grant HE1356/70-1.

K.~D. acknowledges funding from the Methusalem grant METH/24/012 at KU Leuven. This research has used observations obtained at the Mercator Observatory which receives funding from the Research Foundation – Flanders (FWO) (grant agreement I000325N and I000521N).

A.~B. was supported by the Deutsche Forschungsgemeinschaft (DFG) through grant GE2506/18-1. 

R.~R. acknowledges support from Grant RYC2021-030837-I funded by MCIN/AEI/ 10.13039/501100011033 and by “European Union NextGeneration EU/PRTR”. This work was partially supported by Spanish MINECO grant PID2023-148661NB-I00.

M.~V. acknowledges support from the FONDECYT Regular N° 1211941 and N° 1250525.

A.~D.~R acknowledges support from ANID-Subdirección de Capital Humano/Doctorado Nacional/2025-21250519
 
E.~A.~R acknowledges support from ANID-Subdirección de Capital Humano/Doctorado Nacional/2025-21250458

A.~Bob acknowledges support from the Australian Research Council (ARC) Centre of Excellence for Gravitational Wave Discovery (OzGrav), through project number CE230100016.

D.~B.~P. acknowledges support from ANID-Subdirección de Capital Humano/Doctorado Nacional/2025-21250735
  
This project has received funding from the European Research Council under the European Union’s Horizon 2020 research and innovation programme (Grant agreement numbers 101002408).

Based on observations collected at the European Southern Observatory under ESO programme(s) 116.28ZZ.002 and 117.2AET.002. 

Based on observations collected with the Goodman spectrograph at the Southern Astrophysical Research Facility (SOAR) at Cerro Pachon, Chile, under the programme allocated by the Chilean Telescope Allocation Committee (CNTAC), no: 2023B, 2024A and 2025A.
This work has made use of data from the European Space Agency (ESA) mission \textit{Gaia} (https://www.cosmos.esa.int/gaia), processed by the \textit{Gaia} Data Processing and Analysis Consortium (DPAC, https://www.cosmos.esa.int/web/gaia/dpac/consortium). Funding for the DPAC has been provided by national institutions, in particular the institutions participating in the \textit{Gaia} Multilateral Agreement.
\end{acknowledgements}

\bibliography{Bibliography}
\bibliographystyle{aa}

\begin{appendix}
\section{Individual systems}
\label{app:individual_systems}
Here we comment on individual systems, selected either for their scientific interest or for the ambiguity of their orbital solutions. 
The window-function and convergence checks described in Sect.~\ref{determination_of_the_orbital_solutions} were applied to every system in the sample and were passed in all cases. Figure \ref{fig:mass_radius_baraffe} shows the mass-radius relation of the companions showing reflection effects analysed in this work with the theoretical predictions of \citet{Baraffe_2015A&A...577A..42B} shown for comparison.

\paragraph{GALEX~J070134.1-671739} was first catalogued as an sdB in \citet{Geier_2017}. Here we report it as an sdB$+$WD with an orbital period of P$=2.5438798(2)$~days fitting the RV and photometric data together. A clear Doppler boosting signature is seen in its LC. The WD has a high minimum mass of $0.75$~\msun, together with the hot subdwarf mass of $0.57^{+0.10}_{-0.09}$\msun\ this system sits near the Chandrasekhar limit at $1.32$~\msun\ total mass. This system will merge in $\sim1500$~Gyr and therefore contributes negligibly to the SN~Ia rate calculated in Sect. \ref{birthrates} and will become a massive double WD binary. 

\vspace{-10pt}

\paragraph{TYC\,6800-72-1} is a newly identified iHe-sdOB and the only RV variable with an enriched helium atmosphere (-1 < \logy < 0.6) in the 500~pc sample. The object is also the fourth most luminous hot subdwarf in the sample, and stands apart from other stars in the HRD at $\mathrm{log}\,L \approx 2.3$ and effective temperature $\approx$ 38~000~K (see the HRD in Fig.~\ref{fig:mass_m2}). We find the orbital period to be relatively long at 3.6001(6)~days corresponding to a minimum WD companion mass of $0.55^{+0.05}_{-0.06}$~\msun. Two further small aliases are present in the power spectrum, but both coincide with peaks in the window function calculated from 22 epochs over a baseline of $623$~days and are therefore likely sampling artefacts.

\vspace{-10pt}

\paragraph{LS~III+48~50} has been identified as having an emission line in H$\alpha$ \citep{Kohoutek_1997AAHam..11.....K}. The 19 INT and NOT spectra obtained in this work do not cover H$\alpha$, though no other emission lines are detected between $3600 - 5500$ \AA.
The periodic signal in TESS resembles Doppler boosting; however, the period of 6.19~days does not align with any plausible RV solution and is therefore discarded. 
The RV power spectrum exhibits a clear but multi-modal peak; several aliases are visible, though with a strong preference for the central peak at P$=0.72180^{+0.00333}_{-0.00005}$~days. We therefore report the confidence interval around the main peak of the posterior, to which the MCMC sampler converges with an integrated autocorrelation time of $\sim1.6$ steps, yielding well over $10^{3}$ independent samples. The 19 randomly sampled spectra cover a baseline of 790~days. None of the secondary peaks above $4\sigma$ coincides with a significant peak of the spectral window function; they are instead aliases of the adopted orbital solution.

\vspace{-10pt}

\paragraph{TYC\,4563-2614-1} was first identified as a blue object in \citet{Abramyan_1994Ap.....37..224A} and classified as an sdB in Paper~I. Using high-resolution FIES and HERMES spectra its orbital period is $P=4.3969(3)$~days to high confidence, returning a fairly low minimum companion mass of $0.15\pm0.02$~\msun. As no photometric variability is present in its LC, and the minimum detectable mass is above the minimum mass derived from the RV curve, the nature of the companion remains ambiguous. The system may host a low-mass dM/BD or He-WD companion, or the system is oriented at a high inclination. 

\vspace{-10pt}

\paragraph{TYC~5737-1693-1} is a bright star (Gmag$\sim$11.6 mag) that was first identified as an sdB in Paper~I. Two peaks above the $4\sigma$ threshold are present in the power spectrum at P$=11.46(1)$~days and P$=3.06(5)$~days. We find a high probability of $99.9$\% in favour of the 11-day solution, which we adopt; the well-converged chains (integrated autocorrelation time of ${\sim}1.5$ steps) confirm that this preference reflects the posterior itself rather than incomplete sampling of the shorter-period mode. The companion has a very low minimum mass of $0.07\pm0.03$~\msun\ -- the lowest in the sample and its nature is also unknown. This system and TYC~4563-2614-1 are the only two with very low semi-amplitudes at orbital periods longer than one day (annotated in the central-left panel of Fig.~\ref{fig:combined_period_analysis}) and may either be oriented at high inclinations or indeed host very low-mass companions, which would be difficult to explain in theory \citep{Han_2002}. 

\vspace{-10pt}

\paragraph{FBS~2253+335} was classed as a blue object in the FBS \citep{Abramyan_1994Ap.....37..224A} and later found to be an sdB by \citet{Lei_2018ApJ...868...70L}. Here, we determine an orbital period of $P=21.651(6)$~days with a probability of approximately $100$\% of being the true period, making it the longest in the 500~pc sample. A further cluster of peaks is seen above the $4\sigma$ threshold, but upon inspection of the window function these were found to correspond to the one-day sampling alias. Several other small peaks are present in the power spectrum, which may stem from the long baseline of over 2500~days sampled by 22 epochs. The integrated autocorrelation time of ${\sim}1.8$ steps nevertheless indicates that these modes were efficiently explored by the sampler, and a supplementary Gelman-Rubin statistic of $\hat{R}<1.001$ across all parameters is consistent with full convergence. The identified WD companion has a minimum companion mass of $0.50\pm0.05$~\msun. 

\vspace{-10pt}

\paragraph{UCAC4~550-140677} was first identified as an sdO using LAMOST DR5 spectra \citep{Lei_2018ApJ...868...70L}, though it also appears in the catalogue of candidate WDs by \citet{Fusillo_2019MNRAS.482.4570G}, likely due to its high surface gravity of  \logg\ $=6.2\pm0.2$ (Paper~II). As with LS~III+48~50 above, several strong aliases appear at $\sim \pm 0.2$ days around the dominant peak owing to the relatively sparse phase coverage, though none of these aliases coincides with a significant peak of the spectral window function of the 12 observation epochs. Each peak has a comparable probability of being the correct one; therefore, with no other strong peaks exceeding $4\sigma$, we report broad uncertainties for this system to encompass $68$\% of the probability range around the highest peak (the shaded blue regions in the periodogram inset). For this system, the Gelman-Rubin statistic reports $\hat{R}<1.001$ across all parameters, which verifies convergence.
Here we confirm this system as an sdO$+$WD with an orbital period of P$=3.5(2)$~days and a minimum WD companion mass of $0.38^{+0.03}_{-0.04}$~\msun. We note that adopting any of the neighbouring aliases in place of the highest peak would shift the minimum companion mass by at most ${\sim}4.3$\%, which is reflected in the reported companion mass uncertainty.

\vspace{-10pt}

\paragraph{PG~2337+070} was discovered during the PG survey \citep{Green_1986} and is included in the first catalogue of spectroscopically identified hot subdwarf stars \citep{Kilkenny_1988}. Two alias peaks are seen in the periodogram at $\sim$9.29 and 8.77~days, with probabilities of $\sim58$\% and $34$\%, respectively. Neither coincides with a significant peak of the window function, though numerous aliases are again present owing to the 15 epochs spread over a baseline of more than 3650~days. We adopt the higher-probability peak, to which the MCMC sampler converges, but report broad uncertainties that encompass both solutions: $P=9.3^{+0.4}_{-0.5}$~days. We note that the peak at 8.77~days is still a significant alias. Since the two aliases differ in companion mass by only $2.6$\%, this has a negligible effect on the derived minimum mass of $0.44\pm0.04$~\msun.

\vspace{-10pt}

\paragraph{UCAC4~507-015874} was first spectroscopically identified as an sdOB in Paper~I and is a pulsator \citep{Baran_2023A&A...669A..48B, Uzundag_1_2024AA...684A.118U}.
We find an orbital period of $P=7.291(3)$~days with high probability, an integrated autocorrelation time of $\sim1.9$ steps, and good convergence with $\hat{R}<1.0001$. Cross-checking with the window function reveals very strong aliasing, particularly around 1 day and orbital periods beyond 10 days, despite a reasonable coverage of 18 epochs over 644~days. The main peak does not align with these aliases, and remaining peaks in the power spectrum are well-explained by the window function peaks.

\vspace{-10pt}

\paragraph{TYC~2405-1118-1} was first identified as an sdB by \citet{Lei_2018ApJ...868...70L}, and we confirm it here as an sdB$+$WD. A wide cluster of peaks is seen around ${\sim}1.6$~days in the periodogram, arising from the baseline of over 3422~days, though the central peak carries a high probability of ${\sim}80$\%. Given the degree of aliasing, we adopt uncertainties reflecting a $3\sigma$ ($99.7$\%) probability interval around this peak and report P$=1.59^{+0.09}_{-0.06}$~days. The minimum companion mass varies by only ${\sim}2.9$\% across the cluster, and the remaining groups of peaks above $4\sigma$ are aliases of it.

\vspace{-10pt}

\paragraph{CD-22~9142} is a bright sdO star present in the first catalogue of hot subdwarfs \citep{Kilkenny_1988}, though is confirmed as an sdO$+$WD binary for the first time in this work. We find a relatively long orbital period of P$=8.7787(1)$~days. The very long baseline of ${\sim}9000$~days produces strong structure in the window function, most notably near 1~day, which propagates into the power spectrum as a dense forest of aliases. The adopted period, however, does not coincide with any of these features. The WD companion has a minimum mass of $0.64^{+0.05}_{-0.06}$~\msun. 

\vspace{-10pt}

\paragraph{UCAC4~349-002972} was first identified as an sdO in Paper~I. 
Although several peaks lie above the $4\sigma$ threshold which are close together in period space, the highest carries an ${\sim}82$\% probability of being the correct one. We adopt wide uncertainties reflecting a $3\sigma$ ($99.7$\%) probability interval around this peak, each encompassing two further solutions of ${\sim}8$\% probability, and report an orbital period of P$=7.10^{+0.04}_{-0.77}$~days. The minimum mass of the WD companion is $0.44^{+0.04}_{-0.05}$~\msun\ and we note that selecting the alias at $\sim6.3$~days lowers this by up to $5\%$.

\begin{figure}
    \centering
        \includegraphics[width=0.9\linewidth]{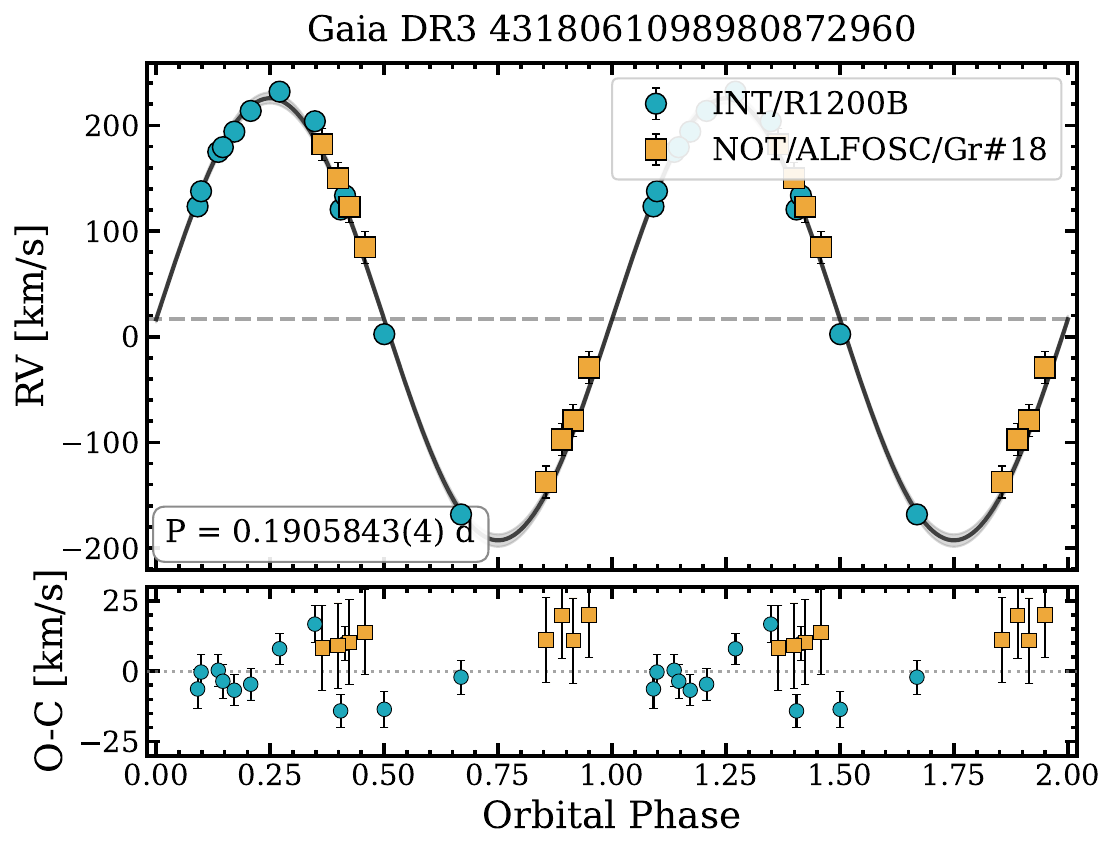}
        \includegraphics[width=0.9\linewidth]{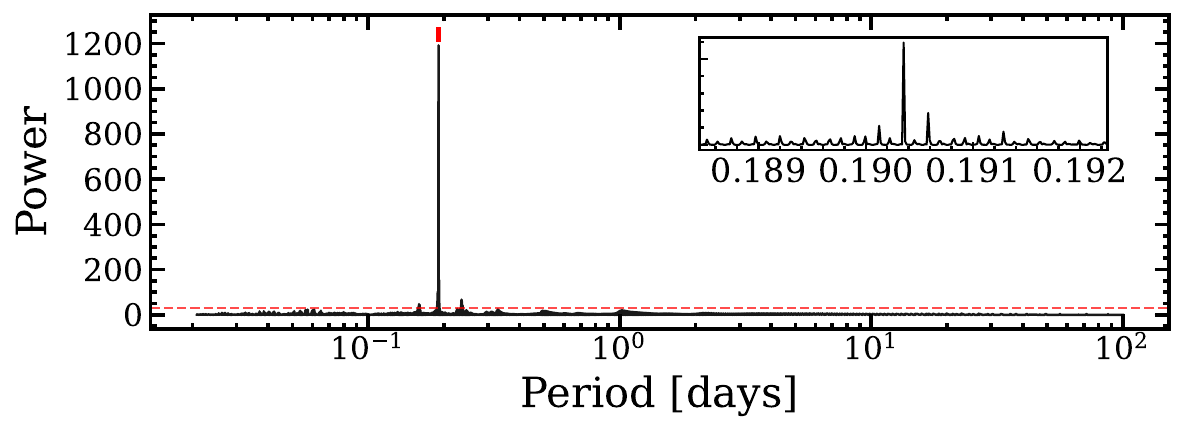}
    \caption{RV curve and power spectrum of a newly identified and solved sdA (pre-ELM) in this work with an orbital period of $0.1905842(3)$~days. Its atmospheric parameters are \teff$=11958\pm150$~K, \logg$=5.4\pm0.08$~dex, and \logy$=-2.5\pm0.2$, and its stellar parameters from its SED are $0.30\pm0.05$~\msun, $0.182\pm0.005$~\rsun, and $0.22\pm0.03$~\lsun.}
    \label{fig:new_sdA}
\end{figure}

\paragraph{Gaia DR3 4318061098980872960}
\label{new_sdA} is a newly identified sdA-class hot subluminous object in this work (see Fig.~\ref{fig:new_sdA}) and has an orbital period of $P=0.1905843(4)$~days. 
It has a primary mass of $0.30\pm0.06$ \msun\ and \teff$=$11~960$\pm150$ K and \logg$=$5.39$\pm0.07$ dex, likely associating it with a pre-low mass WD nature. It is not included and analysed among the hot subdwarfs in this paper, but is included here for completeness. This population will be investigated in more detail in a future paper looking at the population of pre-WDs in 500~pc. 

\begin{figure}
    \centering
        \includegraphics[width=0.9\linewidth]{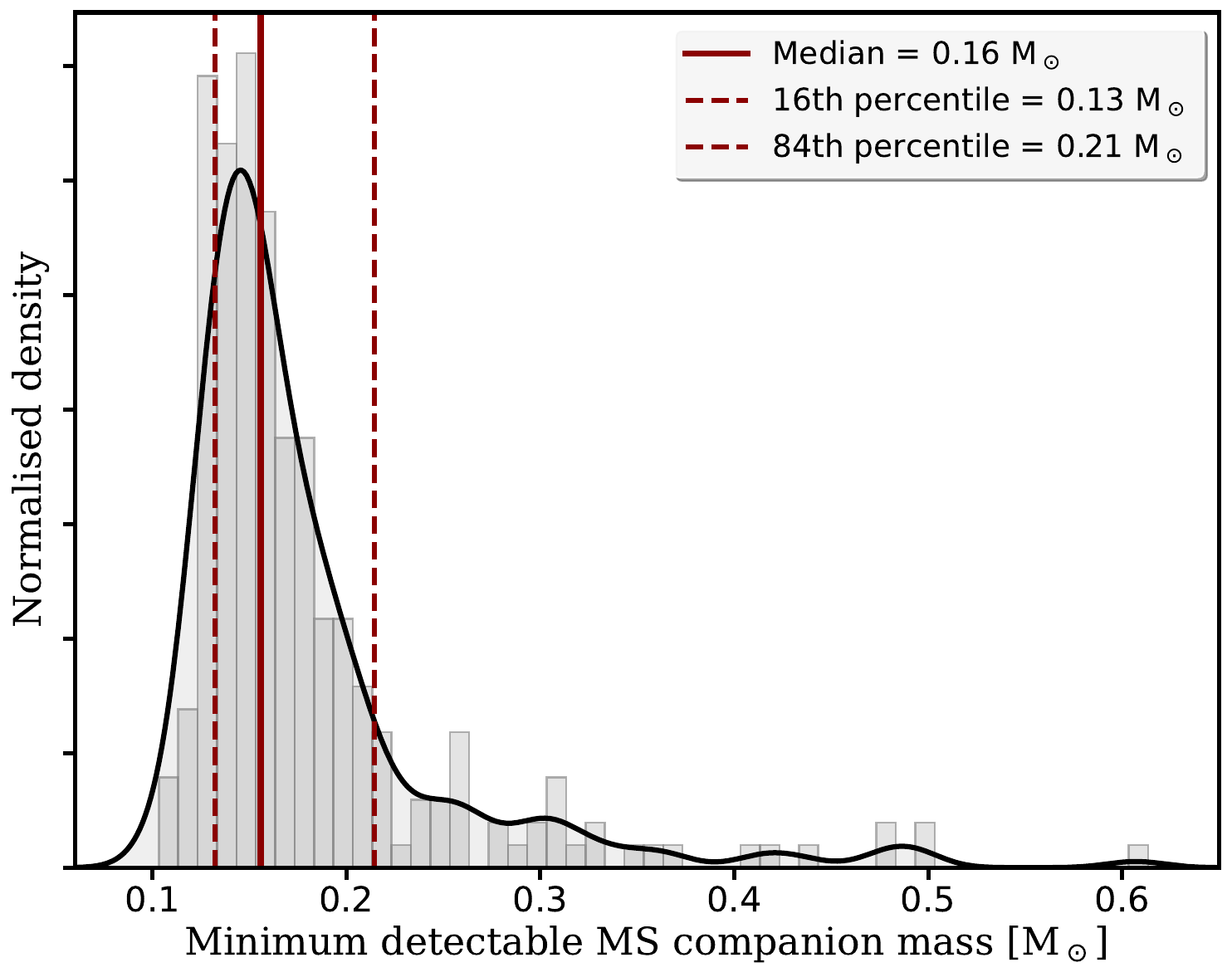}
    \caption{Distribution of the minimum detectable MS companion mass for all stars in the sample (Sect.~\ref{the_nature_of_the_companion}). The histogram is overlaid with a KDE (black curve). The vertical solid and dashed lines mark the median and 16th/84th percentiles, respectively. The distribution reflects the sensitivity of the SED fitting
    method to low-mass MS companions of hot subdwarfs, which varies from star to star depending on the available photometry and properties of the primary.}
    \label{fig:det_mass}
\end{figure}

\begin{figure*}
    \centering
    \begin{minipage}{0.33\linewidth}
        \includegraphics[width=\linewidth]{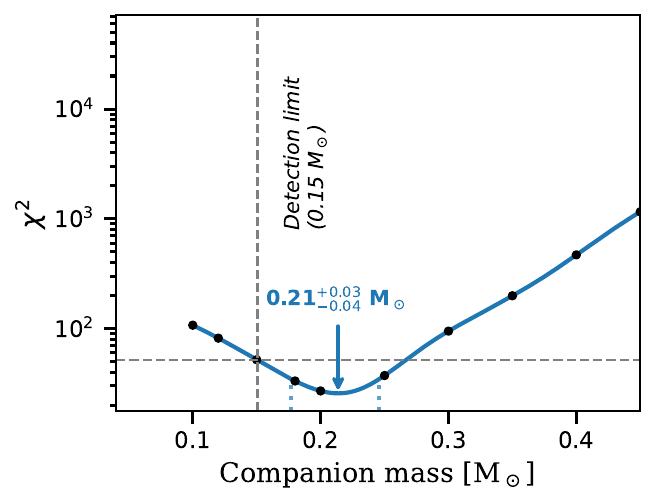}
    \end{minipage}
    \hfill
    \begin{minipage}{0.33\linewidth}
        \includegraphics[width=\linewidth]{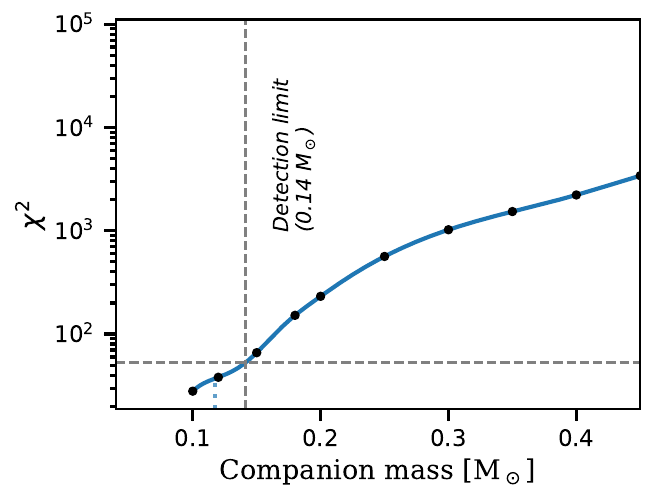}
    \end{minipage}
    \hfill
    \begin{minipage}{0.30\linewidth}
        \includegraphics[width=\linewidth]{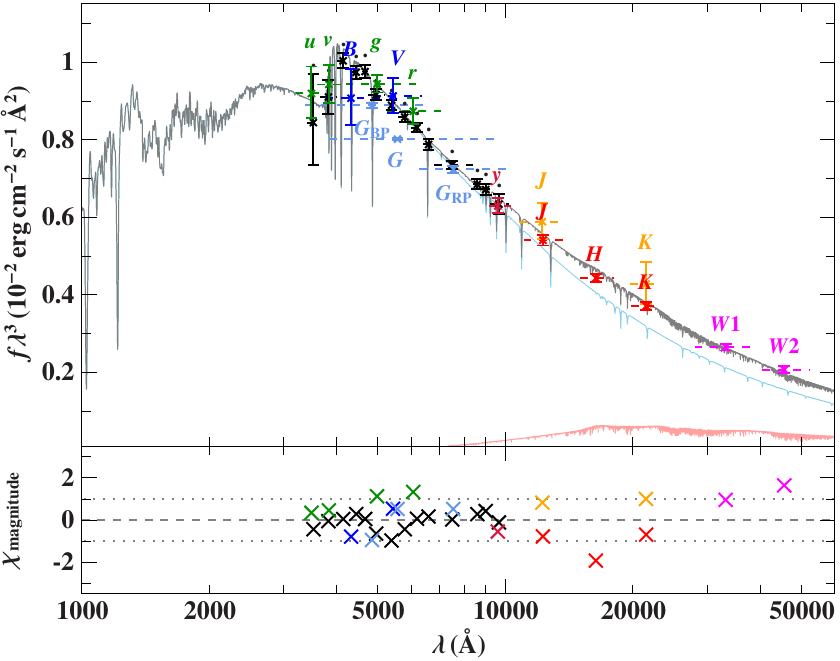}
    \end{minipage}
    \caption{\textit{Left and middle:} Reduced $\chi^2$ of the single-star SED fit as a function of forced MS companion mass ($0.1$--$1.5$~\msun) for EC~01578-1743 and GSC~00141-01628, respectively
    (Sect.~\ref{the_nature_of_the_companion}). The horizontal dashed line marks the $3\sigma$ detection threshold ($p = 0.003$) above which the SED fit is statistically rejected. The vertical dashed line indicates the corresponding minimum detectable companion mass. For GSC~00141-01628 the $\chi^2$ minimum lies below this limit, implying a compact (likely WD) companion. For EC~01578-1743 the minimum (blue arrow) lies above the threshold at $0.21$~\msun, consistent with the known dM companion \citep{Schaffenroth_2023_2}. \textit{Right:} SED of EC~01578-1743 showing the IR excess produced by the companion.}
    \label{fig:sed_min_det_mass}
\end{figure*}

\begin{figure}
    \centering
        \includegraphics[width=0.9\linewidth]{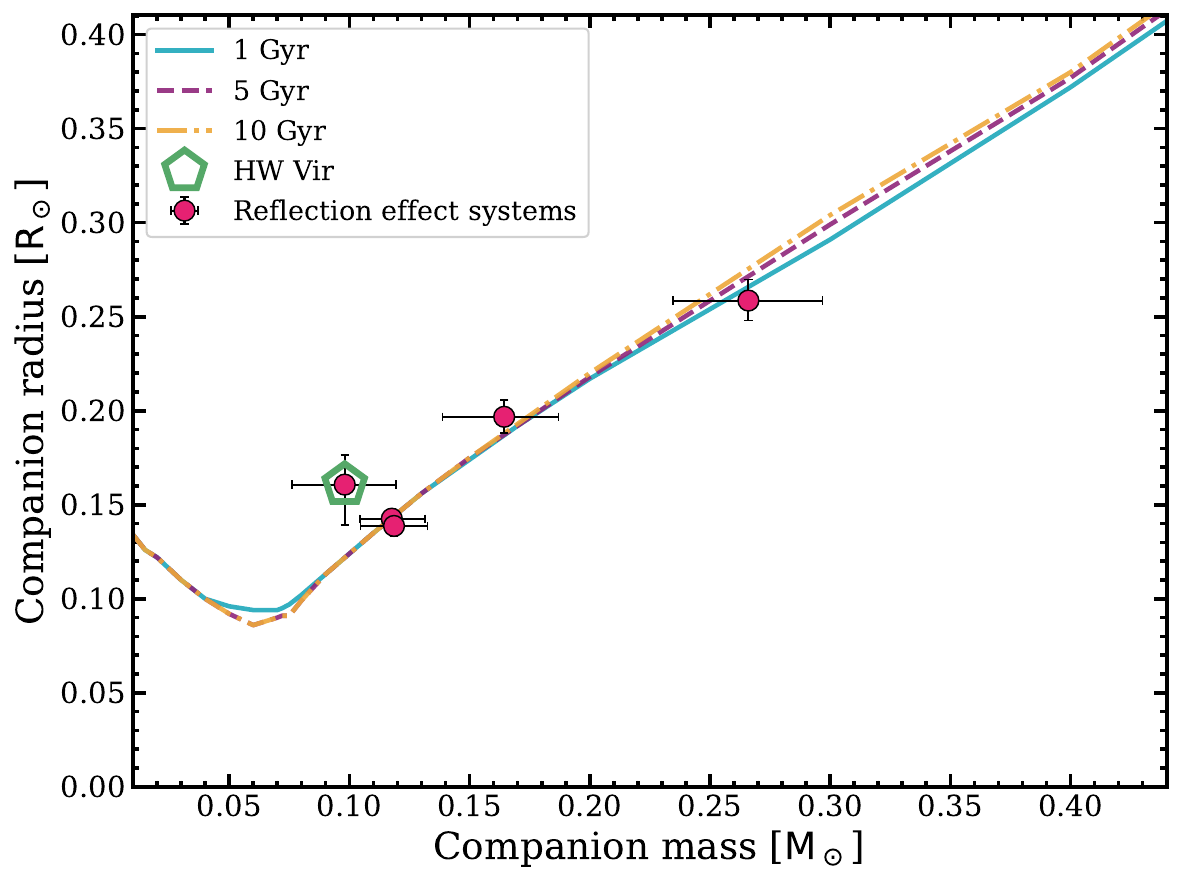}
    \caption{Mass-radius relation of the companions with the reflection effects analysed in this work. Theoretical calculations for ages 1, 5, and 10~Gyr from \citet{Baraffe_2015A&A...577A..42B} are shown for comparison.}
    \label{fig:mass_radius_baraffe}
\end{figure}

\subsection{Eccentric orbital solutions}
\label{appendix_eccentricities}

Close hot subdwarf binaries are expected to have circular orbits, as the CE formation channel efficiently circularises them \citep{Preece_2018MNRAS.481..715P}. Using high resolution spectra from the Fiber-fed
Extended Range Optical Spectrograph (FEROS), \citet{Edelmann_2005AA...442.1023E} detected small eccentricities of 0.02 -- 0.06 for five systems (see figure 5 in that work), while \citet{Geier_2011AA...526A..39G} found no detectable eccentricities in a medium-resolution sample --  though their upper limits of 0.15 to 0.3 are too high to have recovered the small values reported by \citet{Edelmann_2005AA...442.1023E}. 

In this work, we also explored eccentric orbital solutions by fitting the eccentricity $e$ (allowed to vary up to $e=0.3$), and the argument of periastron $\omega$. We adopted the parametrisation of \citet{Ford_2006ApJ...642..505F} ($\sqrt e$ cos $\omega$, $\sqrt e$ sin $\omega$) to avoid the strong correlation between $e$ and $\omega$ near $e=0$. 
Here, we confirm the results of \citet{Edelmann_2005AA...442.1023E} and conclude that  PG~1232-136, [CW83]~1419$-$09, and PG~0133+114 appear to be on eccentric orbits with $e=0.057\pm0.003$, $0.027\pm0.007$, and $0.022\pm0.016$, respectively, using our newly obtained spectra combined with FEROS spectra downloaded from the ESO archive. 
In all other cases, the solutions tended towards $e=0$, consistent with circular orbits. We therefore adopt circular solutions throughout, though we note that this does not rule out eccentricities up to $e\sim0.1$ for the vast majority of systems, as the RV uncertainties in this work are insensitive to these small eccentricities.

\begin{figure*}[p]
    \centering
    \begin{minipage}[t]{0.22\linewidth}
        \centering
        \includegraphics[width=0.95\linewidth]{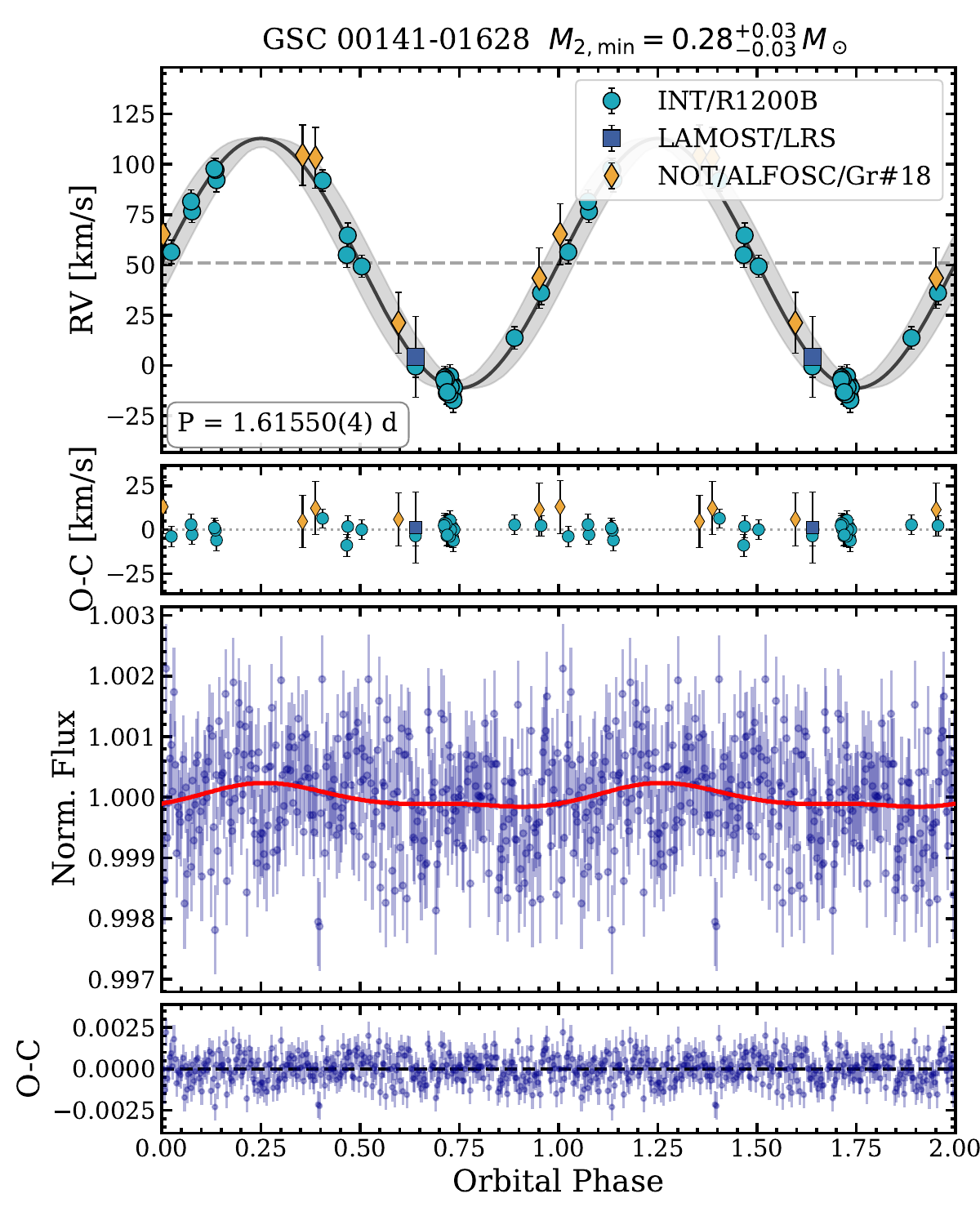}\\
    \end{minipage}
    \hspace{0.005\linewidth}
    \begin{minipage}[t]{0.22\linewidth}
        \centering
        \includegraphics[width=0.95\linewidth]{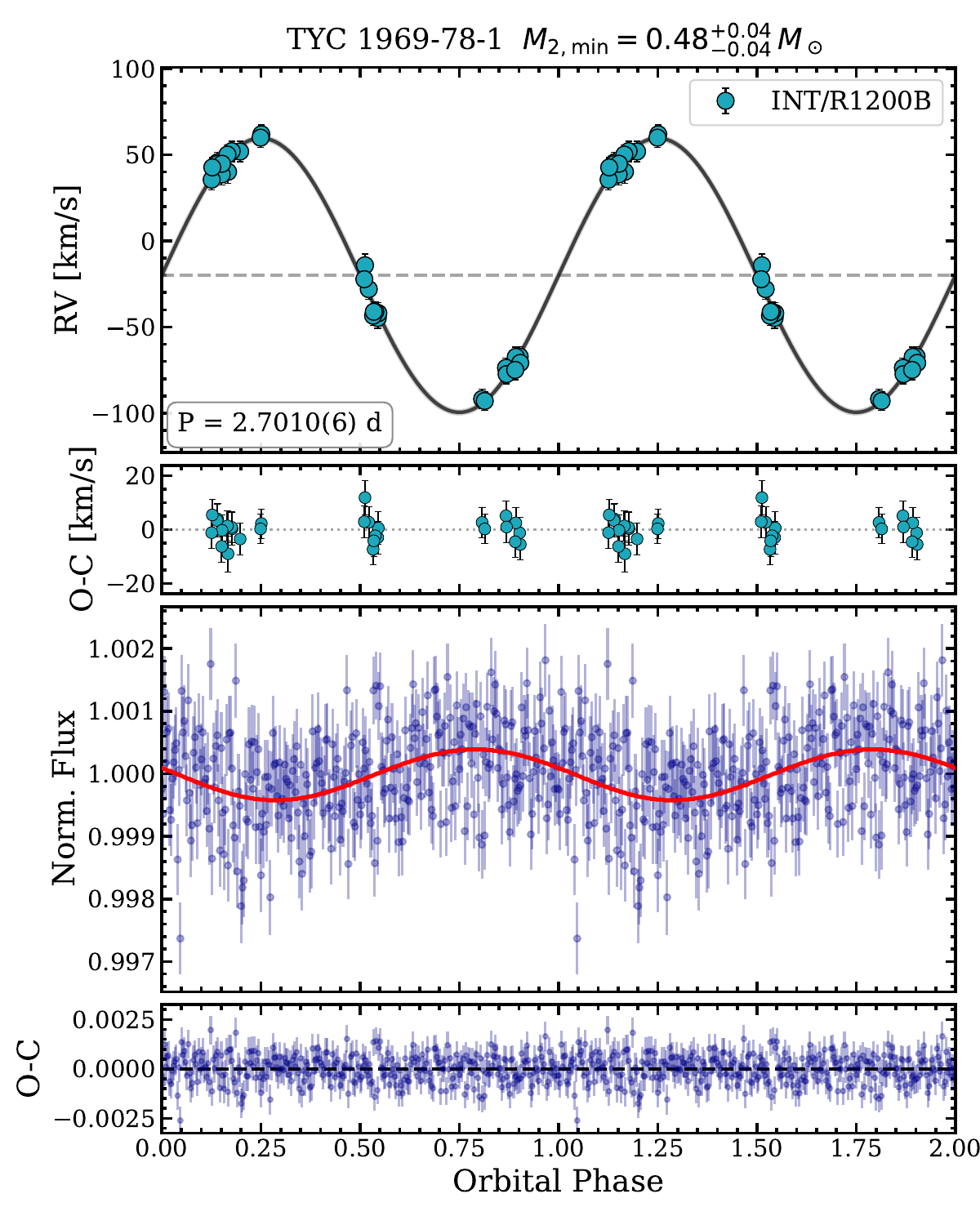}\\
    \end{minipage}
    \hspace{0.005\linewidth}
    \begin{minipage}[t]{0.22\linewidth}
        \centering
        \includegraphics[width=0.95\linewidth]{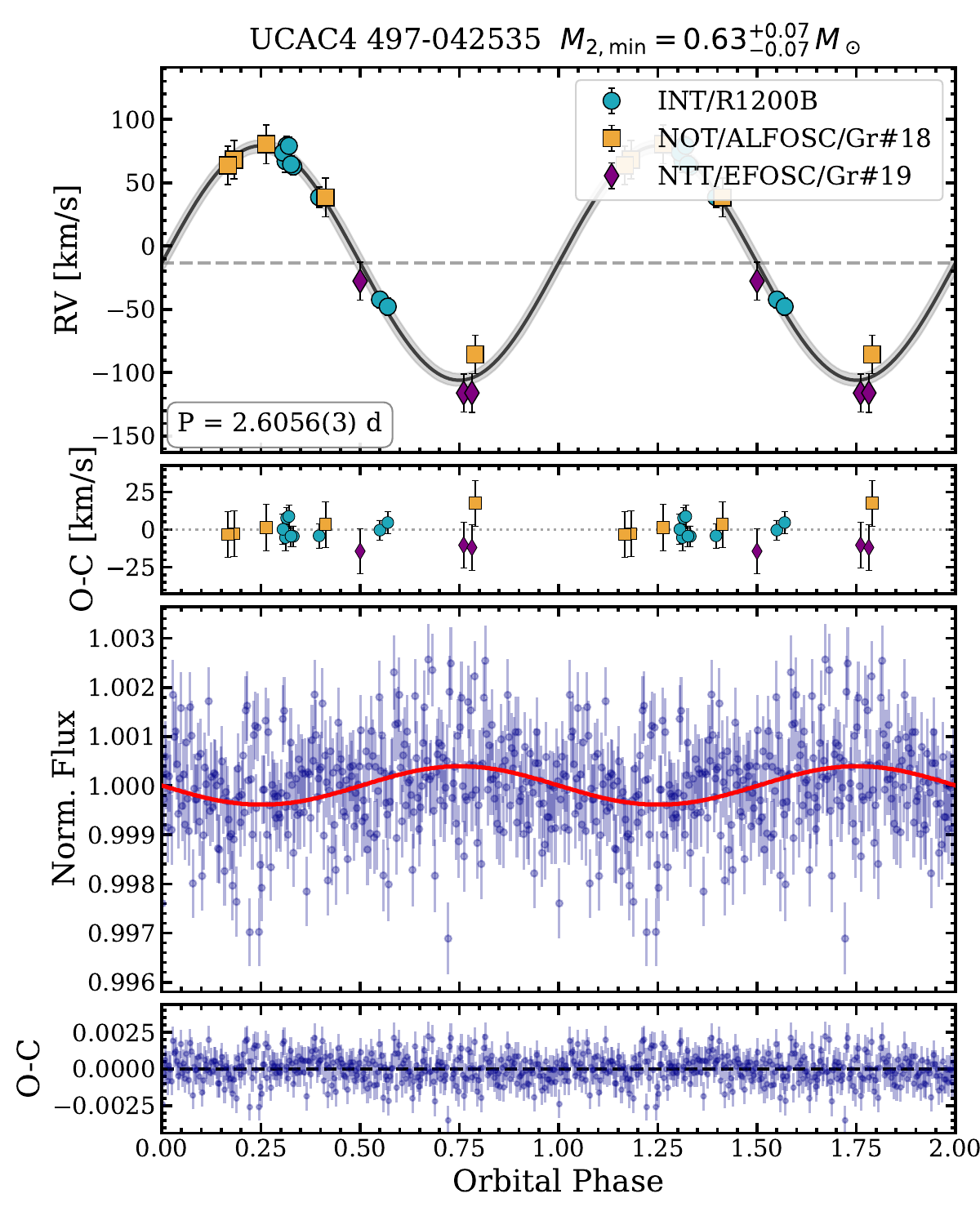}\\
    \end{minipage}
    \hspace{0.005\linewidth}
    \begin{minipage}[t]{0.22\linewidth}
        \centering
        \includegraphics[width=0.95\linewidth]{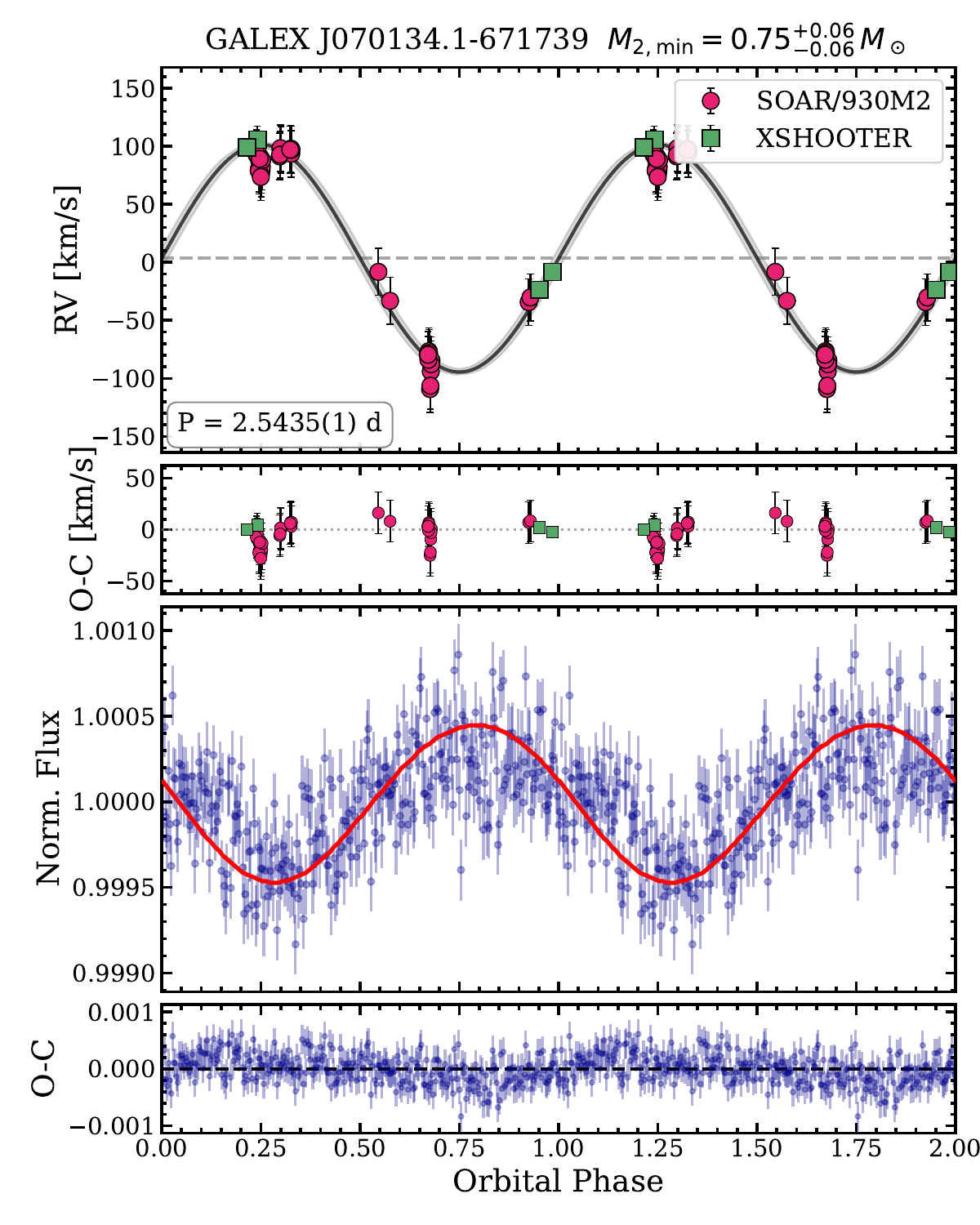}\\
    \end{minipage}
    \vspace{1pt}
\end{figure*}
\begin{figure*}[p]
    \centering
    \begin{minipage}[t]{0.22\linewidth}
        \centering
        \includegraphics[width=0.95\linewidth]{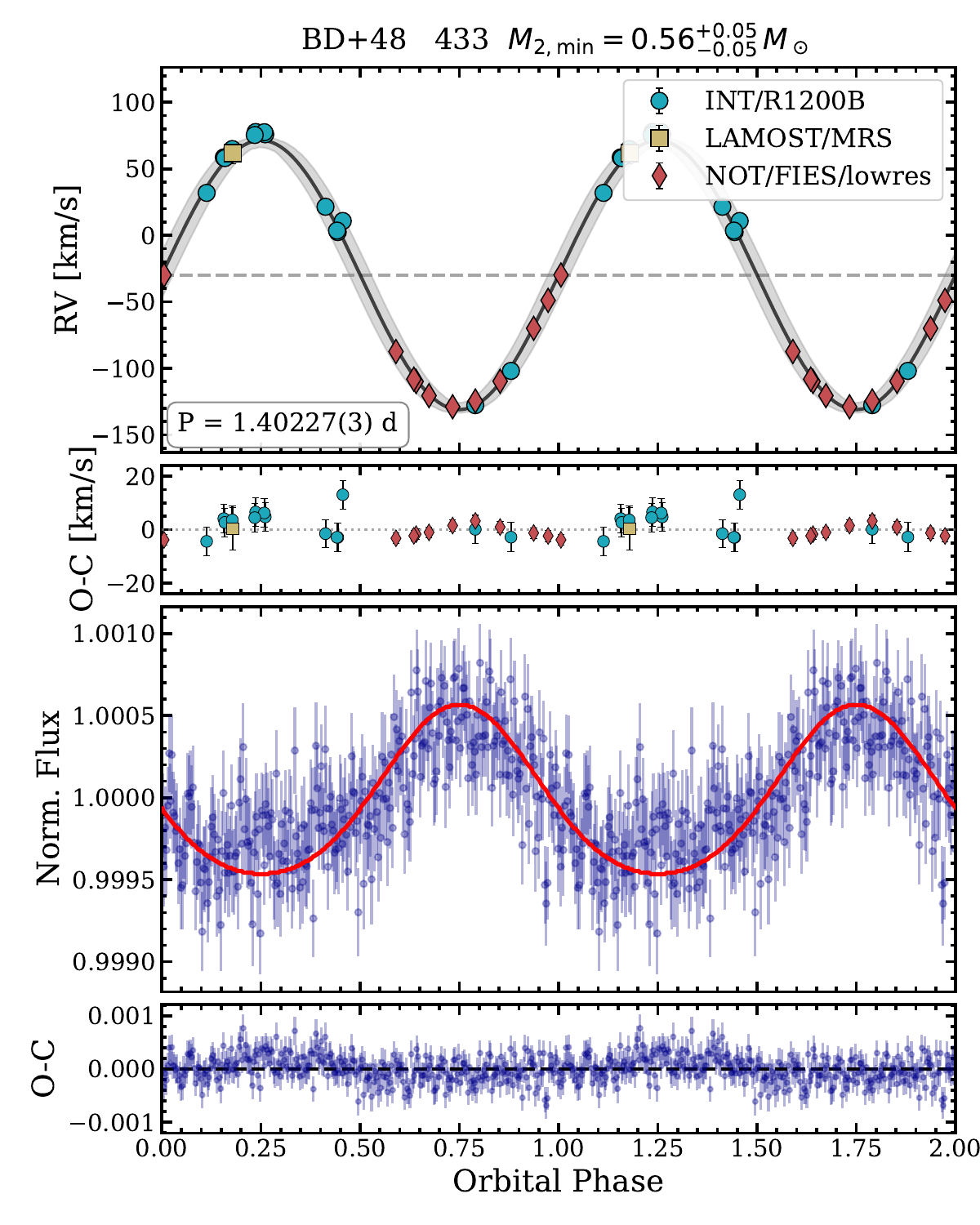}\\
    \end{minipage}
    \hspace{0.005\linewidth}
    \begin{minipage}[t]{0.22\linewidth}
        \centering
        \includegraphics[width=0.95\linewidth]{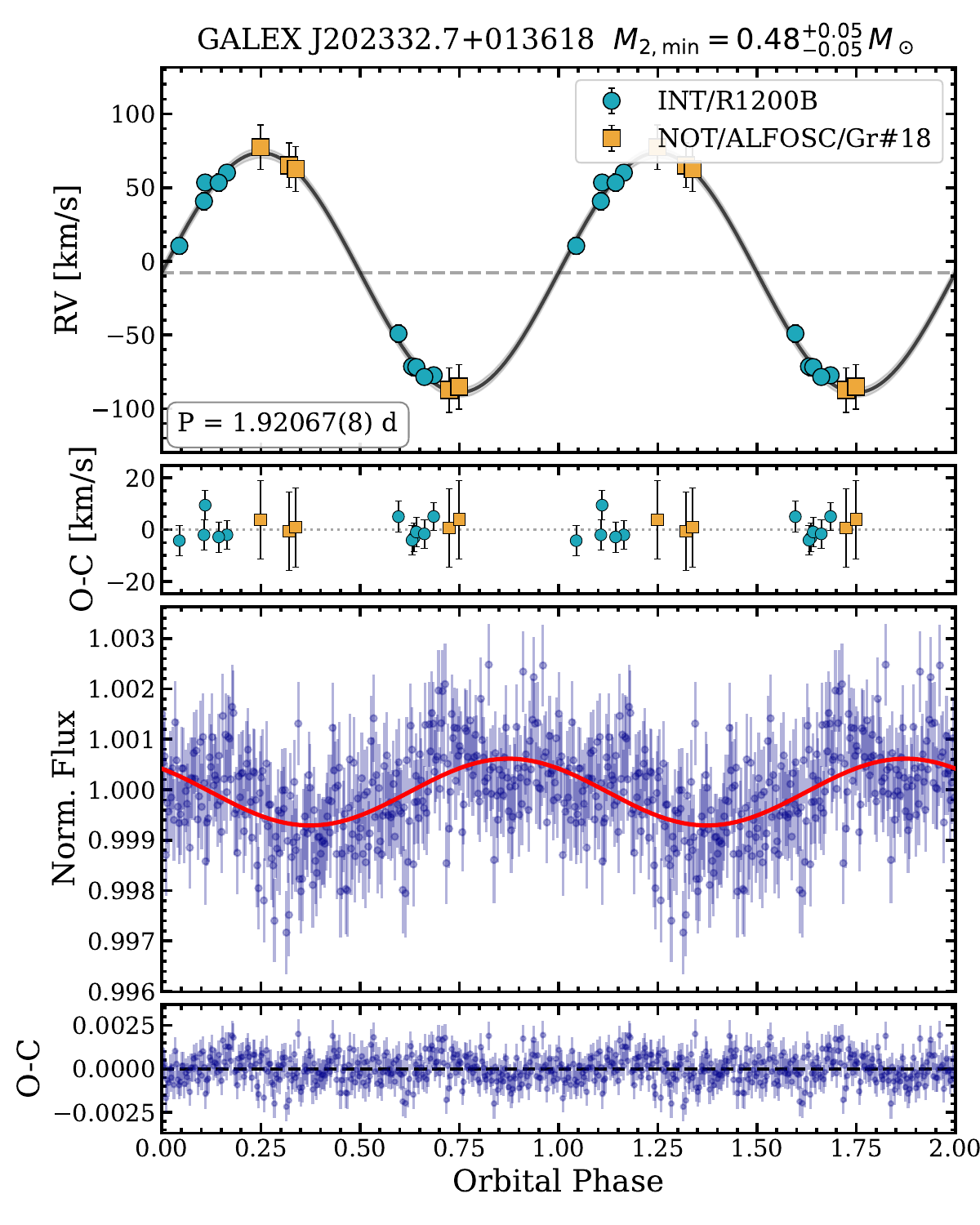}\\
    \end{minipage}
    \hspace{0.005\linewidth}
    \begin{minipage}[t]{0.22\linewidth}
        \centering
        \includegraphics[width=0.95\linewidth]{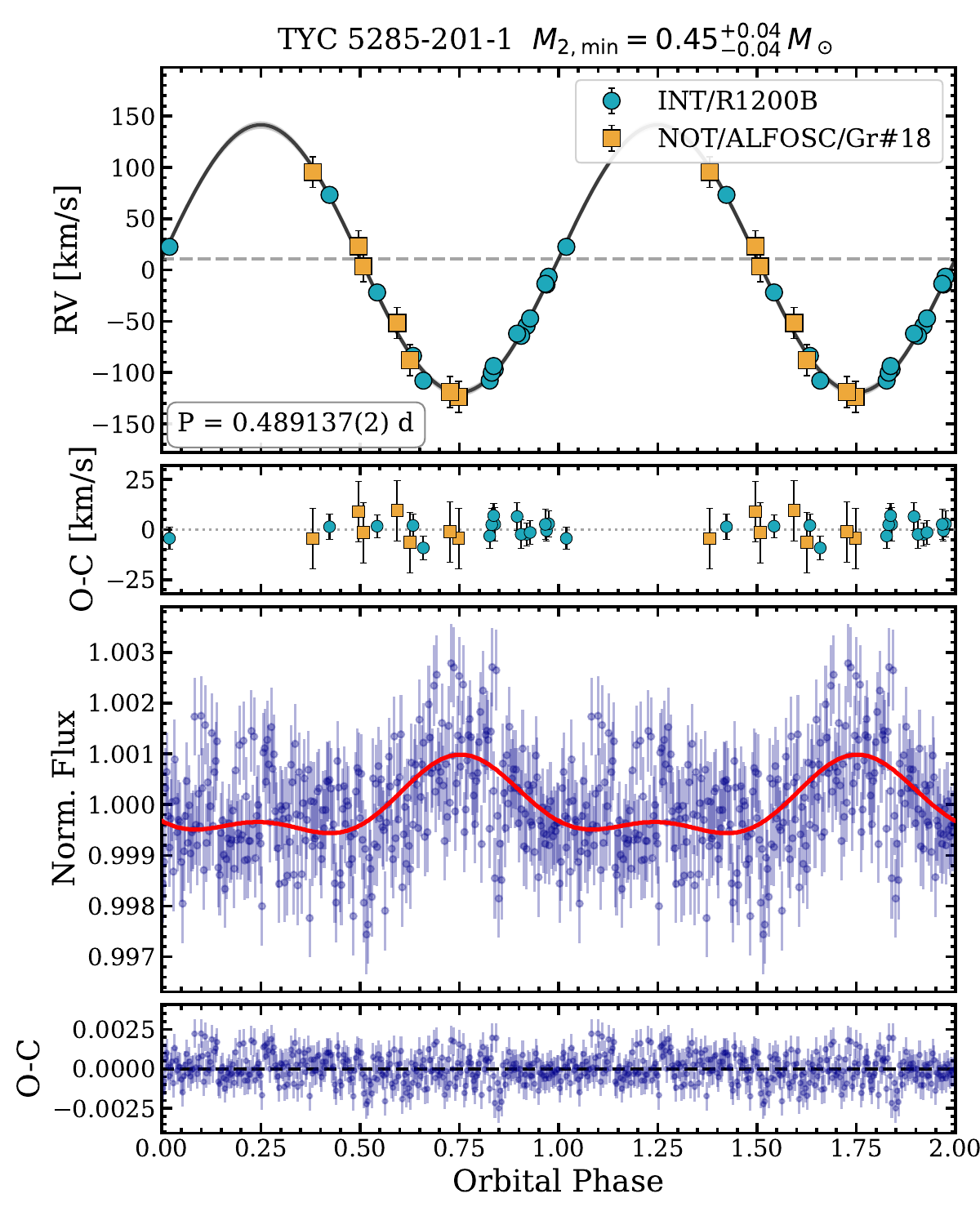}
    \end{minipage}
    \hspace{0.005\linewidth}
    \begin{minipage}[t]{0.22\linewidth}
        \centering
    \end{minipage}
    \vspace{1pt}
    \caption{RV curves (top panels) and phase-folded LCs (bottom panels) for systems where a low-amplitude photometric signal was detected within $\pm10\%$ of the spectroscopic period. The photometric variability is attributed to Doppler boosting, so no system parameters are constrained from the LCs. The red model is shown purely to illustrate the phase-coherent variability.}
    \label{fig:rv_lc_power}
\end{figure*}

\begin{figure*}[p]
    \centering
    \begin{minipage}[t]{0.22\linewidth}
        \centering
        \includegraphics[width=0.95\linewidth]{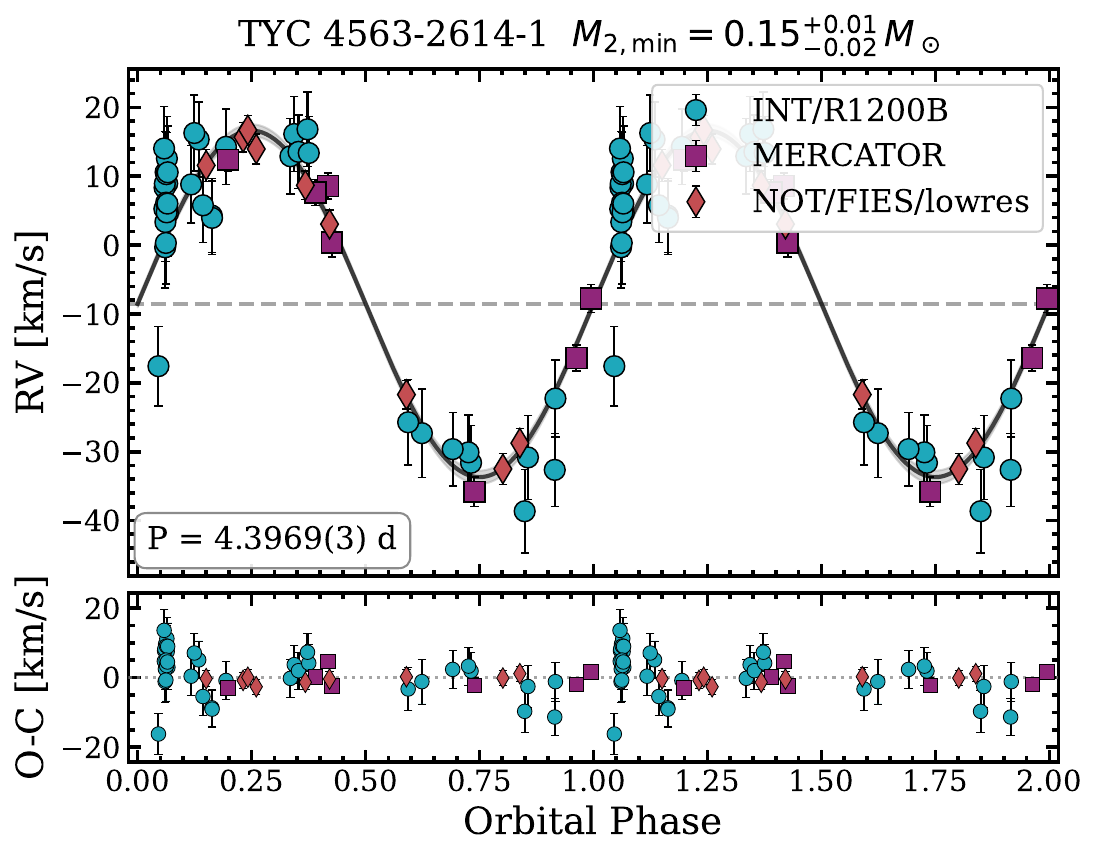}\\
        \includegraphics[width=0.93\linewidth]{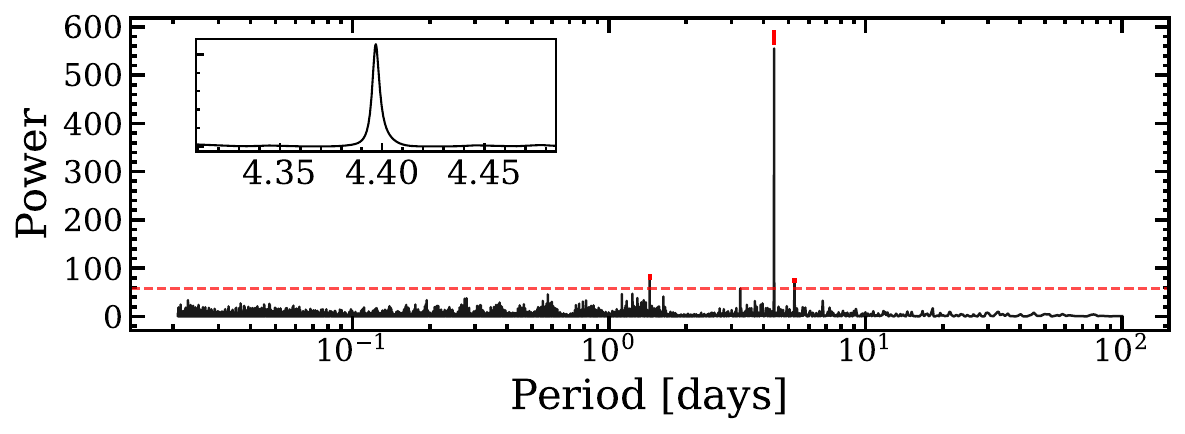}
    \end{minipage}
    \hspace{0.02\linewidth}
    \begin{minipage}[t]{0.22\linewidth}
        \centering
        \includegraphics[width=0.95\linewidth]{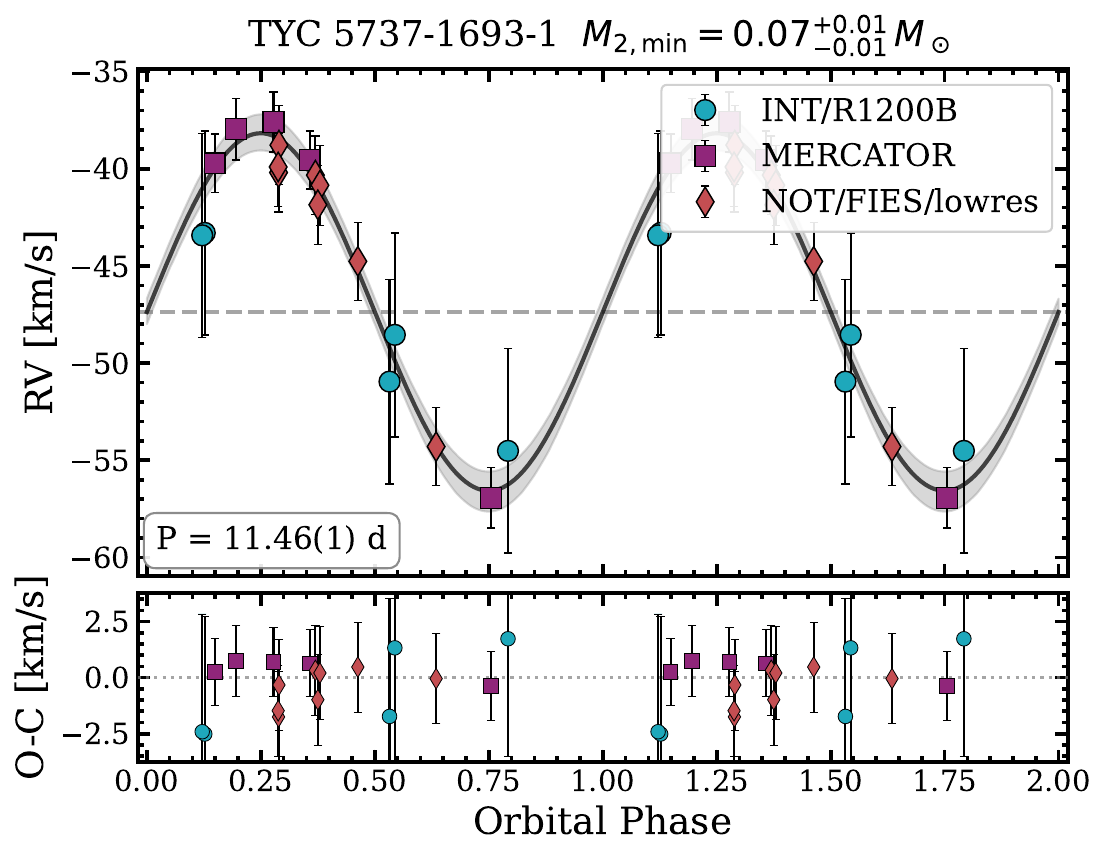}\\
        \includegraphics[width=0.93\linewidth]{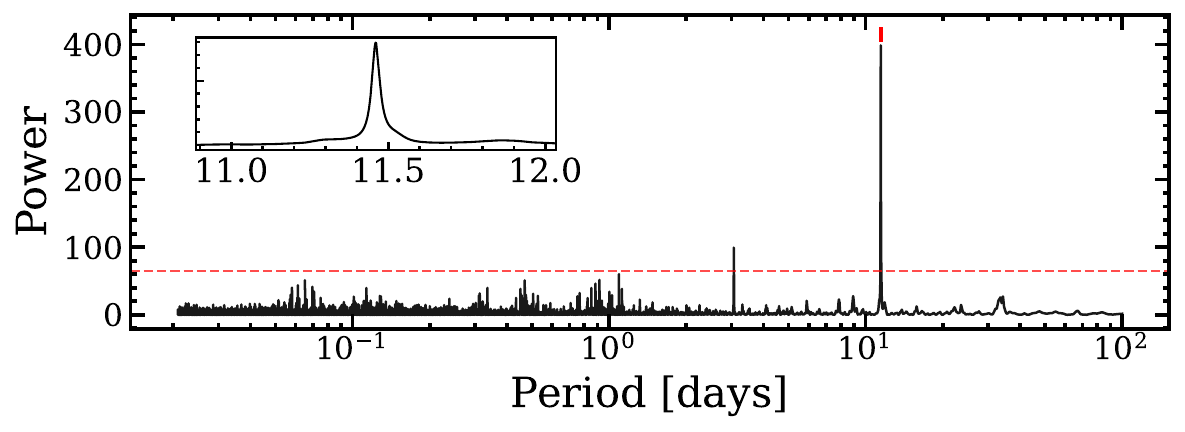}
    \end{minipage}
    \hspace{0.02\linewidth}
    \begin{minipage}[t]{0.22\linewidth}
        \centering
        \includegraphics[width=0.95\linewidth]{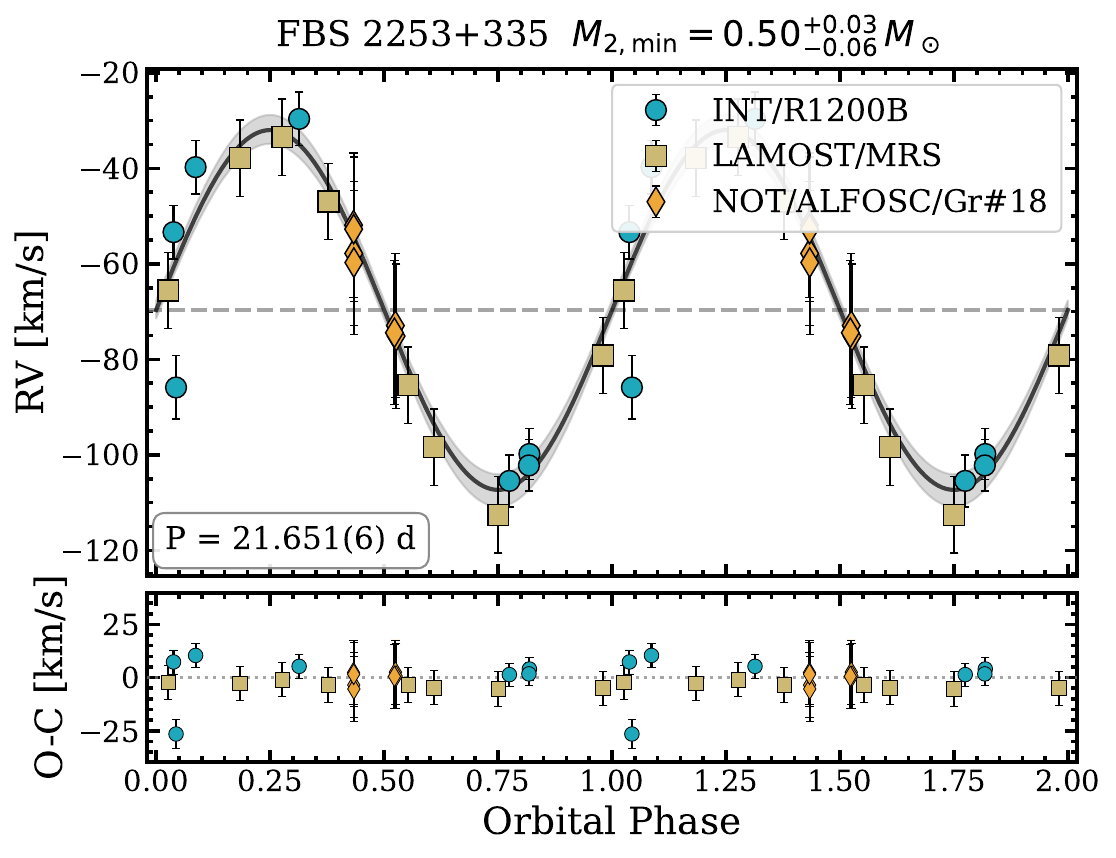}\\
        \includegraphics[width=0.93\linewidth]{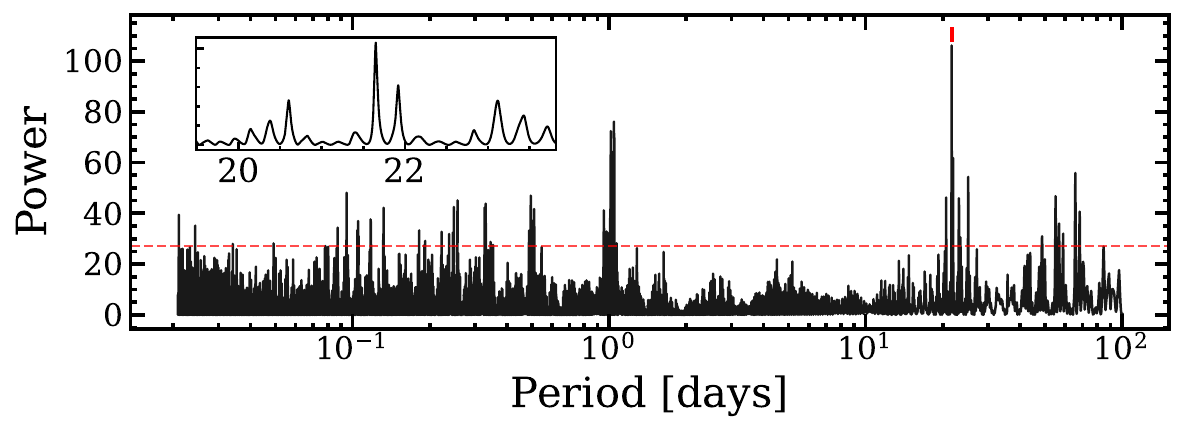}
    \end{minipage}
    \hspace{0.02\linewidth}
    \begin{minipage}[t]{0.22\linewidth}
        \centering
        \includegraphics[width=0.95\linewidth]{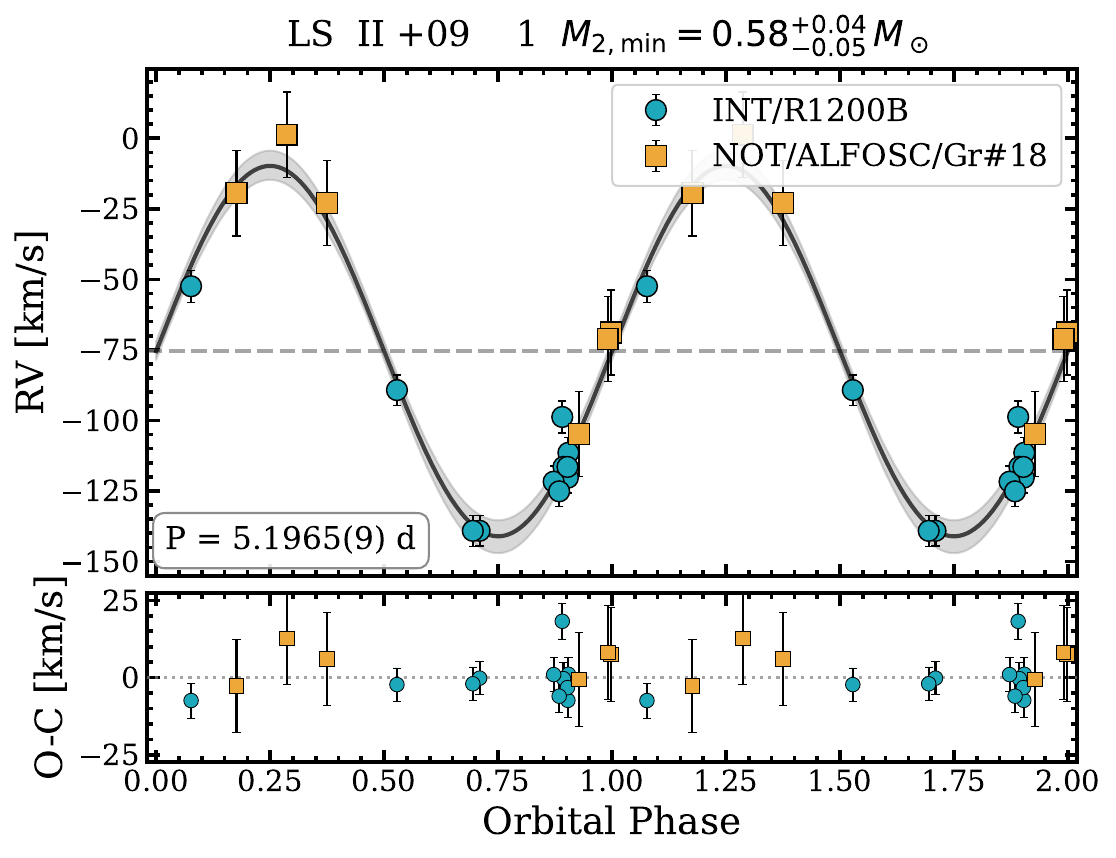}\\
        \includegraphics[width=0.93\linewidth]{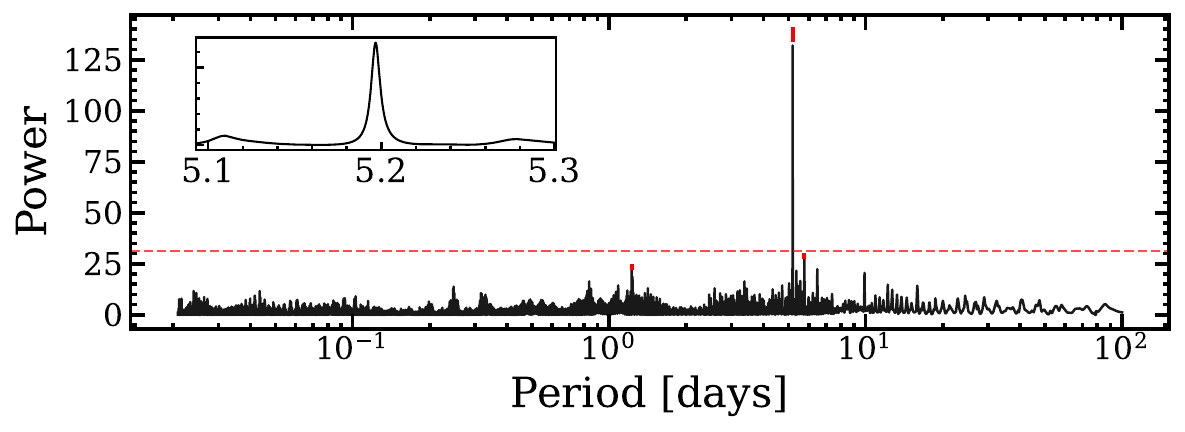}
    \end{minipage}

    \vspace{1pt}

\end{figure*}

\begin{figure*}[p]
    \centering
    \begin{minipage}[t]{0.22\linewidth}
        \centering
        \includegraphics[width=0.95\linewidth]{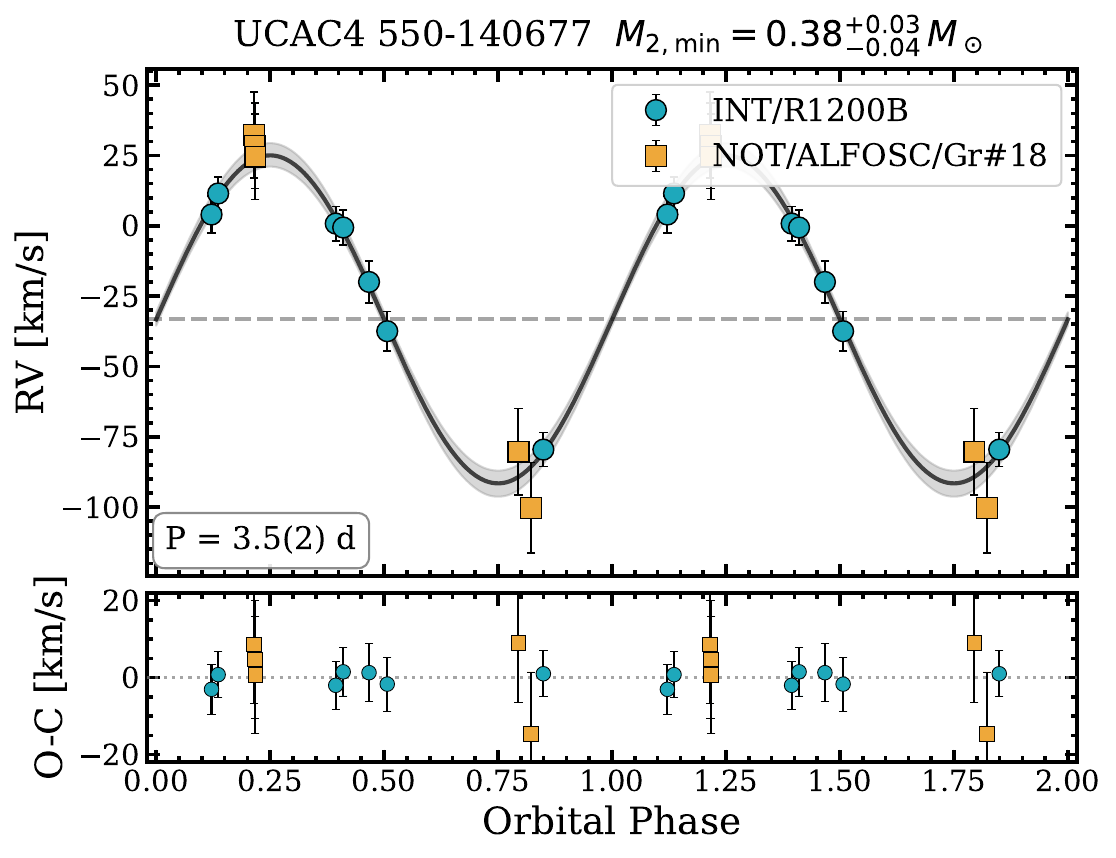}\\
        \includegraphics[width=0.93\linewidth]{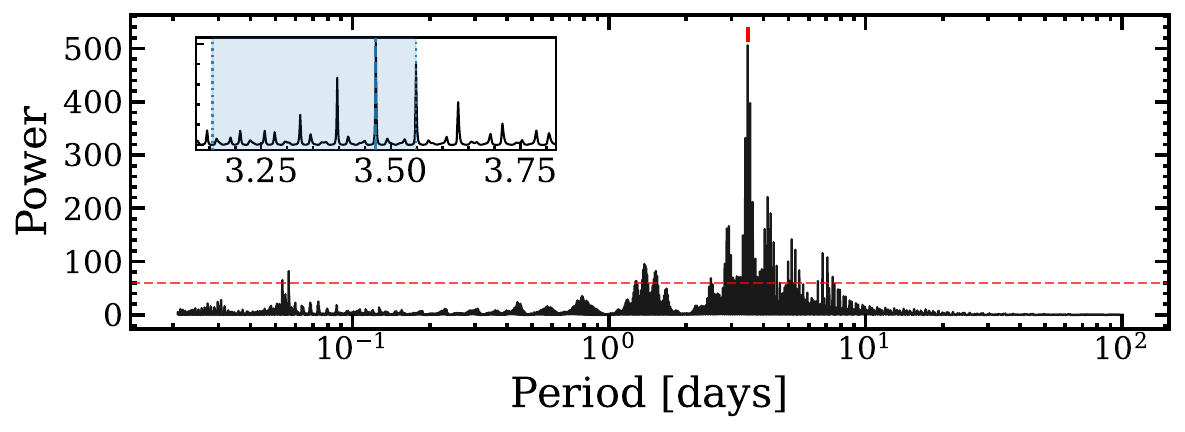}
    \end{minipage}
    \hspace{0.02\linewidth}
    \begin{minipage}[t]{0.22\linewidth}
        \centering
        \includegraphics[width=0.95\linewidth]{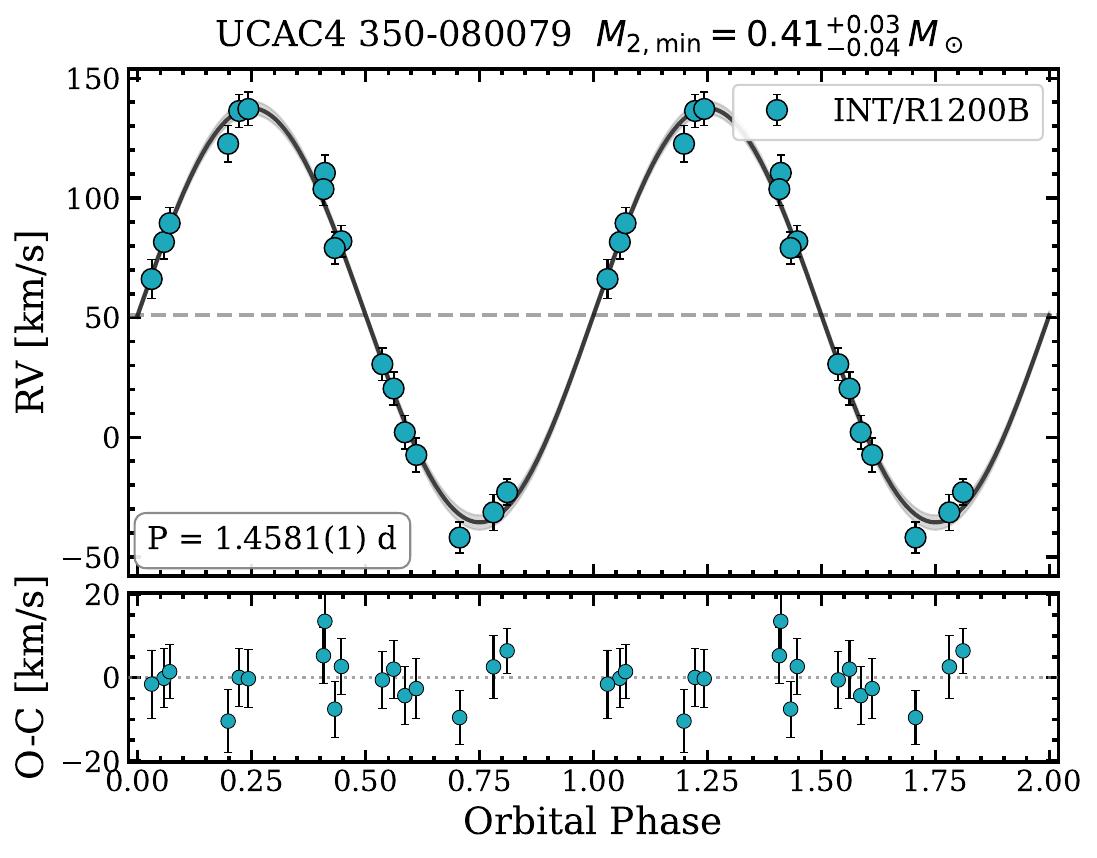}\\
        \includegraphics[width=0.93\linewidth]{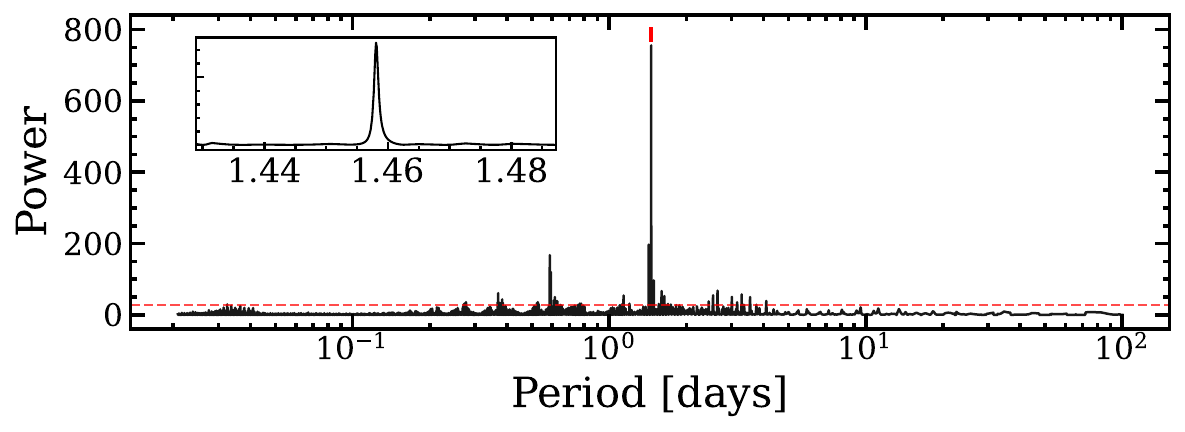}
    \end{minipage}
    \hspace{0.02\linewidth}
    \begin{minipage}[t]{0.22\linewidth}
        \centering
        \includegraphics[width=0.95\linewidth]{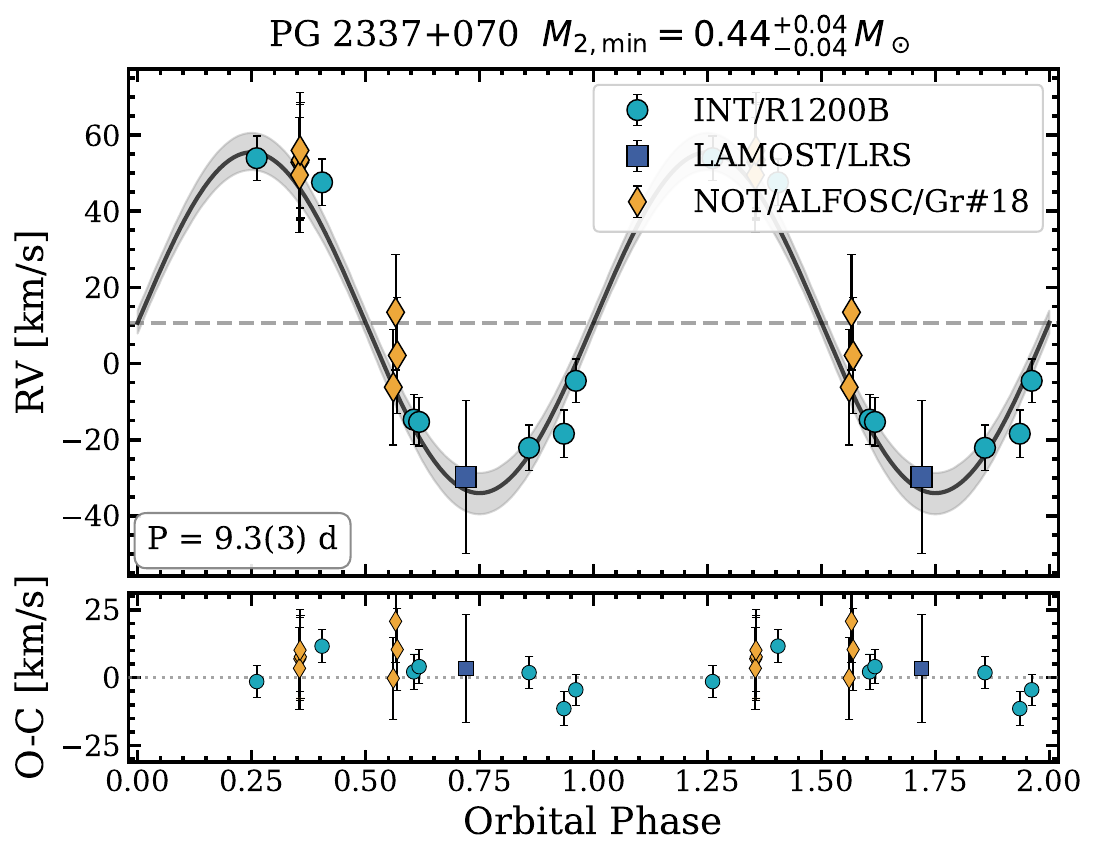}\\
        \includegraphics[width=0.93\linewidth]{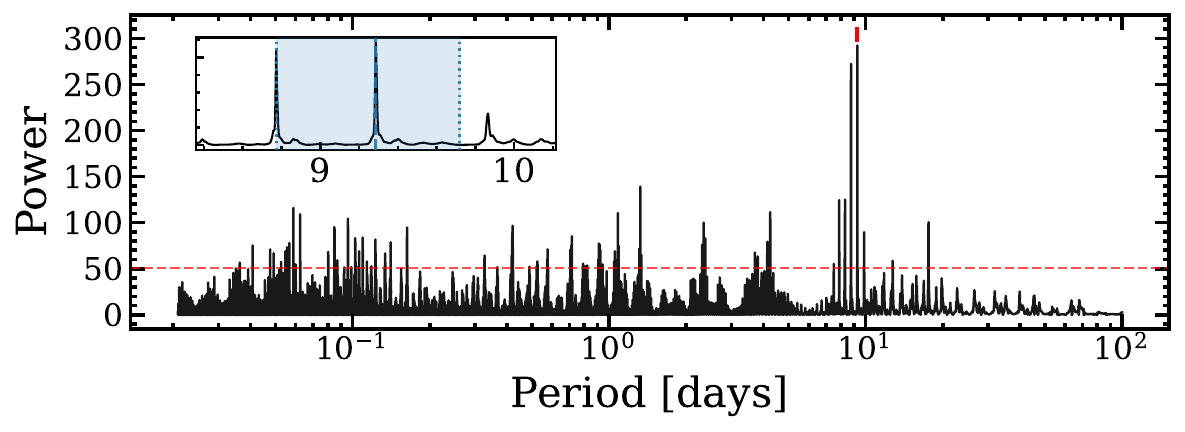}
    \end{minipage}
    \hspace{0.02\linewidth}
    \begin{minipage}[t]{0.22\linewidth}
        \centering
        \includegraphics[width=0.95\linewidth]{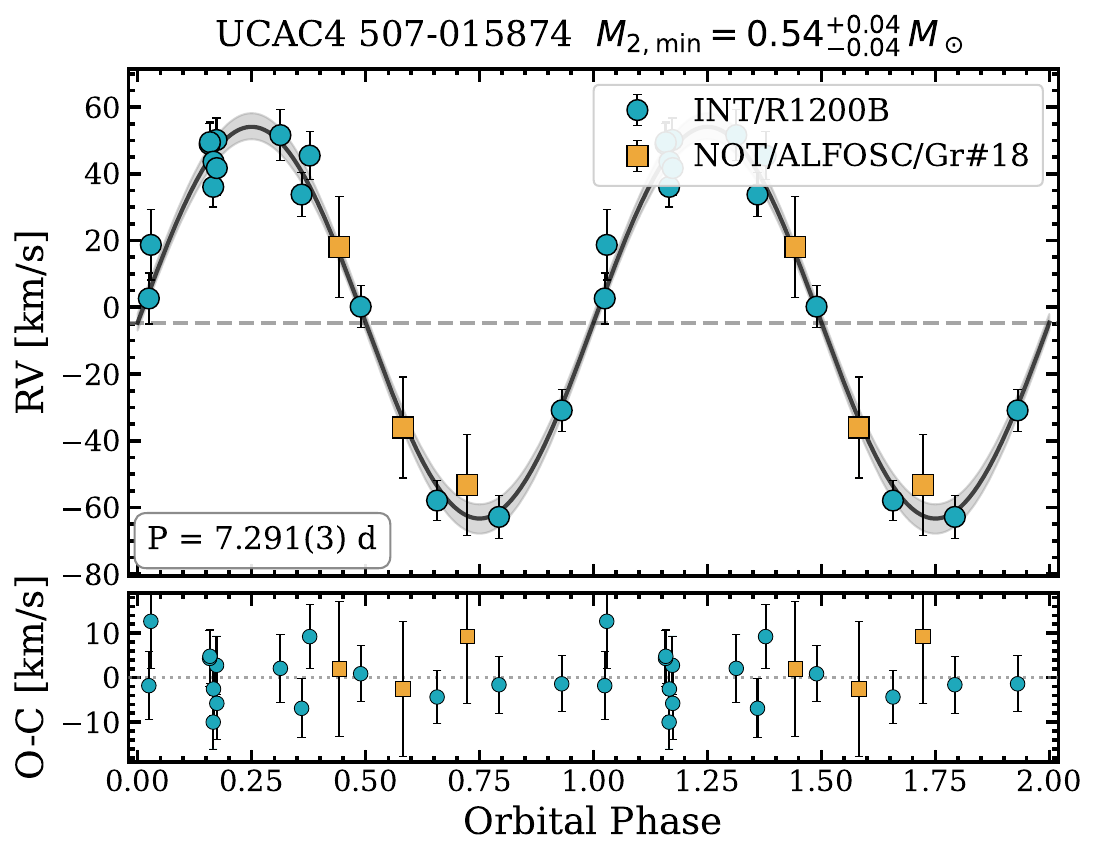}\\
        \includegraphics[width=0.93\linewidth]{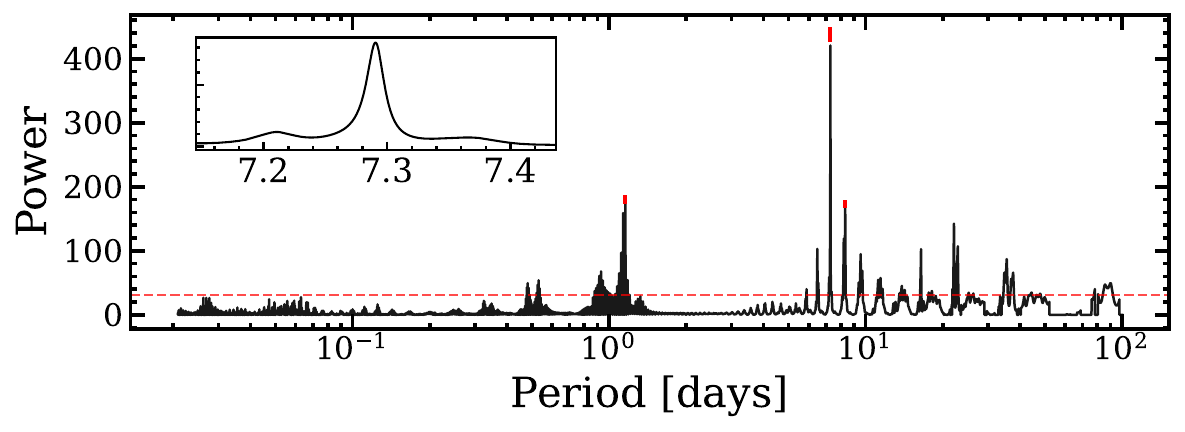}
    \end{minipage}

    \vspace{1pt}

\end{figure*}

\begin{figure*}[p]
    \centering
    \begin{minipage}[t]{0.22\linewidth}
        \centering
        \includegraphics[width=0.95\linewidth]{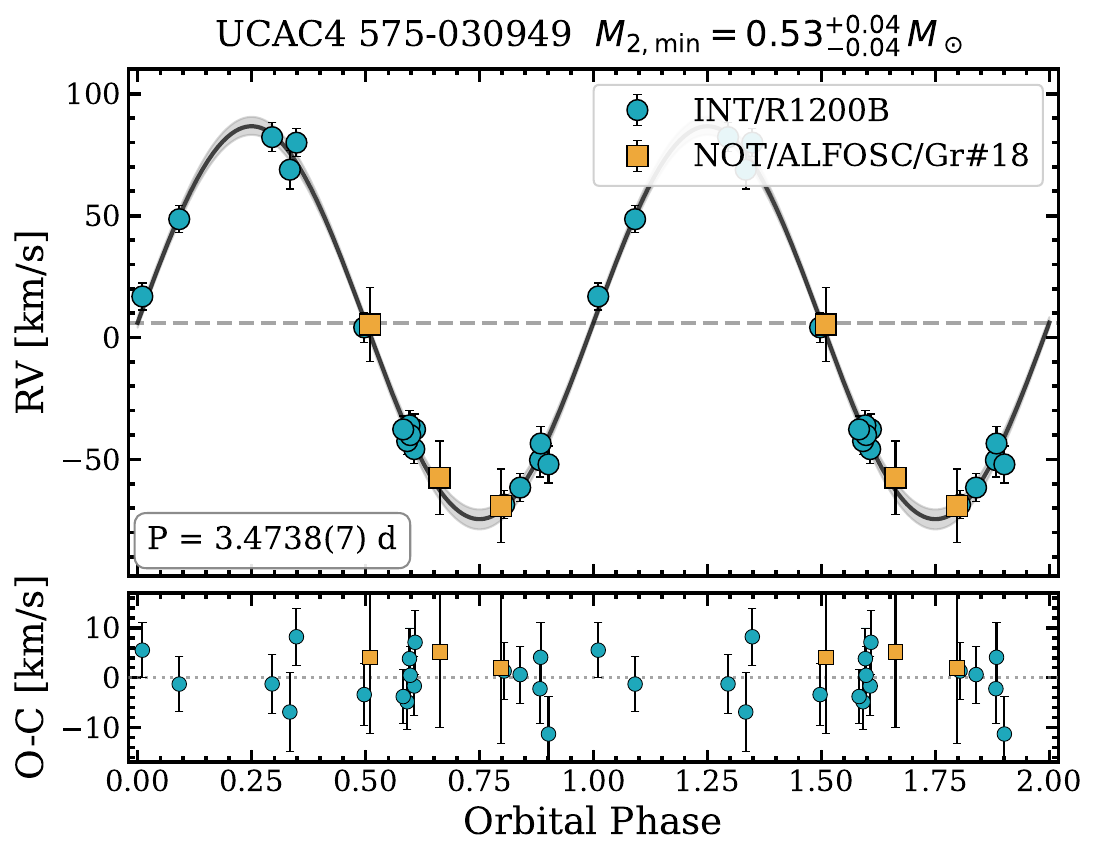}\\
        \includegraphics[width=0.93\linewidth]{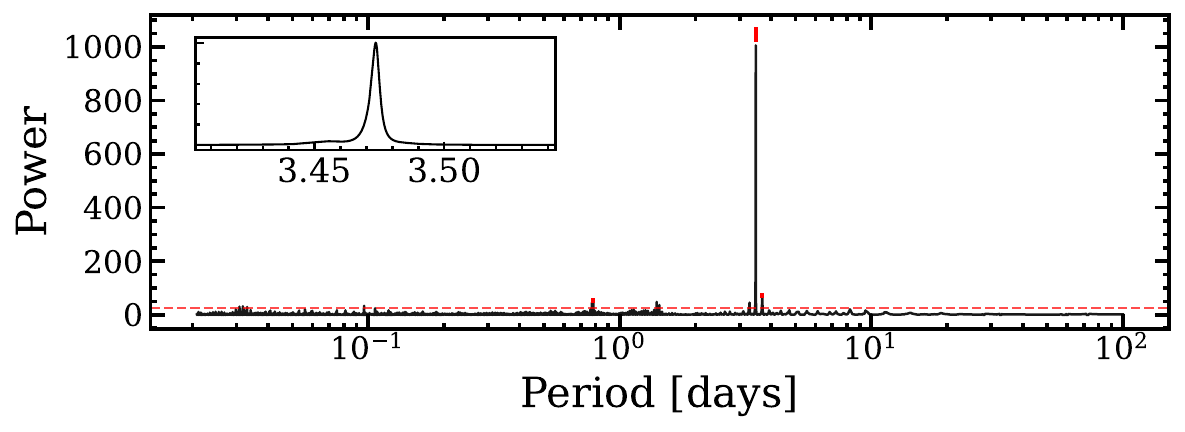}
    \end{minipage}
    \hspace{0.02\linewidth}
    \begin{minipage}[t]{0.22\linewidth}
        \centering
        \includegraphics[width=0.95\linewidth]{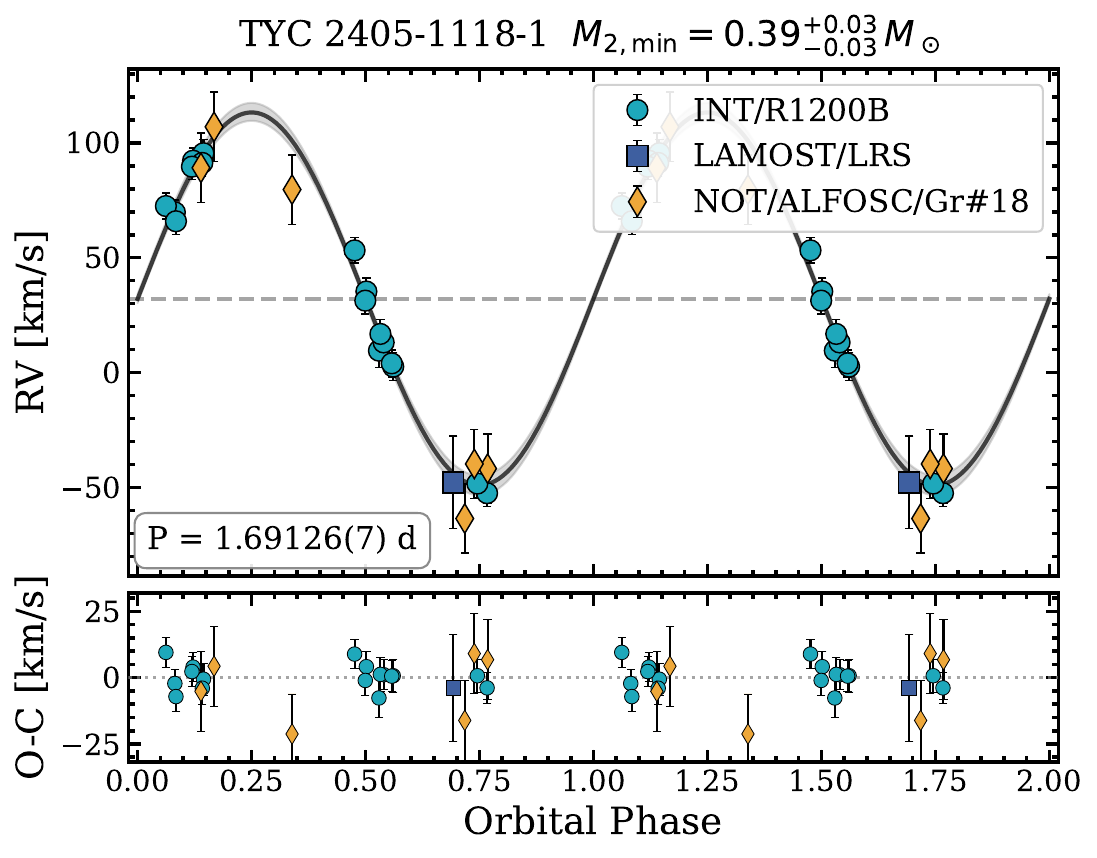}\\
        \includegraphics[width=0.93\linewidth]{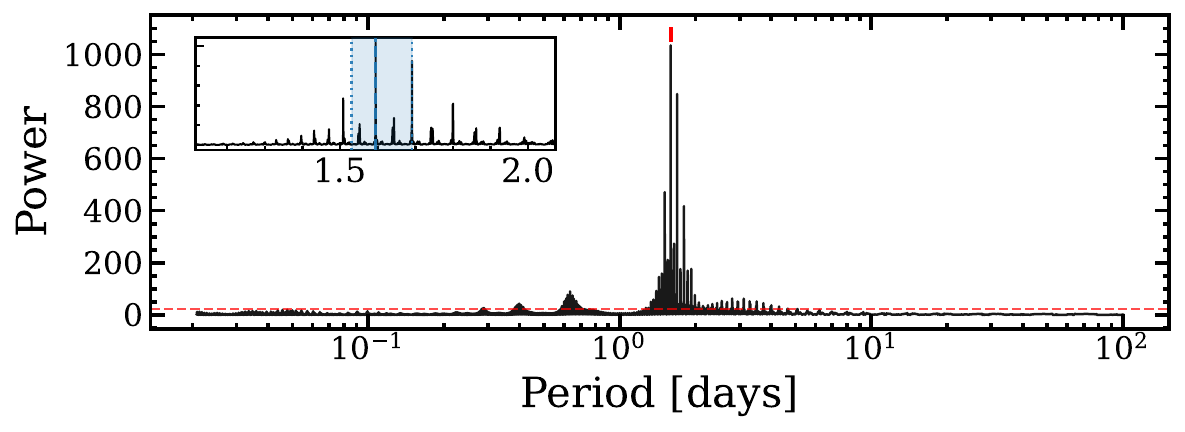}
    \end{minipage}
    \hspace{0.02\linewidth}
    \begin{minipage}[t]{0.22\linewidth}
        \centering
        \includegraphics[width=0.95\linewidth]{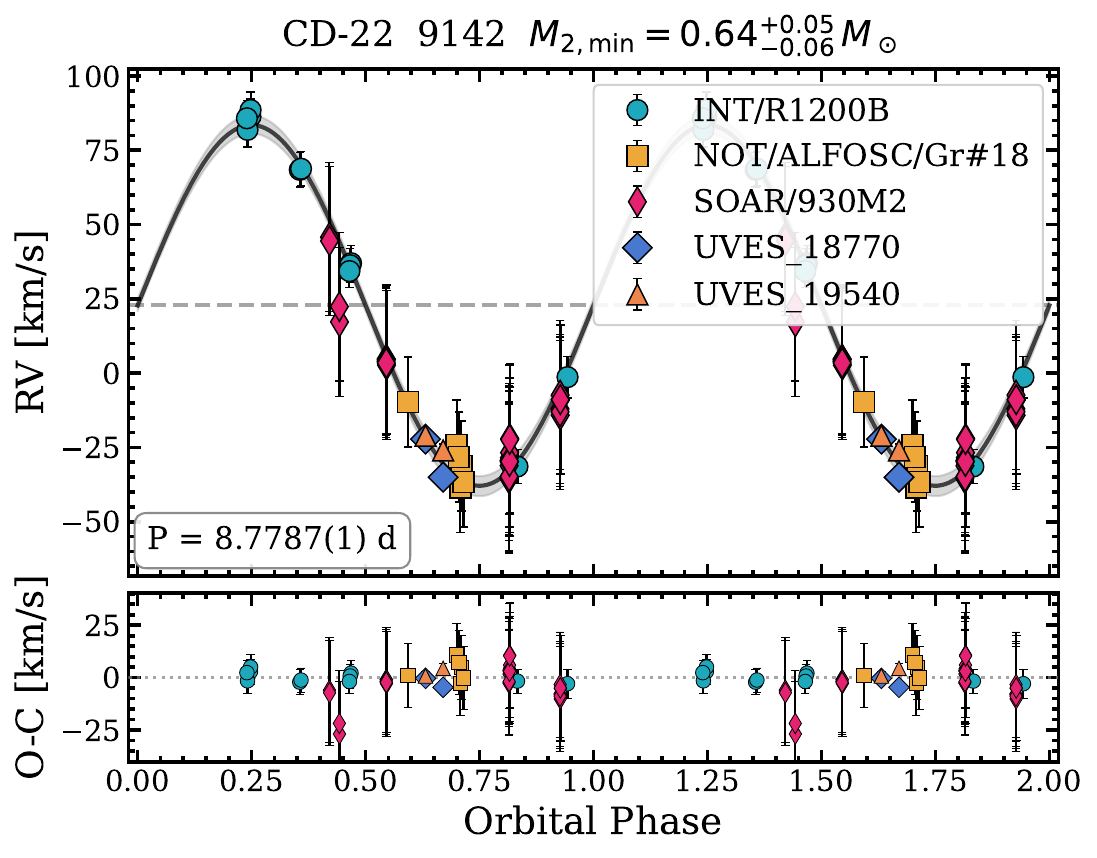}\\
        \includegraphics[width=0.93\linewidth]{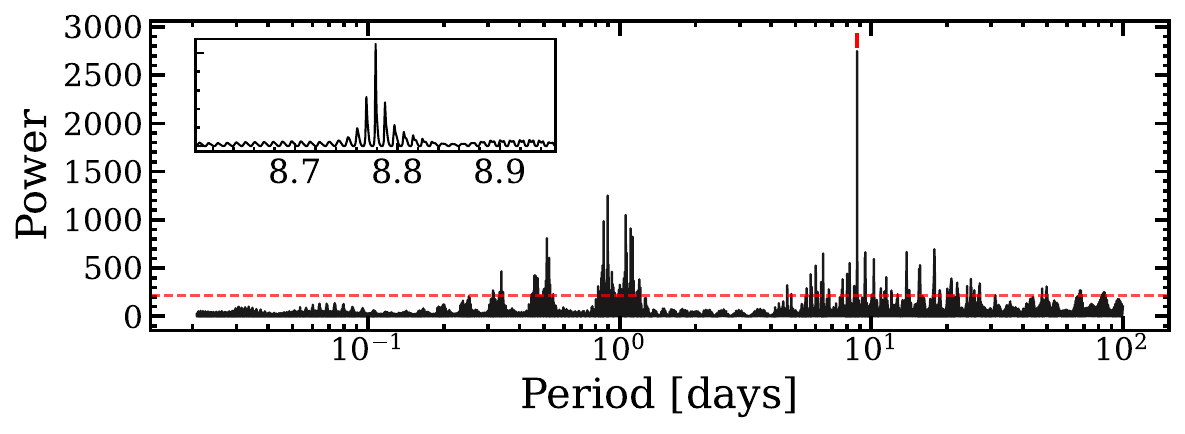}
    \end{minipage}
    \hspace{0.02\linewidth}
    \begin{minipage}[t]{0.22\linewidth}
        \centering
        \includegraphics[width=0.95\linewidth]{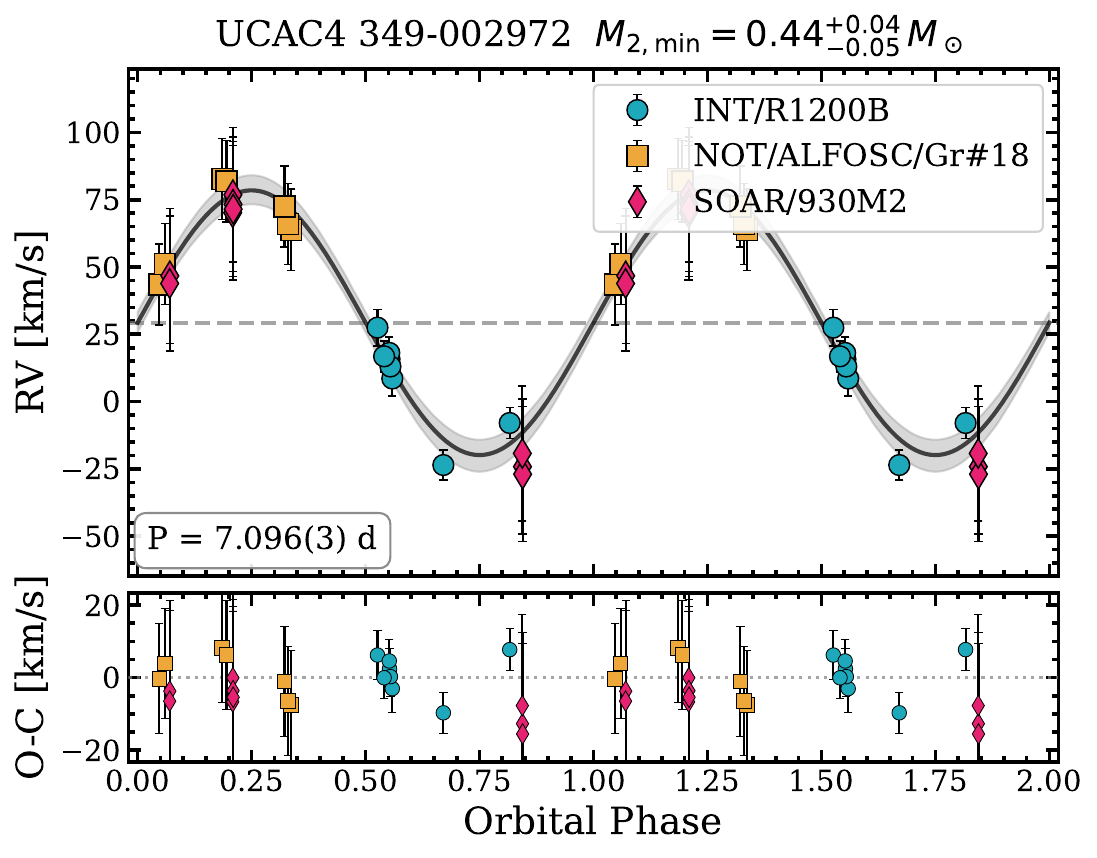}\\
        \includegraphics[width=0.93\linewidth]{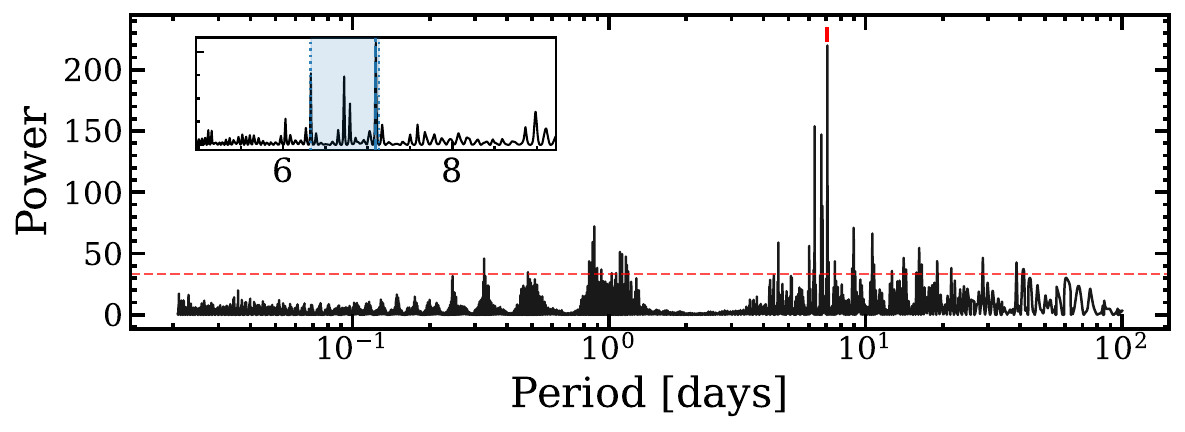}
    \end{minipage}

    \vspace{1pt}

\end{figure*}

\begin{figure*}[p]
    \ContinuedFloat
    \centering
    \begin{minipage}[t]{0.22\linewidth}
        \centering
        \includegraphics[width=0.95\linewidth]{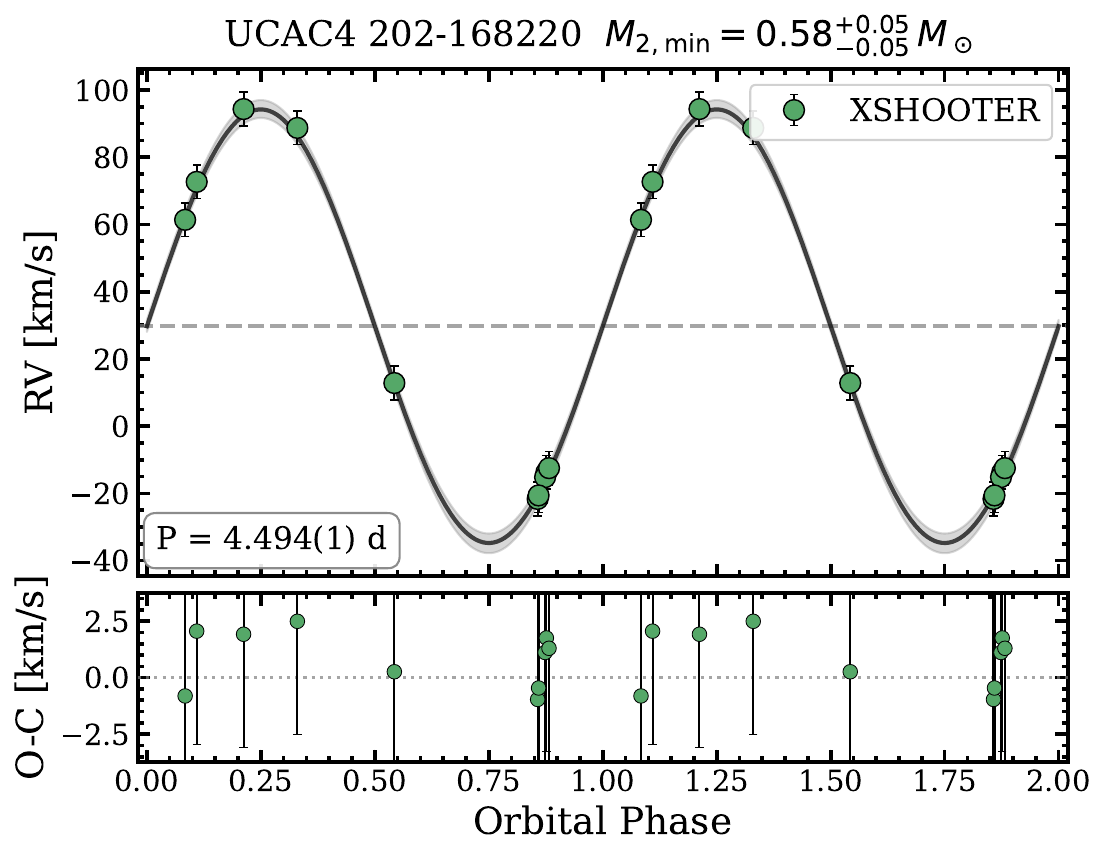}\\
        \includegraphics[width=0.93\linewidth]{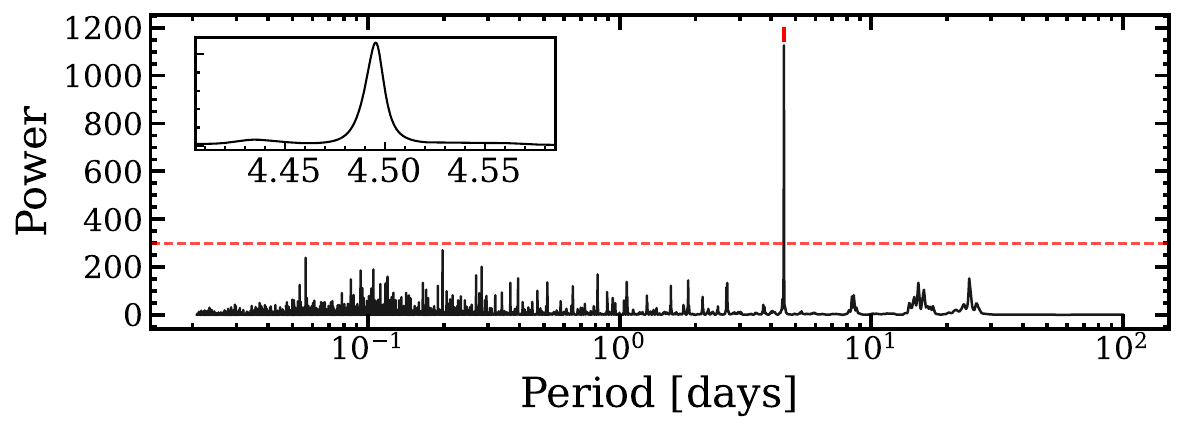}
    \end{minipage}
    \hspace{0.02\linewidth}
    \begin{minipage}[t]{0.22\linewidth}
        \centering
    \end{minipage}
    \hspace{0.02\linewidth}
    \begin{minipage}[t]{0.22\linewidth}
        \centering
    \end{minipage}
    \hspace{0.02\linewidth}
    \begin{minipage}[t]{0.22\linewidth}
        \centering
    \end{minipage}
    \vspace{1pt}
    \caption{RV curves (top) and power spectra (bottom) for sources without LC data.}
    \label{fig:rv_power}
\end{figure*}

\begin{table*}[!]
\centering
\footnotesize
\caption{Light curve fitting results for newly identified short-period binary systems, divided by companion type.}
\label{tab:lc_fitting}
\resizebox{\textwidth}{!}{
\begin{tabular}{@{}llccccccccc@{}}
\toprule\toprule
    Name & Class & $P$ & $K_1$ & $i$ & $q$ & $M_1$ & $R_1$ & $M_2$ & $R_2$ & $a$ \\
    & & [days] & [\kms] & [$\degr$] & & [\msun] & [\rsun] & [\msun] & [\rsun] & [\rsun] \\[0.1cm]
\midrule
    LS~IV~+09~2 & sdB+MS & $0.225154$ & $48.48 \pm 5.14$ & $76.3 \pm 1.3$ & $0.21 \pm 0.02$ & $0.472 \pm 0.122$ & $0.226 \pm 0.011$ & $0.098 \pm 0.022$ & $0.161 \pm 0.019$ & $1.292 \pm 0.088$ \\[0.1cm]
    TYC~4470-864-1 & sdB+MS & $0.467982$ & $42.69 \pm 2.77$ & $80.2 \pm 4.8$ & $0.22 \pm 0.02$ & $0.525 \pm 0.080$ & $0.234 \pm 0.006$ & $0.118 \pm 0.014$ & $0.143 \pm 0.005$ & $2.189 \pm 0.073$ \\[0.1cm]
    ATO~J029.0051$+$40.0561 & sdB+MS & $0.193658$ & $54.23 \pm 3.32$ & $73.8 \pm 4.0$ & $0.21 \pm 0.01$ & $0.561 \pm 0.080$ & $0.213 \pm 0.006$ & $0.119 \pm 0.014$ & $0.139 \pm 0.006$ & $1.238 \pm 0.044$ \\[0.1cm]
    2MASS~J15292631$+$7011543 & sdB+MS & $0.199646$ & $71.07 \pm 6.52$ & $71.2 \pm 4.8$ & $0.30 \pm 0.03$ & $0.548 \pm 0.091$ & $0.234 \pm 0.007$ & $0.164 \pm 0.024$ & $0.197 \pm 0.009$ & $1.284 \pm 0.051$ \\[0.1cm]
    NSVS~842188 & sdB+MS & $0.194934$ & $107.36 \pm 6.32$ & $70.5 \pm 4.8$ & $0.50 \pm 0.03$ & $0.528 \pm 0.085$ & $0.200 \pm 0.006$ & $0.266 \pm 0.031$ & $0.259 \pm 0.011$ & $1.310 \pm 0.047$ \\[0.1cm]
\midrule
    TYC~7691-3990-1 & sdB+WD & $0.074264$ & $172.18 \pm 5.76$ & $87.6 \pm 14.7$ & $0.63 \pm 0.02$ & $0.426 \pm 0.075$ & $0.186 \pm 0.004$ & $0.285 \pm 0.016$ & $0.019 \pm 0.001$ & $0.658 \pm 0.012$ \\[0.1cm]
    HD~191351 & sdB+WD & $0.089163$ & $135.25 \pm 16.99$ & $31.7 \pm 2.9$ & $1.13 \pm 0.21$ & $0.499 \pm 0.094$ & $0.151 \pm 0.002$ & $0.562 \pm 0.061$ & $0.012 \pm 0.001$ & $0.857 \pm 0.011$ \\[0.1cm]
    TYC~3135-86-1 & sdB+WD & $0.248894$ & $167.58 \pm 26.31$ & $60.6 \pm 12.3$ & $1.30 \pm 0.34$ & $0.442 \pm 0.084$ & $0.356 \pm 0.006$ & $0.574 \pm 0.070$ & $0.003 \pm 0.001$ & $1.674 \pm 0.019$ \\[0.1cm]
\bottomrule
\end{tabular}
}
\end{table*}

\begin{table*}[!]
\tiny
\caption{50 newly identified close binaries and orbital parameters.}
\label{table_new_binaries}
\renewcommand{\arraystretch}{1.2}
\centering
\begin{tabular}{cccccccc}
\toprule\toprule
Name & Class & Number of spectra & Period & M$_{2,min}$ & K & $\gamma$ & logp \\
 &  & & [days] & [\msun] &  [\kms] & [\kms] &  \\
\midrule
\multicolumn{7}{c}{\textit{Solved systems}} \\
\midrule
TYC~7691-3990-1 & sdB+WD & 20 & $0.0742719^{+0.0000023}_{-0.0000026}$ & $0.28^{+0.02}_{-0.02}$ & $173.3^{+4.8}_{-4.8}$ & $8.7^{+3.5}_{-3.5}$ & $-287.88$ \\
HD~191351 & sdB+WD & 73 & $0.08916264^{+0.00000035}_{-0.00000037}$ & $0.24^{+0.02}_{-0.02}$ & $135.8^{+4.0}_{-3.9}$ & $-25.9^{+2.7}_{-2.8}$ & $-129.94$ \\
2200231588777454208 & sdB+WD & 15 & $0.1744159^{+0.0000039}_{-0.0000031}$ & $0.77^{+0.05}_{-0.05}$ & $260.6^{+3.8}_{-4.0}$ & $-10.7^{+2.8}_{-2.7}$ & $-300.00$ \\
ATO~J029.0051+40.0561 & sdB+MS & 14 & $0.1936585223^{+0.0000000038}_{-0.0000000038}$ & $0.104^{+0.010}_{-0.010}$ & $54.4^{+3.8}_{-3.8}$ & $11.6^{+2.5}_{-2.5}$ & $-113.95$ \\
TYC~4542-482-1 & sdB+MS & 13 & $0.1949341811^{+0.0000000030}_{-0.0000000015}$ & $0.24^{+0.02}_{-0.02}$ & $104.8^{+7.9}_{-7.8}$ & $-0.4^{+4.7}_{-4.7}$ & $-32.19$ \\
2MASS~J15292631+7011543 & sdB+MS & 14 & $0.1996438628^{+0.0000000039}_{-0.0000000038}$ & $0.16^{+0.01}_{-0.01}$ & $71.8^{+6.2}_{-6.2}$ & $-15.8^{+3.3}_{-3.3}$ & $-59.66$ \\
LS~IV+092 & sdB+MS & 22 & $0.2251545^{+0.0000033}_{-0.0000033}$ & $0.094^{+0.009}_{-0.010}$ & $47.6^{+3.1}_{-3.1}$ & $14.5^{+2.3}_{-2.3}$ & $-121.14$ \\
TYC~3135-86-1 & sdB+WD & 15 & $0.248904^{+0.000010}_{-0.000010}$ & $0.48^{+0.04}_{-0.04}$ & $167.3^{+2.9}_{-2.8}$ & $-11.8^{+2.8}_{-2.7}$ & $-300.00$ \\
TYC~4470-864-1 & sdB+MS & 12 & $0.4679803377^{+0.0000000010}_{-0.0000000010}$ & $0.11^{+0.01}_{-0.01}$ & $42.8^{+0.0}_{-0.0}$ & $-4.7^{+0.0}_{-0.0}$ & $-66.14$ \\
TYC~5285-201-1 & sdB+WD & 23 & $0.4891373^{+0.0000041}_{-0.0000041}$ & $0.46^{+0.04}_{-0.04}$ & $130.1^{+5.3}_{-5.2}$ & $10.9^{+3.4}_{-3.5}$ & $-242.77$ \\
TYC~769-1149-1 & sdOB+WD & 18 & $0.598552625^{+0.000000015}_{-0.000000024}$ & $0.77^{+0.06}_{-0.06}$ & $158.4^{+2.5}_{-2.5}$ & $35.9^{+2.3}_{-2.3}$ & $-300.00$ \\
LS~III+48~50 & sdOB+MS/WD & 19 & $0.72197^{+0.00340}_{-0.00065}$ & $0.18^{+0.02}_{-0.02}$ & $47.6^{+2.4}_{-2.3}$ & $-4.4^{+1.8}_{-1.8}$ & $-80.22$ \\
5881540795158796416 & sdB+WD & 18 & $0.810387^{+0.000024}_{-0.000026}$ & $0.39^{+0.03}_{-0.04}$ & $102.7^{+3.3}_{-3.3}$ & $77.4^{+1.9}_{-1.9}$ & $-256.05$ \\
BD+48~433 & sdB+WD & 25 & $1.4022588^{+0.0000053}_{-0.0000077}$ & $0.57^{+0.04}_{-0.04}$ & $101.9^{+1.1}_{-1.1}$ & $-29.9^{+0.8}_{-0.8}$ & $-300.00$ \\
UCAC4~350-080079 & sdB+WD & 17 & $1.45811^{+0.00013}_{-0.00012}$ & $0.41^{+0.04}_{-0.04}$ & $86.5^{+2.5}_{-2.4}$ & $51.0^{+1.7}_{-1.8}$ & $-258.32$ \\
TYC~2405-1118-1 & sdB+WD & 24 & $1.595^{+0.097}_{-0.064}$ & $0.38^{+0.03}_{-0.03}$ & $80.2^{+2.7}_{-2.7}$ & $26.2^{+1.5}_{-1.5}$ & $-228.92$ \\
GSC00141-01628 & sdB+WD & 35 & $1.615515^{+0.000084}_{-0.000056}$ & $0.28^{+0.02}_{-0.03}$ & $62.0^{+2.4}_{-2.3}$ & $50.9^{+1.8}_{-1.8}$ & $-300.00$ \\
TYC~497-63-1 & sdB+WD & 16 & $1.92069^{+0.00015}_{-0.00015}$ & $0.48^{+0.04}_{-0.04}$ & $81.5^{+5.1}_{-4.8}$ & $-7.7^{+2.5}_{-2.4}$ & $-272.53$ \\
GALEX~J07015-6717 & sdOB+WD & 43 & $2.54387980^{+0.00000018}_{-0.00000019}$ & $0.74^{+0.05}_{-0.05}$ & $96.1^{+3.4}_{-3.3}$ & $4.1^{+3.2}_{-3.2}$ & $-171.76$ \\
UCAC4~497-042535 & sdO+WD & 17 & $2.60551^{+0.00035}_{-0.00035}$ & $0.65^{+0.06}_{-0.06}$ & $93.1^{+4.1}_{-4.2}$ & $-13.9^{+5.2}_{-5.2}$ & $-137.77$ \\
TYC~1969-78-1 & sdB+WD & 29 & $2.697705^{+0.000064}_{-0.000065}$ & $0.48^{+0.04}_{-0.04}$ & $78.5^{+2.3}_{-2.3}$ & $-20.3^{+1.6}_{-1.6}$ & $-300.00$ \\
LAMOST~J213129.05+195157.0 & sdO+WD & 12 & $3.471^{+0.077}_{-0.315}$ & $0.39^{+0.03}_{-0.04}$ & $58.3^{+3.9}_{-3.8}$ & $-33.3^{+2.5}_{-2.5}$ & $-43.87$ \\
UCAC4~575-030949 & sdB+WD & 20 & $3.47378^{+0.00064}_{-0.00069}$ & $0.53^{+0.03}_{-0.04}$ & $80.6^{+3.3}_{-3.4}$ & $6.1^{+2.3}_{-2.3}$ & $-238.10$ \\
TYC~6800-72-1 & iHe-sdOB+WD & 22 & $3.60007^{+0.00066}_{-0.00064}$ & $0.55^{+0.06}_{-0.05}$ & $65.3^{+3.0}_{-3.0}$ & $-21.7^{+2.1}_{-2.1}$ & $-228.59$ \\
TYC~4563-2614-1 & sdOB+MS/WD & 54 & $4.39687^{+0.00031}_{-0.00031}$ & $0.15^{+0.01}_{-0.02}$ & $25.2^{+0.8}_{-0.8}$ & $-8.5^{+0.6}_{-0.6}$ & $-278.33$ \\
UCAC4~202-168220 & sdB+WD & 23 & $4.49386^{+0.00094}_{-0.00154}$ & $0.59^{+0.05}_{-0.05}$ & $64.5^{+2.2}_{-2.2}$ & $29.7^{+1.9}_{-1.9}$ & $-177.76$ \\
LS~II+091 & sdB+WD & 17 & $5.19652^{+0.00093}_{-0.00092}$ & $0.58^{+0.04}_{-0.05}$ & $65.7^{+4.6}_{-4.6}$ & $-75.4^{+3.4}_{-3.4}$ & $-60.79$ \\
UCAC4~349-002972 & sdO+WD & 24 & $7.096^{+0.036}_{-0.771}$ & $0.44^{+0.05}_{-0.05}$ & $49.7^{+3.8}_{-3.7}$ & $30.3^{+2.8}_{-2.7}$ & $-26.54$ \\
UCAC4~507-015874 & sdOB+WD & 18 & $7.2906^{+0.0030}_{-0.0032}$ & $0.54^{+0.04}_{-0.05}$ & $58.7^{+3.5}_{-3.4}$ & $-4.6^{+2.6}_{-2.6}$ & $-109.82$ \\
CD-22~9142 & sdO+WD & 47 & $8.77866^{+0.00013}_{-0.00012}$ & $0.64^{+0.06}_{-0.06}$ & $61.1^{+2.1}_{-2.1}$ & $22.8^{+1.7}_{-1.7}$ & $-300.00$ \\
PG~2337+070 & sdB+WD & 15 & $9.29^{+0.43}_{-0.51}$ & $0.46^{+0.03}_{-0.04}$ & $45.4^{+3.2}_{-3.3}$ & $11.6^{+2.1}_{-2.1}$ & $-37.92$ \\
PG~1610+529 & sdB+WD & 35 & $10.9554^{+0.0042}_{-0.0038}$ & $0.63^{+0.05}_{-0.05}$ & $54.2^{+2.5}_{-2.4}$ & $-37.0^{+1.4}_{-1.4}$ & $-105.46$ \\
TYC~5737-1693-1 & sdB+MS/WD & 18 & $11.4604^{+0.0097}_{-0.0094}$ & $0.069^{+0.006}_{-0.007}$ & $9.2^{+0.7}_{-0.7}$ & $-47.4^{+0.7}_{-0.7}$ & $-25.46$ \\
FBS~2253+335 & sdB+WD & 22 & $21.6505^{+0.0060}_{-0.0058}$ & $0.49^{+0.04}_{-0.04}$ & $37.7^{+2.7}_{-2.7}$ & $-69.7^{+2.0}_{-2.0}$ & $-50.23$ \\
\midrule
\multicolumn{7}{c}{\textit{Unsolved systems - high resolution spectroscopic follow up required}} \\
\midrule
LAMOST~J044847.21-040016.9 & sdB+MS/WD & 19 & $-$ & $-$ & $-$ & $-$ & $-7.49$ \\
2MASS~J02065617+1438585 & sdB+MS/WD & 16 & $-$ & $-$ & $-$ & $-$ & $-8.91$ \\
UCAC4~174-038075 & sdB+MS/WD & 71 & $-$ & $-$ & $-$ & $-$ & $-26.87$ \\
UCAC4~149-221427 & sdB+MS/WD & 51 & $-$ & $-$ & $-$ & $-$ & $-6.77$ \\
5842817473052628480 & sdB+WD & 51 & $-$ & $-$ & $-$ & $-$ & $-250.93$ \\
6028436785642385152 & sdB+MS/WD & 11 & $-$ & $-$ & $-$ & $-$ & $-46.36$ \\
EC~20106-5248 & sdB+MS/WD & 8 & $-$ & $-$ & $-$ & $-$ & $-21.55$ \\
UCAC4~219-125136 & sdB+MS/WD & 9 & $-$ & $-$ & $-$ & $-$ & $-54.53$ \\
UCAC4~280-113383 & sdB+MS/WD & 10 & $-$ & $-$ & $-$ & $-$ & $-14.14$ \\
CD-39~14181 & sdB+WD & 36 & $-$ & $-$ & $-$ & $-$ & $-125.91$ \\
UCAC4~198-195239 & sdOB+MS/WD & 42 & $-$ & $-$ & $-$ & $-$ & $-154.22$ \\
LAMOST~J180933.32+223059.9 & sdB+MS/WD & 16 & $-$ & $-$ & $-$ & $-$ & $-14.66$ \\
6849135629120266624 & sdB+MS/WD & 3 & $-$ & $-$ & $-$ & $-$ & $-22.29$ \\
LS~IV-132 & sdO+MS/WD & 12 & $-$ & $-$ & $-$ & $-$ & $-21.07$ \\
UCAC4~282-094834 & sdB+MS/WD & 11 & $-$ & $-$ & $-$ & $-$ & $-13.24$ \\
HD~350426 & sdO+MS/WD & 15 & $-$ & $-$ & $-$ & $-$ & $-33.73$ \\
\bottomrule
\end{tabular}
\tablefoot{Uncertainties were derived during the MCMC fitting routine.}
\end{table*}

\begin{table*}[!]
\tiny
\caption{45 known close binaries with published orbital solutions.}
\label{table_known_binaries}
\renewcommand{\arraystretch}{1.2}
\centering
\begin{tabular}{ccccccc}
\toprule\toprule
Name & Class & Period & M$_{2,min}$ & $\gamma$ & $K$ & Reference \\
 &  & [days] & [\msun] & [km s$^{-1}$] & [km s$^{-1}$] & \\
\midrule
CD-30~11223 & sdB$+$WD & $0.048979072 \pm 0.000000004$ & $0.73 \pm 0.04$ & $16.5 \pm 0.3$ & $377.0 \pm 0.4$ & \citet{Geier_2013AA...554A..54G} \\
HD~265435 & sdOB$+$WD & $0.068818489 \pm 0.000000003$ & $0.85 \pm 0.07$ & $8.2 \pm 0.8$ & $343.1 \pm 1.2$ & \citet{Pelisoli_2021NatAs...5.1052P} \\
SBSS~1709$+$535 & sdB$+$WD & $0.075835270 \pm 0.000000005$ & $0.39 \pm 0.03$ & $-36.5 \pm 2.0$ & $222.2 \pm 2.8$ & \citet{Yang_2025AA...693A.322Y} \\
BD-07~3477 & sdB$+$MS & $0.1150 \pm 0.0008$ & $0.15 \pm 0.02$ & $-13.0 \pm 0.8$ & $84.6 \pm 1.1$ & \citet{Kupfer_2015} \\
TYC~3556-3568-1 & sdB$+$dM & $0.12576530 \pm 0.00000002$ & $0.102 \pm 0.010$ & $20.1 \pm 0.3$ & $65.7 \pm 0.6$ & \citet{Ostensen_2010MNRAS.408L..51O} \\
TYC~7709-376-1 & sdB$+$dM & $0.13926940 \pm 0.00000004$ & $0.15 \pm 0.01$ & $40.0 \pm 2.0$ & $81.0 \pm 3.0$ & \citet{Schaffenroth_2013AA...553A..18S} \\
TYC~5977-517-1 & sdB$+$MS & $0.143871 \pm 0.000003$ & $0.15 \pm 0.02$ & $10.0 \pm 2.0$ & $87.0 \pm 2.0$ & \citet{Schaffenroth_2023_2} \\
GALEX~J080510.9-105834 & sdB$+$WD & $0.173703 \pm 0.000002$ & $0.039 \pm 0.004$ & $58.2 \pm 0.9$ & $29.2 \pm 1.3$ & \citet{Kawka_2015} \\
GD~1068 & sdB$+$MS & $0.258101 \pm 0.000003$ & $0.21 \pm 0.02$ & $-23.2 \pm 0.4$ & $86.5 \pm 0.5$ & \citet{Schaffenroth_2023_2} \\
HD~269696/AADor & sdO$+$MS & $0.2615397362 \pm 0.0000000008$ & $0.085 \pm 0.009$ & $1.6 \pm 0.1$ & $40.1 \pm 0.1$ & \citet{Kilkenny_2011MNRAS.412..487K} \\
ClMelotte20488 & sdB$+$dM & $0.26584 \pm 0.00004$ & $0.13 \pm 0.02$ & $70.5 \pm 2.2$ & $59.8 \pm 4.5$ & \citet{Kawka_2010MNRAS.408..992K} \\
CPD-64~481 & sdB$+$MS & $0.277263 \pm 0.000005$ & $0.046 \pm 0.003$ & $94.1 \pm 0.3$ & $23.9 \pm 0.1$ & \citet{Edelmann_2005AA...442.1023E} \\
TYC~4544-2658-1 & sdBV$+$dM & $0.3006 \pm 0.0002$ & $0.093 \pm 0.009$ & $-28.6 \pm 1.2$ & $41.9 \pm 1.3$ & \citet{Silvotti_2022MNRAS.511.2201S} \\
GD~1110 & sdB$+$dM & $0.3128 \pm 0.0007$ & $0.023 \pm 0.002$ & $10.0 \pm 2.0$ & $12.8 \pm 0.1$ & \citet{Schaffenroth_PHL457_2014AA...570A..70S} \\
EC~11575-1845 & sdO$+$dM & $0.32762 \pm 0.00001$ & $0.09 \pm 0.02$ & $18.0 \pm 2.0$ & $50.0 \pm 10.0$ & \citet{Exter_2005MNRAS.359..315E} \\
GALEX~J2205-3141 & sdB$+$MS & $0.3415430 \pm 0.0000008$ & $0.11 \pm 0.01$ & $-25.1 \pm 1.0$ & $46.0 \pm 1.0$ & \citet{Kawka_2015} \\
Feige~36 & sdB$+$WD & $0.35386 \pm 0.00006$ & $0.43 \pm 0.05$ & $-0.8 \pm 0.9$ & $134.6 \pm 1.3$ & \citet{Moran_1999MNRAS.304..535M} \\
PG~1232-136 & sdB$+$WD & $0.3630 \pm 0.0003$ & $0.37 \pm 0.03$ & $4.1 \pm 0.3$ & $129.6 \pm 0.0$ & \citet{Kupfer_2015} \\
$[$CW83$]$1419-09 & sdO$+$WD & $0.41780 \pm 0.00002$ & $0.36 \pm 0.04$ & $42.3 \pm 0.3$ & $109.6 \pm 0.4$ & 
\citet{Edelmann_2005AA...442.1023E} \\
FBS~2347$+$385 & sdB$+$WD & $0.462516 \pm 0.000005$ & $0.25 \pm 0.02$ & $2.0 \pm 1.0$ & $87.9 \pm 2.2$ & \citet{Kawka_2010MNRAS.408..992K} \\
KUV~16256$+$4034 & sdB$+$WD & $0.47760 \pm 0.00008$ & $0.082 \pm 0.009$ & $-90.9 \pm 0.9$ & $38.7 \pm 1.2$ & \citet{Copperwheat_2011MNRAS.415.1381C} \\
PG~1544$+$488 & He-sdOB$+$He-sdB & $0.496 \pm 0.002$ & -- & $-25.5 \pm 0.4$ & $86.6 \pm 0.5$ & \citet{Jeffery_2014MNRAS.440.2676S} \\
PG~1519$+$640 & sdB$+$WD & $0.540291430 \pm 0.000000003$ & $0.12 \pm 0.01$ & $0.1 \pm 0.4$ & $42.7 \pm 0.6$ & \citet{Edelmann_2005AA...442.1023E} \\
Feige~11 & sdB$+$WD & $0.569899 \pm 0.000001$ & $0.33 \pm 0.03$ & $7.3 \pm 0.2$ & $104.7 \pm 0.4$ & \citet{Geier_2008AA...477L..13G} \\
HD~188112 & sdB$+$WD & $0.60658584 \pm 0.00000005$ & $0.83 \pm 0.03$ & $26.6 \pm 0.2$ & $188.7 \pm 0.2$ & \citet{Latour_2016AA...585A.115L} \\
PB~9286 & sdB$+$WD & $0.6641 \pm 0.0005$ & $0.35 \pm 0.04$ & $0.0 \pm 5.0$ & $93.9 \pm 5.0$ & \citet{Nemeth_2012MNRAS.427.2180N} \\
JL~82 & sdB$+$MS & $0.73710 \pm 0.00005$ & $0.099 \pm 0.009$ & $-1.6 \pm 0.8$ & $34.6 \pm 1.0$ & \citet{Edelmann_2005AA...442.1023E} \\
V*EQPsc & sdBV$+$dM & $0.80088 \pm 0.00005$ & $0.101 \pm 0.010$ & $25.9 \pm 1.3$ & $34.9 \pm 1.6$ & \citet{Baran_2019MNRAS.489.1556B} \\
EC~02200-2338 & sdB$+$WD & $0.80220 \pm 0.00007$ & $0.39 \pm 0.04$ & $20.7 \pm 2.3$ & $96.4 \pm 1.4$ & \citet{Copperwheat_2011MNRAS.415.1381C} \\
Ton~S183 & sdB$+$WD & $0.82770 \pm 0.00002$ & $0.30 \pm 0.03$ & $50.5 \pm 0.8$ & $84.8 \pm 1.0$ & \citet{Kupfer_2015} \\
EC~21556-5552 & sdB$+$MS/WD & $0.83400 \pm 0.00007$ & $0.27 \pm 0.03$ & $31.4 \pm 2.0$ & $65.0 \pm 3.4$ & \citet{Copperwheat_2011MNRAS.415.1381C} \\
PG~0918$+$029 & sdB$+$WD & $0.876790 \pm 0.000002$ & $0.25 \pm 0.03$ & $104.4 \pm 1.7$ & $80.0 \pm 2.6$ & \citet{Maxted_2001} \\
EC~12408-1427 & sdB$+$WD & $0.902430 \pm 0.000001$ & $0.21 \pm 0.02$ & $-52.2 \pm 1.2$ & $58.6 \pm 1.5$ & \citet{Copperwheat_2011MNRAS.415.1381C} \\
PG~1452$+$198 & sdB$+$WD & $0.964980 \pm 0.000004$ & $0.36 \pm 0.03$ & $-9.1 \pm 2.1$ & $86.8 \pm 1.9$ & \citet{Copperwheat_2011MNRAS.415.1381C} \\
GALEX~J225444.1-551505 & sdB$+$WD & $1.227020 \pm 0.000005$ & $0.39 \pm 0.04$ & $4.2 \pm 2.0$ & $79.7 \pm 2.6$ & \citet{Kawka_2015} \\
PG~0133$+$114 & sdB$+$WD & $1.237870 \pm 0.000003$ & $0.36 \pm 0.04$ & $-0.3 \pm 0.2$ & $82.0 \pm 0.3$ & \citet{Edelmann_2005AA...442.1023E} \\
CD-30~19716 & sdB$+$WD & $1.269780 \pm 0.000002$ & $0.40 \pm 0.03$ & $-2.9 \pm 1.0$ & $92.7 \pm 1.5$ & \citet{Morales_2003MNRAS.338..752M} \\
$[$CW83$]$1735$+$22 & sdO$+$WD & $1.278 \pm 0.001$ & $0.53 \pm 0.05$ & $20.6 \pm 0.4$ & $103.0 \pm 1.5$ & \citet{Edelmann_2005AA...442.1023E} \\
HD~171858 & sdB$+$WD & $1.632800 \pm 0.000005$ & $0.28 \pm 0.02$ & $62.5 \pm 0.1$ & $60.8 \pm 0.3$ & \citet{Edelmann_2005AA...442.1023E} \\
PB~7352 & sdB$+$WD & $3.621660 \pm 0.000005$ & $0.38 \pm 0.03$ & $-2.1 \pm 0.3$ & $60.8 \pm 0.3$ & \citet{Edelmann_2005AA...442.1023E} \\
CD-24~731 & sdOB$+$WD & $5.850 \pm 0.003$ & $0.52 \pm 0.05$ & $20.0 \pm 5.0$ & $63.0 \pm 3.0$ & \citet{Kupfer_2015} \\
PG~1032$+$406 & sdOB$+$WD & $6.7791 \pm 0.0001$ & $0.21 \pm 0.02$ & $24.5 \pm 0.5$ & $33.7 \pm 0.5$ & \citet{Morales_2003MNRAS.338..752M} \\
Feige~108 & sdB$+$WD & $8.74651 \pm 0.00001$ & $0.41 \pm 0.03$ & $45.8 \pm 0.6$ & $50.2 \pm 1.0$ & \citet{Edelmann_2005AA...442.1023E} \\
PG~1619$+$522 & sdOB$+$WD & $15.3578 \pm 0.0008$ & $0.39 \pm 0.03$ & $-52.5 \pm 1.1$ & $35.2 \pm 1.1$ & \citet{Maxted_2001} \\
Ton~13 & sdOB$+$WD & $15.58300 \pm 0.00005$ & $0.41 \pm 0.03$ & $-68.6 \pm 0.6$ & $41.5 \pm 0.8$ & \citet{Copperwheat_2011MNRAS.415.1381C} \\
\bottomrule
\end{tabular}
\tablefoot{Quoted uncertainties are symmetric and were taken directly from the literature.}
\end{table*}

\end{appendix}

\end{document}